\documentclass[aps,prb,superscriptaddress,showpacs,twocolumn,floatfix,amssymb]{revtex4-2}
\usepackage{graphicx,color,physics}
\usepackage{hyperref} % https://tex.stackexchange.com/a/532130
\usepackage{cleveref} 
\usepackage{float}

\usepackage[todonotes={textsize=small},defaultcolor=cyan,authormarkuptext=name,commentmarkup=todo]{changes}
\definechangesauthor[name=Alan]{al}
\definechangesauthor[name=Emilio]{e}
\definechangesauthor[name=Giuseppe]{g}
\definechangesauthor{0}
\definechangesauthor{1}
\definechangesauthor{2}
\definechangesauthor{3}
\definechangesauthor{4}
\definechangesauthor{5}
\definechangesauthor[name={}]{typo}

\let\oldcite\cite
\renewcommand{\cite}[1]{\mbox{\oldcite{#1}}}

\usepackage{siunitx} % Use SI units for numbers
\usepackage{dsfont}

\usepackage[caption=false]{subfig}
\newcommand{\phantomsubfloat}[1]{
	{% apply caption setup only temporarily
		\captionsetup[subfigure]{labelformat=empty}
		\subfloat[][]{#1}
	}%
}

\newcommand{\comsol}{COMSOL Multiphysics\textsuperscript{\textregistered}}
\newcommand{\hc}{\mathrm{h.c}}

\begin{document}
\title{Many-body symmetry-protected zero boundary modes\\ of synthetic photo-magnonic crystals}
%Synthetic Topological Magnonic Crystals}
%Quantum Phase Transitions and Symmetry-Protected Topological edge modes in a lattice Dicke model}

\author{Alan Gardin}
\affiliation{Quantum and Nano Technology Group (QuaNTeG), School of Chemical Engineering, The University of Adelaide, North Terrace Campus, Adelaide, 5000, South Australia, Australia}

\author{Emilio Cobanera}
\affiliation{Department of Physics, SUNY Polytechnic Institute, 
%100 Seymour Ave, 
Utica, NY 13502, USA}
\affiliation{Department of Physics and Astronomy, Dartmouth College, 6127 Wilder Laboratory, Hanover, New Hampshire 03755, USA}

\author{Giuseppe Carlo Tettamanzi}
\affiliation{Quantum and Nano Technology Group (QuaNTeG), School of Chemical Engineering, The University of Adelaide, North Terrace Campus, Adelaide, 5000, South Australia, Australia}

\begin{abstract}
The topological classification of insulators and superconductors, the ``ten-fold way", is grounded on fermionic many-body symmetries and has had a dramatic impact on many fields of physics. Therefore, it seems equally important to investigate a similar approach for bosons as tightly analogous to the fermionic prototype as possible. There are, however, several obstacles coming from the fundamental physical differences between fermions and bosons. Here, we propose a theory of free boson topology (topological classification and bulk-boundary correspondence) protected by bosonic many-body symmetry operations, namely, squeezing transformations, particle number, and bosonic time reversal. We identify two symmetry classes that are topologically non-trivial in one dimension. They include key models like the bosonic Kitaev chain, protected by a squeezing symmetry within our framework, and the celebrated bosonic SSH
model, protected by a squeezing symmetry and particle number. To provide a robust experimental platform for testing our theory, we introduce a new quantum meta-material: photo-magnonic crystals. They are remarkable for their experimental flexibility and natural affinity for displaying band topological physics at microwave frequencies. We engineer a many-body symmetry-protected topological photo-magnonic chain with boundary modes mandated by a Pfaffian invariant. Using an electromagnetic finite-element modelling, we simulate its reflection and transmission and identify experimental signatures of its boundary modes. The experimental tuning of the crystal to its symmetry-protected topological phase is also addressed. Our modelling of the photo-magnonic chain provides a thorough blueprint for its experimental realisation and the unambiguous observation of its exotic physics.
%As a result, we identify seven symmetry classes. 
%Five of these can be regarded as representatives of the Altland-Zirnbauer classes A, AI, BDI, and D, here emerging from many-body symmetry constraints placed on non-Hermitian bosonic dynamical matrices. 
%Two of the symmetry classes are topologically non-trivial in one dimension. They include key models like the the topological quantum amplifier known as the bosonic Kitaev chain (BKC), protected by a squeezing symmetry within our framework, and the celebrated bosonic SSH model, protected by a squeezing symmetry and particle number. 
\end{abstract}
\maketitle
\tableofcontents

\section{Introduction}
\paragraph{Topology in physics.} Topology -- the mathematical theory of 
the properties of a space that do not change under continuous deformations --
was an important but specialised tool in physics till the late 1970s. Up to
that point, it was mostly used for investigating non-linear field theories
in particle physics~\cite{RevModPhys.52.661}
and order parameters in condensed matter physics~\cite{RevModPhys.51.591}. 
However, after the discovery of the integer quantum Hall effect in
the 80s \cite{1986Klitzing}, topological physics took the centre 
of the stage and has remained there ever since~\cite{wilczek1989geometric}.  
Topological quantum physics in particular has now evolved into several branches, 
examining how topological facts can constrain different quantum objects such 
as ground state manifolds, higher-energy eigenstates, band structures, or full 
Hamiltonian operators. The various approaches reflect the different dominant 
features of various physical systems, such as the classical/quantum dichotomy, 
the relative importance of many-body effects (weak as opposed to strong interactions), 
or the dominant symmetries. 

A prototypical example of topological physics is the topological band theory 
developed first for systems of independent, ``free" fermions~\cite{2016Bansil} and
much more recently for \underline{stable} systems of free 
bosons~\cite{Zhou_2020, Chaudhary2021}. 
For independent particles, the many-body Hamiltonians are quadratic in the 
particle creation and annihilation operators and so their diagonalization
can be accomplished by diagonalising first a simpler associated
``single-particle" Bogoliubov-de Gennes ``Hamiltonian''~\cite{Blaizot1986}.  
Topology is encoded in the single-particle band structure which, for
gapped fermions, leads to the topic of topological insulators and superconductors 
\cite{2010Hasan,2011Qi}
and the calculation of topological invariants like Chern/winding numbers in
solid-state physics. For bosons, topology can be associated to the directional
transport and/or amplification of photons or other bosonic quasiparticles
by way of surface modes 
\cite{Shindou2013, Peano2016_PhysRevX.6.041026, Peano2018, 2018McDonald}.

By contrast, another example of topological physics, the fractional quantum Hall states 
(FQHs) \cite{1999Stormer}, exists only because of strong interactions and cannot
be reduced to band topology. FQHs are examples of intrinsic topological order 
\cite{KITAEV20032, KITAEV20062, 2017Wen}. Phases of matter with intrinsic topological 
order are characterised by specific properties such as long-range entanglement, excitations 
with fractional charges and/or exotic quantum statistics (``anyons"), 
and robust ground state degeneracy. From this point of view, systems 
which support non-trivial band topology fit the framework of 
symmetry-protected topologically (SPT) ordered phases of matter \cite{2015Senthil,2016Chiu}.
Here, contrary to intrinsic topological order, the low-energy exciations 
are short-range entangled, have a finite energy gap, and no fractional charges. 
We will not return to the subject of intrinsic topological quantum order
in the following. Our work in this paper fits better
the framework of topological band theory and the theory of
SPT phases of matter.  

\paragraph{Fermionic platforms}
For free fermions and stable free bosons, the universal consequence of 
non-trivial band topology in the bulk are bands of boundary-localised 
states for systems with a termination: this is the celebrated bulk-edge 
correspondence~\cite{2016Chiu, Ozawa2019}. These surface bands 
have all kinds of exotic transport properties and, in addition, 
after properly including symmetry protection,
they are provably robust against small 
continuous deformations, bulk disorder, and boundary perturbations. 
For this reason, they are strong candidates for becoming integrated
into quantum devices. In fact, the quantum Hall effect supports
the SI standard for resistance~\cite{2005Klitzing} and is an
integral part of fully commercialised technologies. Another example
is the still speculative Majorana qubit~\cite{Kitaev_2001}. Both of
these applications are sensible proposals because the boundary states mandated
by bulk topology are, in addition, symmetry protected. 

%\paragraph{Fermionic protecting symmetries of the tenfold way.}
Thus, it is natural to focus one's attention on the identification of the protecting symmetries.
Consider the most investigated case of the tenfold way~\cite{Kitaev_2009, Ryu_2010}. 
It can, and is often, described as a topological classification  of
insulators and superconductors based on three possible symmetry constraints defined
abstractly at the level of the Bogoliubov-deGennes Hamiltonian 
(namely time-reversal symmetry, particle-hole symmetry, and chiral symmetry).
Depending on the symmetry constraints met by a prototype Hamiltonian of the system 
considered, along with the space dimensionality, the periodic table of the tenfold way~\cite{PhysRevB.55.1142} predicts which kind of topological invariant 
(winding number $\mathbb Z$, Pfaffian $\mathbb Z_2$) characterises the band topology
of the system. The associated bulk-boundary correspondences are derived
by way of a set of mathematical results collectively known as 
``index theorems"~\cite{bleecker2013index}. The formulation
in terms of abstract single-particle symmetry constraints is powerful but it
leaves open the question of how and why these constraints arise in physical 
systems. Hence, the tenfold way classification and associated bulk-boundary correspondences
has been reformulated in terms 
of four many-body symmetries \cite{2016Kennedy, Alldridge2020, 2017Alase}: 
fermion number, spin $j=1/2$ rotations, (spinful) time reversal, and an anti-unitary 
transformation that exchanges fermionic creation and annihilation operators. 
These four many-body symmetries, in suitable combinations, yield the three 
characteristic single-particle symmetries. 
Grounding symmetry protection on many-body symmetries has significant experimental 
advantages: it makes diagnosing protection possible with less control over the 
microscopic Hamiltonian.  

\paragraph{Bosonic platforms.}
In this paper we investigate in depth
the the manifestations of band topology, 
physical symmetry protection, and bulk-boundary correspondences 
in photo-magnonic systems. As we just saw, for free fermions, 
these three ingredients come together perfectly thanks to the 
classical index theorems. For free bosons, the situation is far 
more complex and currently open-ended. Experimentally, topology 
has been observed to play a role in 
photonic lattices \cite{2020Carusotto,2020Clark,2023Wang,2024Qian,2025Parto}, 
circuit QED \cite{2024Busnaina}, and cavity opto-mechanics \cite{2024Slim}. 
These systems naturally display non-hermitian physics~\cite{Ashida02072020} 
and have motivated and stimulated 
research in non-hermitian topology~\cite{Ashida02072020,2022Dinga,2023Okuma}. 
In addition, magnons -- the bosonic quasi-particle associated 
with spin waves~\cite{2009Stancil} -- are also actively investigated in the context of bosonic 
and non-hermitian topology \cite{2020Malki,2023YuZouZengRaoXia,2023ZhuoKangManchonCheng}.

In spite of all this work, cavity magnonic systems \cite{2021Rameshti}, 
notable for demonstrating the hybridisation of photons with magnons,
have not been investigated from the point of view of
band topology. This is despite their natural tendency for non-hermitian physics, 
notably due to the flexibility in engineering dissipative coupling \cite{2020Wang} 
and exceptional points (EPs) \cite{2017Harder,2025Lambert}. 
One basic challenge is scaling up the methods and techniques 
of cavity magnonics to crystal-like arrays. This is one of the 
challenges addressed in this paper. We call the resulting
multi-cavity system a photo-magnonic crystal. One can think 
of it as a new kind of quantum meta-material. 

Photo-magnonic crystals have two natural features that promise 
interesting topological band physics. First, they naturally mix very 
``heavy" excitations, the magnons, with very mobile excitations, 
the photons. (Unlike the magnons, the photons can leak between cavities.)
Second, it is possible to engineer synthetic background gauge fields 
in each unit cell of a photo-magnonic crystal \cite{Gardin2023}, as 
experimentally demonstrated in Ref.~\cite{2023Gardin}. 
In this light, photo-magnonic crystals are strongly reminiscent of 
topological Kondo insulators~\cite{PhysRevLett.104.106408} and other
heavy-fermion systems with non-trivial band topology. This is both
exciting and concerning because it suggests that the existing tools 
of non-Hermitian and/or bosonic topological physics and bosonic 
topological physics may not fit photo-magnonic crystals well. Rather,
in analyzing photo-magnonic crystals, one would hope to deploy a tight 
bosonic analogue of the tenfold way plus a physical set of protecting 
symmetries. Unfortunately, it is not clear what this ``bosonic analogue of
the tenfold way"
should be. Fermions and bosons are, after all, fundamentally different. 
For instance, bosonic systems have no ``Fermi sea'', and can naturally 
exhibit dynamics akin to those found in non-hermitian systems without dissipation \cite{Flynn2020}. 
Hence, it has been hard to ground the topological physics of weakly-interacting 
bosons on many-body symmetries. 

One contender for a comprehensive theory of bosonic
band topology is the topological classification of general non-hermitian 
systems \cite{PhysRevX.9.041015}. Unfortunately, this classification 
(into 38 classes) does not mesh well with the bulk-boundary correspondence.
To quote Ref.~\cite{Ashida02072020}, ``[...] a [topological]
classification [of non-Hermitian systems] often fails to correctly predict
the presence of topological edge states because the break-down of the 
bulk-boundary correspondence is rather common in non-Hermitian systems."
%, where up to 38 topological classes were found. (I think
% the final number was 40 after an erratum came out)
Other contenders focus exclusively on dynamically stable or even just thermodynamically
stable free boson systems \cite{Peano2018, Zhou_2020, Chaudhary2021}.
There are a couple of problems with these approaches. First,
stability constraints do not mesh well with index theorems, but
without them there is no good way to uncover \underline{robust} 
bulk-boundary correspondences. (Ref.~\cite{Peano2018} manages to 
bypass this problem by zooming in on two-dimensional systems). 
Second, by imposing the stiffer constraint of thermodynamic 
stability~\cite{Peano2018, Zhou_2020}, one sacrifices the tight 
link between topological transitions and quantum phase transitions, which is a 
hallmark of the fermionic tenfold way; see Ref.~\cite{PhysRevB.102.125127} 
for a more detailed discussion. (This problem does affect Ref.~\cite{Peano2018}.)
With hindsight,
it is surprisingly tricky to preserve this link for bosons; see 
Ref.~\cite{Mariam2025} for more information. 

%Trying to find the best path through this obstacle course, 
In this paper, we propose a framework which balances band topology, bulk-boundary
correspondences, and physical symmetry protection for free bosons
by a) introducing a symmetry classification built directly on 
physical many-body symmetries, and b) 
by invoking index theorems for both identifying 
the topologically  non-trivial symmetry classes and establishing
the associated bulk-boundary correspondences (in one dimension for simplicity;
we will  explore higher dimensions in a forthcoming publication). 
As a result, within our framework, the bulk-boundary
correspondence never breaks down even though bosonic systems can be
effectively non-Hermitian~\cite{Flynn2020}, the protecting symmetries 
are explicit physical many-body symmetries from the start, and we need not
make any stability assumptions. Quantum amplifiers like the bosonic Kitaev chain \cite{2018McDonald} fit our picture just as well as dynamically stable systems.  

\paragraph{Outline and reading guide.}
To summarize, there are two important problems we aim to address 
in this paper:
first, establish by way of careful simulation 
the experiment feasibility of photo-magnonic crystals; and
second, develop an optimal theory, rooted on physical 
symmetries and index theory, for investigating
their topological physics and symmetry protection. We begin by 
introducing this new  experimental platform of bosonic
physics in~\cref{sec:photo-magnonic-crystals}. 
Taking the example of a one-dimensional photo-magnonic crystal, we show that 
their design and experimental signatures can be characterized with remarkable 
accuracy (compared to real experiments) using electromagnetic finite-element 
modelling. Furthermore, the essential physics of these systems can also be 
studied within a quantum Hamiltonian formalism, with excellent agreement with 
classical modelling. 
In~\cref{sec:topological-crystal}, according to our intuition about photo-magnonic 
crystals, we find that a topologically mandated mode localised on the edge of the 
chain can appear by modifying the boundary conditions of the crystal. 
Further theoretical analysis explains the presence of this edge mode in topological
terms, but numerical investigations show that it is not protected by simple
symmetries. Not knowing what the protecting symmetries are, there is no way 
to know how the boundary mode will respond to various perturbations.

To address this problem, we first review the existing approaches for topological 
classifications of free bosons in \cref{sec:context-strategy}.
This allows us to establish and motivate our strategy, which consists of dropping 
the constraint of stability (this is how we manage to deploy index
theorems) and focus on physical bosonic symmetries as the foundation
of topological classifications. Next, we develop a many-body symmetry classification 
of bosonic systems in \cref{sec:topological-classification}, based on three many-body 
symmetry operations: physical bosonic time reversal, particle number, and a squeezing
operation. An important result is that we find two symmetry classes that are topologically 
non-trivial in one dimension. The symmetry class associated with a squeezing operation 
is characterised by a \textit{winding number} invariant.
The symmetry class associated with both a squeezing and particle number symmetry 
is characterised by a $\mathbb Z_2$-valued \textit{Pfaffian} invariant. 
For suitable parameter regimes, these classes cover not only the one-dimensional 
photo-magnonic chain, but also two important bosonic topological models -- the bosonic 
Su-Schrieffer-Heeger (SSH) model~\cite{2024Qian} and the bosonic Kitaev 
chain~\cite{2018McDonald} -- which have both been experimentally realised.

Next, \cref{subsec:topological-classification-summary} provides a concise overview 
of the main results of \cref{sec:topological-classification}, highlighting how the 
formalism can be used in practice. This is illustrated by the fully worked-out
examples of the bosonic  Kitaev chain and the bosonic SSH model. Finally, 
in \cref{sec:dicke-chain-revisited}, 
we focus on the symmetry-protected phase of the 1D photo-magnonic crystal of 
\cref{sec:photo-magnonic-crystals}, and discuss how to experimentally realise it. 
We conclude in \cref{sec:conclusion} with a summary of the results and ideas
for future work.

We propose two reading paths, depending on the background and interest of the reader. 
An experimentalist already familiar with cavity magnonics can skim through \cref{sec:photo-magnonic-crystals}, and refer to \cref{eq:H0:bosonic} and \cref{fig:schematic:H0} to learn
about the photo-magnonic crystal. They can then read about topological aspects 
in \cref{sec:topological-crystal,sec:context-strategy}. Instead of reading the entirety of \cref{sec:topological-classification}, we recommend only reading the summary in 
\cref{subsec:topological-classification-summary}. The remainder of the manuscript can 
be read normally. For a theorist working in topological physics, it is possible to 
jump directly to \cref{subsec:topological-crystal:effective-hamiltonian}, and proceed 
linearly from there. The main objective of the preceding sections is to show that the 
effective Hamiltonian introduced in \cref{subsec:topological-crystal:effective-hamiltonian} 
is a very good model for true experiments, and explains why the proposed experimental 
platform has attractive features for topological physics.

\section{\label{sec:photo-magnonic-crystals} 
Photo-magnonic crystals}
In this section we introduce the simplest photo-magnonic crystal in one dimension, maintaining a close contact with the established 
experimental capabilities of cavity magnonics. 
In particular, we simulate our experimental proposal using an electromagnetic finite-element software, and the input-output formalism based on a Hamiltonian formalism.
As expected, both modelling techniques concur, and provide an excellent estimation of realistic experimental observables.
Thus, these methods allows modelling and engineering different aspects of the crystal.
%which reveals the presence of edge modes depending on the boundary 
%conditions. In order to investigate this edge mode theoretically, 
%we move to a Hamiltonian formalism. 
%Combined with the input-output formalism, we reproduce the results of the classical electromagnetic simulations, but with a quantum model.
%This allows us to justify the presence of this edge mode, and study its response to a smooth deformation of the crystal.

%In this section, we demonstrate how the Dicke chain can be implemented using a chain of cavity magnonics systems. We recall that we introduced the basics of a a single cavity magnonics unit cell in \cref{subsec:cavity-magnonics}, which naturally realises a Dicke model. Thus, here we detail the coupling of neighbouring cavities with each other, resulting in a Dicke chain. This proposal allows implementing the normal phase of the anisotropic Dicke model with $g=\Delta$. In the next section, we will detail additional ingredients to potentially realise the anisotropic version with $g \neq \Delta$, as well as higher-dimensional topological lattice models.

\subsection{\label{subsec:cavity-magnonics} Fundamentals of cavity magnonics} 
The subject of cavity magnonics is the interaction between the electromagnetic
field confined to a cavity and a biased ferromagnet in the cavity~\cite{2021Rameshti}. 
These hybrid systems often operate in the microwave frequency range. 
Hence, the photonic degrees of freedom can be designed to specification using standard microwave engineering techniques~\cite{2011Pozar}. 
The ferromagnet is commonly approximated as a macrospin undergoing Larmor precession in response to a tunable static magnetic field $\vb{H}_0$; the role of this tunable field then is to bias the macrospin to lie along its direction. 
These features make cavity magnonics systems highly tunable for a large variety of experiments.

\subsubsection{Theory of magnon-photon interaction}
In this section we provide a concise technical introduction to cavity magnonics systems. A classical formalism can be employed to model these systems, based on coupling Maxwell's equations with the Landau-Lifschitz-Gilbert equation \cite{2021Macedo}, or lumped element models \cite{2021Rameshti}. 
Here, we will detail the quantum Hamiltonian approach, because it is the formalism used for the topological theory developed in \cref{sec:topological-classification}. We first discuss separately the confined light degrees of freedom and their magnonic counterpart, provided by the ferromagnetic insertion. 
Next, we investigate the interaction between the two systems.
Finally, we describe briefly the complete experimental setup.

\paragraph{Cavity modes.}
Let us confine the electromagnetic field to a cavity of volume $V_{ph}$, 
also called a resonator, whose dimensions are chosen to position the low excitation energy electromagnetic modes in the microwave frequency range (for us, a few GHz). 
Each cavity mode $j$ is associated to a bosonic annihilation operator $a_{j}$~\footnote{Contrary to refs.~\cite{Gardin2023,2023Gardin}, we use  $a_{j}$ to denote the cavity modes instead of $c_{j}$.} that removes photons of frequency $\omega_{a,j}/2\pi$ from the cavity. 
The quantised magnetic field, in the Schrödinger picture, is
given by the formula~\cite{2019Flower}
\begin{equation}
    \label{eq:quantised-magnetic-field}
       \vb{b}(\vb{r}) = \sum_{j}\sqrt{\frac{\hbar}{2\epsilon_{0}\omega_{a,j}} }(a_{j}+ a_{j}^\dagger) \nabla \times \vb{u}_{j}(\vb{r}),
\end{equation}
where $\vb{u}_j $ captures the spatial dependence of the cavity modes. The frequency, 
$\omega_{a,j}/2\pi$, and the spatial dependence, $ \vb{u}_{j} $, of each cavity mode 
$j$ is determined by the geometry of the cavity. They can be calculated numerically using a 
finite-element modelling software such as \comsol,
see for instance Ref.~\cite{2020Bourhill}.

The dynamics of the confined electromagnetic degrees of freedom, by themselves, is controlled by the electromagnetic Hamiltonian 
\begin{equation}
    H_{EM}=
\int_{V_{ph}}d^3\vb{r}\,\left(
\frac{\epsilon_0}{2}
\vb{e}^2+\frac{1}{2\mu_0}\vb{b}^2 \right).
\end{equation}
After discarding the energy of the vacuum, it boils down to
\begin{equation}
    H_{EM}=\sum_j\hbar \omega_{a,j}a_{j}^\dagger a_{j},
\end{equation}
and so the electromagnetic modes behave like a system of decoupled quantum harmonic oscillators. 
We recall that, assuming the walls of the cavity to be perfect electric conductors, with $\vu{n}$ a unit vector normal to the cavity walls, the electromagnetic field satisfies the conditions
\begin{align}
    \vu{n} \cdot \vb{e} = \rho_s, \quad \vu{n} \times \vb{e} = 0\\
    \vu{n} \cdot \vb{b} = 0, \quad \vu{n} \times \vb{b} = j_s
\end{align}
where $\rho_s$ and $j_s$ are the electric surface charge density and current density, respectively \cite{2011Pozar}.

\paragraph{Magnon modes.}
The magnon degrees of freedom enter the picture by including magnetically ordered matter in the cavity. 
The insertion of choice is a small sphere of yttrium-iron garnet (YIG), a ferrimagnetic insulator. 
The microscopic spins in the YIG sphere are biased by an applied static, classical magnetic field 
$\vb{H}_0 = H_{0} \vu{z}$. We are not including the cavity quantum magnetic field yet.  
While several spin waves, that is, magnon modes can exist for a magnetised sphere~\cite{2009Stancil}, the Kittel mode associated with the uniform precession of all the microscopic spins is the one that is the easiest to control. 
For a spherical body of ferromagnetic matter, the frequency of the Kittel mode is related to the biasing magnetic field by the formula~$\omega_{m} = \gamma \mu_{0} \abs{\vb{H}_0}$. 
The energy of the Kittel mode is described by the Hamiltonian
$
H_{Kittel} = -\omega_{m} S_{z},$
where $S_{z}$ is the $\vu{z}$ 
component of the macrospin, that is,
total spin, operator $\vb{S}$.

For a ferromagnetic material 
hosting a large number $N_{s}$ of spins with (dimensionless) value of the spin $s$, the maximum value of the macrospin $S_{max}=N_{s}s$ is large enough to warrant the use of the Holstein-Primakoff~\cite{1940Holstein} representation for the spin algebra in terms of canonical bosons. In this representation,
if $m$ is a bosonic annihilation operator, and $S^{\pm}$ are the macrospin raising and lowering operators, then 
\begin{align}
    \label{eq:holstein-primakoff}
    S_z &= \hbar (S- m^\dagger m),\\
    S^+ &= S_{x} + iS_{y}= \hbar \sqrt{2S - m^\dagger m}\, m \simeq \hbar \sqrt{2S}\, m,\\
    S^- &= S_{x} - iS_{y}=\hbar m^\dagger\,
     \sqrt{2S - m^\dagger m}
    \simeq \hbar \sqrt{2S}\, m^\dagger.    
\end{align}
The expansion to leading order in $1/2S$
yields the Holstein-Primakoff approximation of the 
magnetic Hamiltonians by way of a free boson Hamiltonian. 

The replacement of angular momentum operators with bosonic creation and destruction operators is justified if $\expval{m^\dagger m} \ll 2S$. 
This condition is readily satisfied for macroscopic samples of YIG due to its remarkably high spin density, to bet, $n_{s}=4.22 \times 10^{27}$ m$^{-3}$ \cite{2020Bourhill} Suitably grouped, the microscopic constituents of the YIG can be described as a Heisenberg ferromagnet of spins $s=\frac{5}{2}$. Hence, for the YIG spheres we consider throughout, which typically have a radius $r \simeq 0.1$ mm, we find that $2S = 3.5 \times 10^{16}$.
Hence, $2S \gg \expval{m^\dagger m}$ in the thermal state all the way to room temperature. 
This is the reason why, after discarding the energy of the magnonic vacuum, the magnon mode hosted by a YIG sphere is well-modelled by the Hamiltonian 
\begin{equation}
    H_{\text{Kittel}}=- \omega_m S_z \approx \hbar \omega_{m} m^\dagger m.
\end{equation}

\paragraph{Magnon-photon coupling in the microwave frequency range.}
Generally, magnons and photons interact through the Zeeman or the Faraday effect. 
The Zeeman effect is the dominant one in the microwave range, favoured by the dimensions of the cavities under discussion. 
The Zeeman effect is described by the effective interaction Hamiltonian 
\begin{equation}
    H_I =  \int_{V_{ph}} \dd[3]{\vb{r}}\, \gamma\vb{s}(\vb{r}) \cdot \vb{b}(\vb{r})= 
    \gamma\frac{  \vb{S} }{V_m} \cdot \int_{V_m} \dd[3]{\vb{r}}\,  \vb{b}(\vb{r}),
\end{equation}
where $V_m$ is the volume of the YIG sphere, 
$\vb{b}$ is the quantised magnetic field
in the cavity, see 
Eq.~\cref{eq:quantised-magnetic-field}, 
and $\vb{s}$ is the spin density. 
Setting  
$ \vb{s}(\vb{r})\approx\vb{S}/V_m $
throughout the YIG sphere and zero elsewhere amounts to describing the magnetic degrees of freedom in terms of the Kittel magnon mode only.  

In what follows, we will assume that the cavity has been designed so that the $\vu{z}$ component of the confined magnetic field $\vb{b}$ is negligible (this will be the case for our experimental proposal). 
In addition, we will similarly assume that there is one electromagnetic mode that is both well separated in frequency from other modes and couples the strongest to the Kittel magnon model. 
Then, as far as the interaction between photons and magnons is concerned, we can approximate the cavity quantised magnetic field in terms of this one mode $a$ as 
\begin{equation}    
    \vb{b}(\vb{r}) \approx \sqrt{ \frac{\hbar}{2\epsilon_{0}\omega_{a}} } (a + a^\dagger ) \nabla \times \vb{u}_a(\vb{r})
\end{equation}
The next step is to rewrite the macrospin in terms of spin raising/lowering operators.
%$S^\pm = S_x \pm i S_y$, 
Letting $\beta_x$ and $\beta_{y}$ stand for $\beta_x= (\grad \times \vb{u}_a) \cdot \vu{x}$ and $\beta_{y} = (\grad \times \vb{u}_a) \cdot \vu{y}$, one finally obtains that
\begin{align}
    H_I 
    %&= \frac{\gamma}{2V_m} \sqrt{ \frac{\hbar}{2\epsilon_{0}\omega_{c}} } \int_{V_m} \dd[3]{r} \qty[ \qty(S^+ +S^-)b_x -i\qty(S^+-S^-)b_y] (c+c^\dagger)\\
    =
    %\nonumber
    %&\frac{\gamma}{2V_m} \sqrt{ \frac{\hbar}{2\epsilon_{0}\omega_{a}} } \int_{V_m} \dd[3]{\vb{r}} \qty[S^{+}\qty(\beta_{x}-i\beta_{y}) + \hc] (a+a^\dagger)=\\
     \qty(\widetilde{g} S^{+} + \widetilde{g}^* S^{-})\qty(a+a^\dagger),
\end{align}
provided the coupling strength is given by the formula
\begin{equation}
    \label{eq:coupling-strength:macro}
    %\frac{G}{2\pi} 
    \widetilde{g}=\frac{\gamma}{2V_m} \sqrt{ \frac{\hbar}{2\epsilon_{0}\omega_{a}} } \int_{V_{m}} \dd[3]{\vb{r}} \qty(\beta_{x}(\vb{r})-i \beta_{y}(\vb{r})).
\end{equation}

\paragraph{The effective Hamiltonian.}
Within our setup, the effective Hamiltonian of the cavity with the magnetic insertion is the sum of the interaction Hamiltonian and the Hamiltonians of the electromagnetic and matter degrees of freedom by themselves. 
It is interesting to stop short of the Holstein-Primakoff approximation for a second. 
Then, the cavity plus insertion effective Hamiltonian is 
\begin{equation}
    \label{eq:magnon-photon-hamiltonian:angular}
    H = \hbar\omega_{a}a^\dagger a -\omega_{m}S_{z} + \qty(\widetilde{g} S^{+} + \widetilde{g}^{*} S^{-})\qty(a+a^\dagger),
\end{equation}
having left out
all the photon modes that do not couple appreciably to the Kittel magnon modes.
This Hamiltonian is precisely the celebrated Dicke model~\cite{PhysRev.93.99}. However, the Holstein-Primakoff approximation holds very naturally for our setup. Hence, in this paper, we will take the effective Hamiltonian to be 
\begin{equation}
    \label{eq:magnon-photon-hamiltonian:bosonic}
    H = \hbar\omega_{a}a^\dagger a + \hbar \omega_{m} m^\dagger m + \hbar (gm + g^* m^\dagger)(a+a^\dagger)
\end{equation}
in terms of the coupling strength 
\begin{equation}
    \label{eq:coupling-strength}
    \frac{g}{2\pi} =\sqrt{2S} \frac{\widetilde{g}}{2\pi}.
    %\frac{\gamma}{2V_m} \sqrt{ \frac{\hbar S}{\epsilon_{0}\omega_{c}} } \int_{V_{m}} \dd[3]{\vb{r}} \qty(b_{x}(\vb{r})-i b_{y}(\vb{r})).
\end{equation}

\paragraph{Cavity magnon polaritons.}
The normal modes of \cref{eq:magnon-photon-hamiltonian:bosonic} are hybrid photon-magnon quasi-particles. 
In the context of cavity magnonics, they are called cavity magnon polaritons \cite{2019Flower}. 
Let us investigate these quasi-particles in some more detail. We will take the opportunity to also introduce the fundamentals of the formalism used for the topological analysis of \cref{sec:topological-classification}.

Rather than starting directly with \cref{eq:magnon-photon-hamiltonian:bosonic}, let us consider the most general Hamiltonian quadratic in two mutually commuting bosonic modes, $a$ and $b$,
\begin{equation}
\begin{aligned}
    \label{eq:unit-cell-general-qbh}
    \frac{{\cal H}}{\hbar}
    &= \omega_a a^\dagger a + \omega_b b^\dagger b + (g ab^\dagger+ \delta ab + \eta a^2 + \xi b^2 + \hc)
\end{aligned}
\end{equation}
where $\qty[a,a^\dagger]=\qty[b,b^\dagger]=1$.
Defining 
\begin{eqnarray}
    \label{eq:PHI}
    \Phi^\dagger = \qty[a^\dagger, b^\dagger, a, b],
\end{eqnarray}
the quadratic bosonic Hamiltonian (QBH) $\mathcal{H}$ is equivalent, up to a constant energy offset, to $$
\widehat{G}=\frac{1}{2}\Phi^\dagger G\tau_3\Phi.
$$
The matrix $G$ plays a central role.
It is called the bosonic dynamical matrix. Explicitly,
\begin{equation}
\label{eq:hint-G-structure}
    G = \hbar\mqty[
    \mqty(
        \omega_a & g^*\\
        g & \omega_b
    )
    & \mqty(
        \eta^* & \delta^* \\
        \delta^* & \xi^*
    )\\
    \mqty(
        \eta & \delta \\
        \delta & \xi
    ) & 
    \mqty(
        -\omega_a & -\delta^* \\
        -g & -\omega_b
    )
    ] = \mqty[K & -\Delta\\ \Delta^* & -K^*]
\end{equation}
in terms of the blocks 
\begin{equation}
    K = \hbar \mqty[
    \omega_a & g^*\\
    g & \omega_b
    ], \quad \Delta = \hbar\mqty[
        \eta & \delta \\
        \delta & \xi
    ].
\end{equation}
In the remainder of the manuscript, we will refer to $K$ and $\Delta$ as the hopping and pairing matrices, respectively. In quantum optics language, $K$ accounts for the (number-conserving) co-rotating terms, while $\Delta$ describes counter-rotating terms. 

The matrix $G$ is called the dynamical matrix because it captures concisely the Heisenberg equations of motion. To see this, notice on one hand that
\begin{align*}
    \dot{a} 
    &= \frac{i}{\hbar} \qty[\mathcal{H}, a] = i \qty(\omega_a a - g^* b - \delta^* b^\dagger - 2\eta^* a^\dagger),\\
    \dot{b} 
    &= \frac{i}{\hbar} \qty[\mathcal{H}, b] = i \qty(\omega_b b - g a - \delta^* a^\dagger - 2\xi^* b^\dagger),\\
    \dot{a}^\dagger 
    &= \frac{i}{\hbar} \qty[\mathcal{H}, a^\dagger] = i \qty(\omega_a a + g b + \delta^* b^\dagger + 2\eta a^\dagger),\\
    \dot{b}^\dagger
    &= \frac{i}{\hbar} \qty[\mathcal{H}, b^\dagger] = i \qty(\omega_b b + g a + \delta^* a^\dagger + 2\xi b^\dagger).
\end{align*}
On the other hand, recalling Eq.~\eqref{eq:hint-G-structure} and Eq.~\eqref{eq:PHI}, these four equations can be compactly re-written as
\begin{equation}
    \label{eq:HeisenberG}
    \dot \Phi = G \Phi,
\end{equation}
with $G$ defined above. 
%An arbitrary scalar factor on each row of $\Phi$ can be considered to define a bosonic form $\hat{v}=\Phi^\dagger v$, see \cref{eq:dynamics-G} in \cref{subsec:formalism-qbh}.
If follows from Eq.~\eqref{eq:HeisenberG} that the diagonalisation of the matrix $G$ 
determines the spectrum of resonant frequencies (eigenvalues) and the normal modes (eigenvectors) of the system.% pecified by $\mathcal{H}$. 
%Indeed, the eigenvectors describe the hybridisation between the two modes $a$ and $b$, and indicate the hybridisation ratios. As for the eigenvalues, it is worth noting that they come in positive/negative pairs. For the example of two modes considered above, we thus obtain two positive eigenvalues $\omega_\pm$, and two negative eigenvalues$-\omega_\pm$.

Applying this recipe to the magnon-photon Hamiltonian of \cref{eq:magnon-photon-hamiltonian:bosonic}, we can predict the spectrum of the hybridised modes, that is, the cavity magnon polaritons. The hopping and pairing matrices are
\begin{equation}
    K = \hbar \mqty[
    \omega_a & g^*\\
    g & \omega_m
    ], \quad \Delta = \hbar\mqty[
        0 & g \\
        g & 0
    ].
\end{equation}
Then, the eigenvalues of the dynamical matrix $G$ of \cref{eq:hint-G-structure} are \cite{2019Flower}
\begin{equation}
    \omega_\pm^2 = \frac{\omega_a^2 + \omega_m^2}{2} \pm \frac{\sqrt{\qty(\omega_a^2-\omega_m^2)^2 + 16 \omega_a \omega_m \abs{g}^2}}{2},
\end{equation}
and consists of a pair of positive eigenvalues $\omega_\pm$, and another pair of negative eigenvalues $-\omega_\pm$. In the limit of low coupling strength compared to the mode's frequencies, $\abs{g}\ll \omega_a,\omega_c$, we can perform a rotating wave approximation (RWA), and the Hamiltonian of \cref{eq:magnon-photon-hamiltonian:bosonic} reduces to 
\begin{equation}
    \label{eq:magnon-photon-hamiltonian:bosonic:rwa}
    H = \hbar\omega_{a}a^\dagger a + \hbar \omega_{m} m^\dagger m + \hbar (g^* am^\dagger + ga^\dagger m),
\end{equation}
where the counter-rotating terms $am + a^\dagger m^\dagger$ have been neglected. As a result, the pairing matrix $\Delta$ vanishes, and now the Hamiltonian conserves the number of excitations $N = a^\dagger a + m^\dagger m$, since $\qty[N,H]=0$. As a result, the system can be described solely based on the hopping matrix $K$. The eigenvalues of $K$ are
\begin{equation}
    \omega_\pm = \frac{\omega_a + \omega_m}{2} \pm \frac{\sqrt{\qty(\omega_a-\omega_m)^2 + 4 \abs{g}^2}}{2}.
\end{equation}
and correspond to the frequencies of the cavity magnon-polaritons. The eigenvectors 
\begin{align}
    \label{eq:magnon-photon-hamiltonian:polaritons}
    p_\pm & \propto g^* a -(\omega_a-\omega_\pm)m
\end{align}
of $K$ describe the cavity magnon polariton creation/annihilation operators.
Thus, as the magnon's mode frequency $\omega_m/2\pi$ is swept by the static magnetic field $H_0$, we find a standard energy level repulsion, the signature of coherent coupling. Similarly, upon normalising the polariton operators of \cref{eq:magnon-photon-hamiltonian:polaritons}, we find that $\omega_m/2\pi$ effectively tunes the ``weight'' of each operator, i.e. it controls the degree of hybridisation between the photons and magnons.
Let us now discuss the experimental signature of this physics.

%Using the parameters inferred from the finite-element simulations of the unit cell, i.e. $\omega_a/2\pi = 9.998$ GHz and $g/2\pi=112.5$ MHz, we diagonalise $G$ numerically as a function of $\omega_m/2\pi$, and we plot the positive eigenvalues $\omega_\pm/2\pi$ as red dashed lines in \cref{fig:N1:H0:S21}. {\color{blue} make next less cryptic?} Reassuringly, they match the observed resonances of the transmission.
%{\color{blue} Should we include the quasi-particle modes? They are the ``magnonic cavity polaritons" (terrible name)}

\subsubsection{\label{subsec:implementation-Dicke-unit-cell}Standard experimental implementation and modelling}
In this section, we describe the basics elements composing a cavity magnonics experiment. We also show how the latter can be simulated to good accuracy with a classical finite-element modelling of the electromagnetic field, and that the results agree with the Hamiltonian formalism discussed above.

\begin{figure}[t]
    \centering
    \includegraphics[width=\columnwidth]{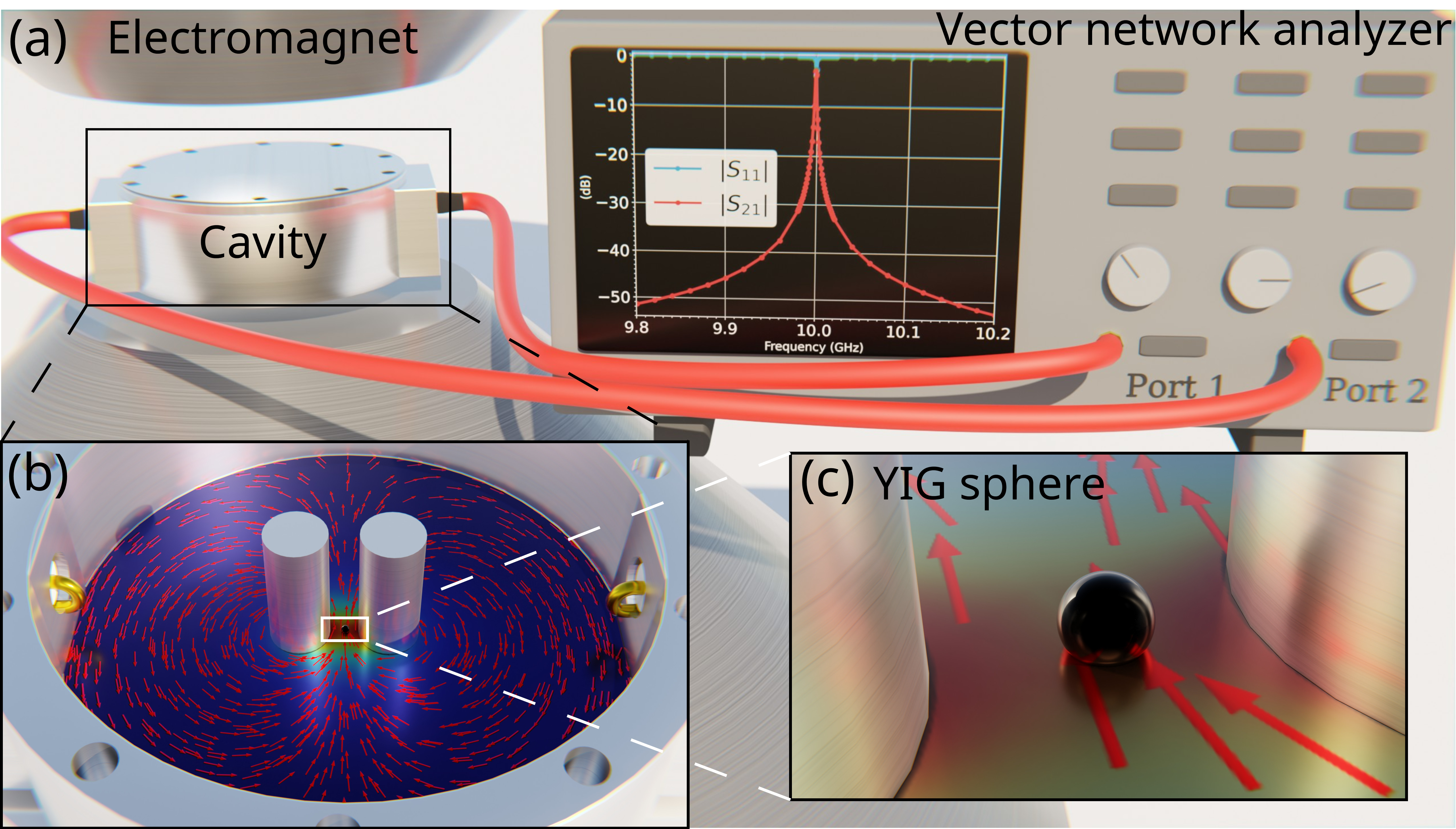}
    \phantomsubfloat{\label{fig:cavity-magnonics-experiment:setup}}
    \phantomsubfloat{\label{fig:cavity-magnonics-experiment:cavity}}
    \phantomsubfloat{\label{fig:cavity-magnonics-experiment:YIG}}
    \vspace*{-2em}
    \caption{(a) Setup of a cavity magnonics experiment, where a microwave cavity (metallic grey box) is mounted inside an electromagnet applying a static magnetic field, and connected to a vector network analyser to monitor reflection and transmission coefficients. (b) Interior of the re-entrant two-post cavity \cite{2014Goryachev}, loaded with a YIG sphere placed between the two posts, where the cavity magnetic field $\vb{b}$ is strongest (indicated by the colour gradient). Two loop antennas on the left and right sides of the cavity couple to the cavity mode. The lid of the cavity is not shown. (c) Zoom on the YIG sphere, acting as a macrospin and coupling to the static applied magnetic field, and the cavity mode's RF magnetic field. }
    \label{fig:cavity-magnonics-experiment}
\end{figure}

The standard experimental setup for a cavity magnonics experiment, shown in \cref{fig:cavity-magnonics-experiment}, consists of
\begin{enumerate}
    \item a cavity loaded with a YIG sphere, see \cref{fig:cavity-magnonics-experiment:cavity,fig:cavity-magnonics-experiment:YIG},
    \item an electromagnet for applying a static magnetic field $\vb{H}_0=H_0 \vu{z}$ to bias the YIG sphere, and     
    \item a vector network analyser (VNA), connected to the cavity, for measuring its reflection and transmission coefficients, the S-parameters of the system.
\end{enumerate}  
The VNA couples with the photons in the cavity through two magnetic loop antennas protruding inside the cavity (see \cref{fig:cavity-magnonics-experiment:cavity}. 
These probe the photon degrees of freedom, and thus the magnon as well, due to photon-magnon coupling.

In general, a microwave cavity hosts many electromagnetic modes. 
In principle, they all couple to the Kittel magnon mode provided the associated magnetic field, integrated over the volume of the YIG sphere, does not vanish, recall~\cref{eq:coupling-strength:macro}). 
Furthermore, the cavity magnetic field should be as uniform as possible within the volume of the YIG sphere if only the Kittel mode, and not higher-energy magnon modes, is to be excited. 
For these reasons, a two-post re-entrant cavity \cite{2014Goryachev} is a flexible option for experiments. 
Its modes can be tuned in frequency, and can be separated by a large frequency gap \cite{2020Bourhill}. 
We focus on the cavity mode shown in \cref{fig:cavity-magnonics-experiment:cavity} because its magnitude is strongest between the two posts of the cavity, as shown by the color gradient. 
Since this RF magnetic field is strongest between the two posts, this is where the YIG sphere is placed to maximise the coupling. 

Indeed, in quantum optics, it is often desirable to reach the strong coupling regime, where the coupling strength $g/2\pi$ exceeds the dissipation rates of each mode. In cavity magnonics, this regime corresponds to $g \gg \kappa_{a},\kappa_{m}$, where $\kappa_{a}/2\pi$ ($\kappa_m/2\pi$) is the dissipation rate of the cavity (magnon) mode.
For three-dimensional cavities, the dissipation rates for both the photon and magnon modes are between a few MHz and a few dozens MHz \cite{2016Wang,2018Wang}. The dissipation rates for the photons depend on intrinsic losses due to the finite conductivity of the cavity walls, but also the coupling to the probes (extrinsic losses) to excite the cavity modes. To limit the former, the cavities are typically machined from oxygen-free copper and then polised to minimise surface roughness. For the Kittel magnon mode, dissipation rates can vary depending on the size of the sample, the purity of the YIG crystal, and how well polished the sphere is
\cite{2021Rameshti}. 

%A useful metric to characterise the light-matter interaction regime is the cooperativity $C = \frac{4g^{2}}{\kappa_{c}\kappa_{m}}$, which allows distinguishing between the weak-coupling regime where $C<1$, and the strong coupling regime where $C > 1$. 
As per \cref{eq:coupling-strength}, the coupling strength scales as $\sqrt{ S } \propto \sqrt{ N } \gg 1$ where $N$ is the number of spin in the YIG sphere, due to the high spin density $n_{s}=4.22 \times 10^{27}$ m$^{-3}$ of YIG \cite{2020Bourhill}. 
Thanks to the $\sqrt{ N } \gg 1$ scaling of the coupling strength $g / 2\pi$, the strong coupling regime is now routinely achieved in experiments with both three-dimensional \cite{2014Zhang,2015Tabuchi,2016Tabuchi,2016HarderHydeBaiMatchHu,2017Harder} and planar cavities \cite{2017Bhoi,2021Rao,2022Zhong,2022Kaffash,2022Ma}. In fact, most cavity magnonics experiments are in the strong coupling regime. With further optimisation of the cavity \cite{2019Flower,2020Bourhill,2021Macedo}, it is possible to reach the ultrastrong coupling regime where $g \gtrsim 0.1 \omega_{a},0.1\omega_{m}$ \cite{2019FriskKockum,2020LeBoite}, see refs \cite{2014Goryachev,2023Bourcin,2023Ghirri,2021Golovchanskiy,2021Golovchanskiya} for experimental demonstrations.

%To illustrate the tunability of cavity magnonics systems, we used \comsol\ to engineer a reentrant cavity \cite{2014Goryachev}
%A typical cavity magnonics setup satisfying these constraints is shown in \cref{fig:cavity-magnonics-experiment:cavity}, i of a . This setup can easily reach the strong coupling regime, which can be experimentally verified through the observation of an anti-crossing in an RF reflection or transmission measurement of the cavity (more details below). Alternatively, it is possible to realise planar cavity magnonics systems for instance using split-ring resonators \cite{2017Bhoi,2022Kaffash,2022Ma,2024Wagle}. However, the lower dimensionality of the cavity leads to higher dissipation rates, which can reach hundreds of MHz. 
%Notably, for moderate ($g \ll  \abs{\omega_{c}-\omega_{m}}$) values of the coupling strength, a rotating wave approximation reduces the magnon-photon coupling to a simple beam-splitter interaction $cm^\dagger + c^\dagger m$.

%In the preceding section, we have shown that a cavity magnonics system naturally realises a Dicke Hamiltonian (see \cref{eq:magnon-photon-hamiltonian:angular,eq:magnon-photon-hamiltonian:bosonic}).

%We now detail a physical implementation based on the two-post-reentrant cavity shown in \cref{fig:cavity-magnonics-experiment}. This will form the ``unit cell'' of a photo-magnonic crystal. 

In our experimental proposal, the walls of the cavity are almost circular in shape; the typical radius is \SI{10}{\milli\meter}. 
In contrast, the walls on either side of the posts are flattened with a length of $l=$ \SI{5}{\milli\meter}, see the left and right sides of \cref{fig:cavity-magnonics-experiment:cavity} or \cref{fig:coupled-cavities:design}. 
This corresponds to an offset of $10-\sqrt{10^2-(5/2)^2} \simeq$ \SI{0.3}{\milli\meter} from the walls of a perfectly circular cavity. The reason for this geometry is that 
it facilitates the coupling between cavities. % as will be discussed further in the next section. 
The height of the cavity is fixed at \SI{5}{\milli\meter} and the gap between the top of the posts and the lid is \SI{300}{\micro\meter}. The posts have a radius of \SI{1}{\milli\meter} and are placed \SI{3}{\milli\meter} apart. The radius of the YIG sphere is \SI{150}{\micro\meter}, and it is shifted in the $\vu{z}$ direction by \SI{100}{\micro\meter} so that it does not touch the bottom of the cavity.

We used \comsol, a finite-element modelling software, to simulate the electromagnetic physics of this cavity;
for details, see the supplementary materials. An eigenmode analysis gives a cavity resonance frequency of $\omega_a/2\pi=9.999$ GHz, which is verified by computing the S-parameters of the cavity without a YIG sphere. 
The results are plotted on the screen of the VNA in \cref{fig:cavity-magnonics-experiment:setup}, and reproduced in \cref{fig:coupled-cavities:s-param:one}. 
Importantly, the reflection (transmission) amplitude show a dip (peak) at $\omega_a/2\pi=9.998$ GHz, confirming the eigenmode analysis.

\begin{figure}[t]
    \centering
    \includegraphics[width=\columnwidth]{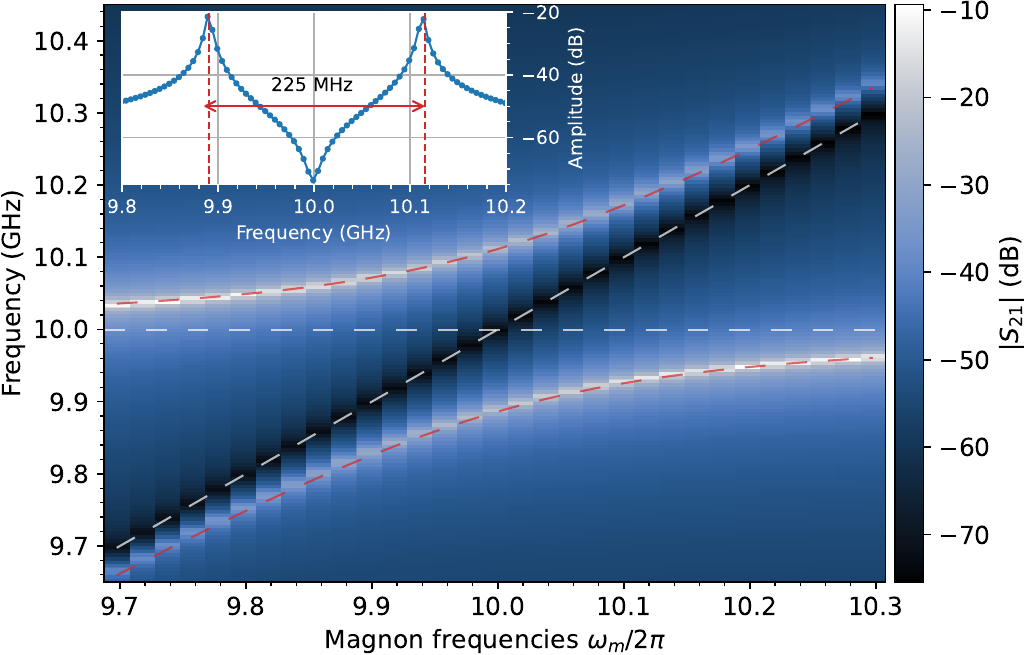}
    \caption{Simulation of transmission amplitude ($\abs{S_{21}}$) of the unit cell of the crystal using COMSOL. The level repulsion signals the strong magnon-photon coupling. The dashed red lines correspond to the spectrum of the Hamiltonian of \cref{eq:magnon-photon-hamiltonian:bosonic} with parameters $\omega_a/2\pi=9.999$ GHz and $g/2\pi=122.5$ MHz. The dashed white lines are the frequencies of the uncoupled photon and magnon modes, $\omega_a/2\pi$ and $\omega_m/2\pi$ respectively. Inset: frequency cut at resonance $\omega_m=\omega_a$ corresponding to the intersection of the two dashed white lines. The frequency spacing of 225 MHz between the two resonances is twice the coupling strength, hence $g/2\pi=122.5$ MHz.}
    \label{fig:N1:H0:S21}
\end{figure}

We now explore the consequences of placing a YIG sphere of radius \SI{150}{\micro\meter} between the posts of the cavity. 
Using the approach of ref \cite{2019Flower,2020Bourhill}, we can use an eigenmode analysis to evaluate numerically the coupling strength $g/2\pi$ of \cref{eq:coupling-strength}, and find $g/2\pi= 131$ MHz. Once again, this can be verified by simulating the transmission amplitude through the cavity using COMSOL, and we obtain the result shown in \cref{fig:N1:H0:S21}, suggesting a coupling strength $g/2\pi=112.5$ MHz. The observed level repulsion as the resonance frequency of the magnon is swept is evidence that the strong coupling regime is being reached. Reassuringly, the observed resonances agree with the spectrum of the Hamiltonian \cref{eq:magnon-photon-hamiltonian:bosonic}, showing that the quantum model agrees with the electromagnetic simulation. 

Note that in actual experiments the linewidths are expected to be broader due to additional loss channels that 
are not included in our simulations (for instance the finite conductivity of the cavity walls). 
Still, the coupling strength is large enough that moderate linewidth broadening will preserve the strong coupling regime. 
We note the presence of an anti-resonance (diagonal black line) in \cref{fig:N1:H0:S21} following the magnon frequency $\omega_m/2\pi$. 
This behaviour is expected and well understood, see ref \cite{2016HarderHydeBaiMatchHu} for a discussion. 
Finally, since the system is symmetric, we have $S_{21}=S_{12}$ and $S_{11}=S_{22}$. 
We have also simulated $S_{11}$ and we also observe an anti-crossing, see the supplementary material.

\subsubsection{Synthetic gauge fields}
\label{subsubsec:phase}
%\paragraph{Coupling phase.}
The magnon-photon coupling strength of \cref{eq:coupling-strength} can be, in principle, complex valued. Writing 
$g = \abs{g}e^{i\varphi}$, we refer to $\abs{g}$ as the magnitude of the coherent coupling, and $\varphi$ as the coupling phase. The coupling phase depends on the direction of the average magnetic field within the magnetic sample as \cite{Gardin2023}
\begin{equation}
    \label{eq:coupling-phase}
    \varphi = \mathrm{arg}\, g = \mathrm{arg} \qty{\int_{V_{m}} \dd[3]{\vb{r}} \left(\beta_{x}(\vb{r})-i 
    %\int_{V_{m}} \dd[3]\vb{r}}
    \beta_{y}(\vb{r})\right)}
\end{equation}
where $\mathrm{arg }\, z$ denotes the argument of the complex number $z$, and we recall that $\beta_x= (\grad \times \vb{u}_a) \cdot \vu{x}$ and $\beta_{y} = (\grad \times \vb{u}_a) \cdot \vu{y}$. 
In a system with a single YIG insertion, the coupling strength can be chosen real-valued (or, building towards \cref{sec:dicke-chain-revisited}, purely imaginary) through a rotation $U=\exp\qty(i\varphi m^\dagger m)$ of the magnon mode, and hence the coupling phase $\varphi = \mathrm{arg}\, g$ %can be ignored
does not influence the energy levels of the system.

However, the generalisation of \cref{eq:magnon-photon-hamiltonian:bosonic} to multiple modes can be used to synthesise a $U(1)$ gauge field through the coupling phases %. In this setup, only the gauge-invariant observables describe physical effects 
\cite{Gardin2023,2023Gardin}. 
With hindsight, combined with the methods that follow, this technology can be used to build lattices which we call photo-magnonic crystals, in which the photons are indirectly coupled to a synthetic gauge field through the magnons. 
This is one of the key capabilities of this new meta-material, which we will investigate further in future publications.

\subsection{\label{subsec:implementation-Dicke-chain} From single cavities
to coupled arrays}
%Experimental implementation of a crystal}

Following what was discussed in the previous sections, we propose to optically couple an array of cavities with one magnetic insertion in each to obtain the simplest instance of a new quantum meta-material: \textbf{the photo-magnonic crystal.} 
The distinctive characteristic of this meta-material is that the photons are highly mobile quasi-particles, hopping between cavities. 
By contrast, the YIG spheres cannot possibly interact directly in any appreciable way, and so the magnons are very heavy quasi-particles. 
The actual quasi-particles of a photo-magnonic crystal result from the hybridisation of these two kinds of radically different (in mobility) quasi-particles. 
Experience with comparable models of fermions suggests that such a setup could lead to a topologically non-trivial band structure. 
As we will see, this is in fact the case already for the simplest one-dimensional photo-magnonic crystal.

Ideally, one would like to be able to describe the connection between the cavities through a photon-photon hopping term of the form
\begin{equation}
    \label{eq:hamiltonian:photonic-tight-binding}
   {\cal H}_{hopping}= -\hbar \sum_j \qty(t a_{j}a_{j+1}^\dagger + t^* a_j^\dagger a_{j+1}),
\end{equation}
where $a_j$ denotes the cavity mode of cavity $j$ interacting strongly with the Kittel magnon mode of that cavity, and $t/2\pi$ is the cavity-cavity coupling. 
In this paper we will focus on clean photo-magnonic crystals and leave disorder for future investigation.

\begin{figure}[t]
    \centering
    \includegraphics[width=\columnwidth]{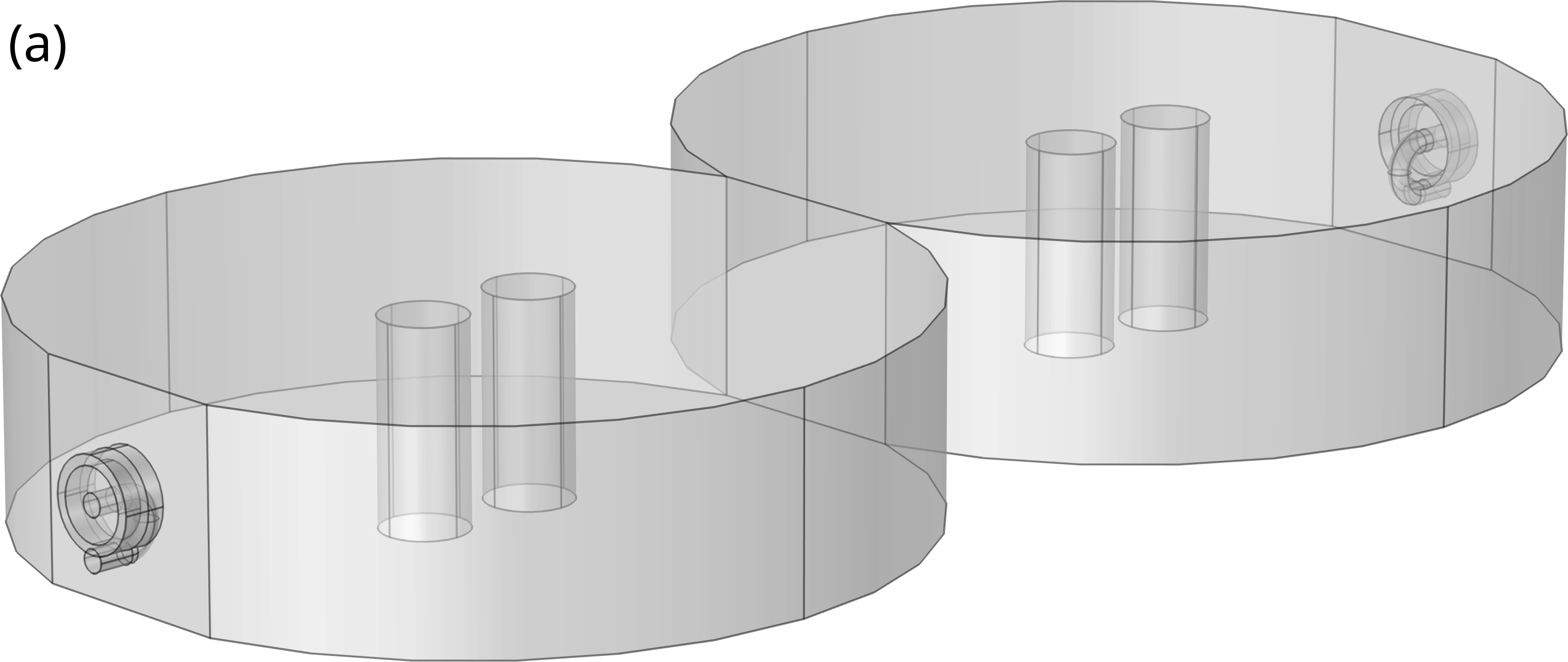}
    \includegraphics[width=\columnwidth]{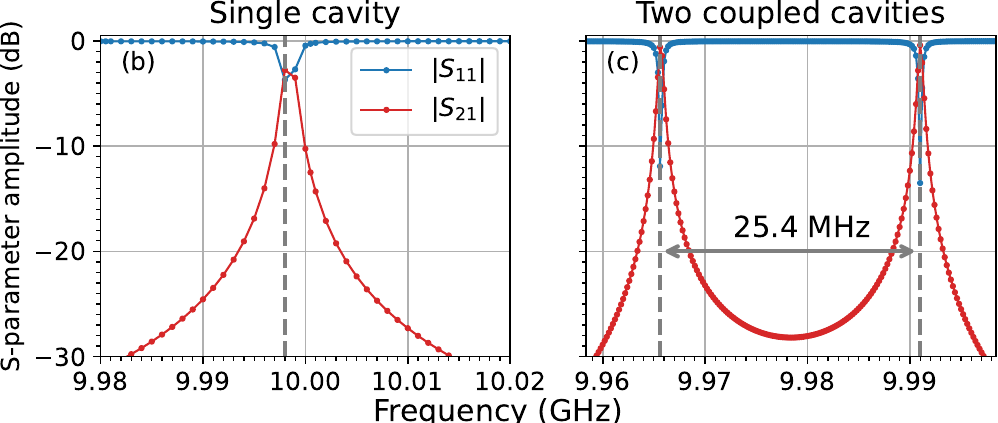}
    \phantomsubfloat{\label{fig:coupled-cavities:design}}
    \phantomsubfloat{\label{fig:coupled-cavities:s-param:one}}
    \phantomsubfloat{\label{fig:coupled-cavities:s-param:two}}
    \vspace{-3em}
    \caption{(a) COMSOL model of two cavities (without YIG spheres) coupled with a rectangular iris. (b) Simulated reflection and transmission parameters of the cavity unit cell without a YIG sphere. The grey vertical line indicates the location of the estimated resonance frequency located at $\omega_a/2\pi=9.998$ GHz. (c) Simulated reflection and transmission parameters of two reentrant cavities coupled through an iris as per (a). The grey vertical line indicates the location of the estimated resonances located at $9.9656$ GHz and $9.991$ GHz.}
    \label{fig:coupled-cavities}
\end{figure}

In practice, the ability to realise this kind of optical coupling depends on the architecture of the cavities. 
In coplanar cavities, based on split ring resonators for instance, the cavity-cavity coupling can be implemented by capacitive coupling, as realised for instance in ref \cite{2024Qian}. 
For the three-dimensional cavities considered here, recall \cref{eq:hamiltonian:photonic-tight-binding}, the coupling can be implemented by an iris, that is, a small aperture on a wall of the cavity; see \cref{fig:coupled-cavities:design}. 
Through the iris, a cavity mode can leak out of the cavity as an evanescent wave, which can then overlap with a similar evanescent wave from the neighbouring cavity. 
Intuitively, a non-vanishing overlap results in the linear tight-binding Hamiltonian of \cref{eq:hamiltonian:photonic-tight-binding} between the neighbouring cavity modes. 
In principle, this aperture can be extended along the crystal direction to create a rectangular waveguide between the cavities. 
Due to the exponential decay of the evanescent wave, the length of such a waveguide allows controlling the magnitude of the photon hopping strength.

\paragraph{Quality control: coupling two re-entrant cavities.}
To confirm this intuition, let us investigate quantitatively a system of two cavities with magnetic insertions as specified in the previous section. 
The cavity modes have resonance frequencies~$\omega_a/2\pi$, so that the photonic part of the Hamiltonian reads
\begin{equation}
    \label{eq:hamiltonian:coupled-cavities}
    \mathcal{H}_p = \sum_{j=1}^{2} \qty[\hbar \omega_{a} a_{j}^\dagger a_{j} - \hbar( ta_{j}a_{j+1}^\dagger + \hc) ]
\end{equation}
according to our heuristic arguments.
Due to the coupling $t/ 2\pi$, the normal modes of \cref{eq:hamiltonian:coupled-cavities} have angular frequencies $\omega_{a} \pm \abs{t}$, i.e. the twofold degeneracy at $\omega_{a}$ is split by $\abs{t}$. Thus, we expect to observe two resonances, one for each mode, separated by a frequency gap of $2\abs{t}/2\pi$.

\paragraph{Generalisation to the 1D photo-magnonic crystal.}
To verify that this simple model is accurate enough, we used COMSOL to model the coupling of two cavities by a rectangular iris as shown in \cref{fig:coupled-cavities:design}. 
The dimensions of the rectangular iris are \SI{5}{\milli\meter} by \SI{5}{\milli\meter}, corresponding to the length of the flat interface times the height of the cavity. 
An eigenmode analysis reveals two modes at $\omega_{a}-\abs{t} = 2\pi \times 9.9657$ GHz and $\omega_{a}+\abs{t} = 2\pi \times 9.991$ GHz, giving $\abs{t} / 2\pi = 12.65$ MHz. 
Furthermore, we simulated the S-parameter of the coupled cavities in \cref{fig:coupled-cavities:s-param:two}. 
We observe a frequency splitting of 25.4 MHz, confirming a photon hopping strength of $\abs{t} / 2\pi = 12.7$ MHz. 
It is worth noting that the resonance frequencies in the coupled cavity setup are not centred around the single-cavity resonance $\omega_{a}/2\pi=9.998$ GHz anymore, but rather around $\omega_a/2\pi=9.9783$ GHz. This can be explained by the small alteration of the cavity geometry due to the presence of the iris: before, a cavity unit cell was perturbed by two probes, symmetrically on either side. Now each cavity mode is now coupled to a probe on one side, and the other cavity on the other.

\begin{figure}[!t]
    \centering
    \includegraphics[width=\columnwidth]{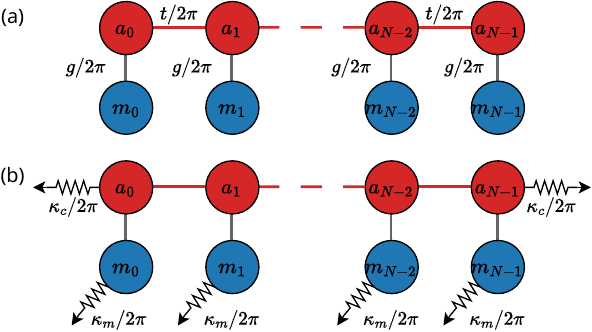}
	\phantomsubfloat{\label{fig:schematic:H0:closed}}
	\phantomsubfloat{\label{fig:schematic:H0:open}}
	\vspace{-2em}
    \caption{Schematic of the proposed 1D photo-magnonic crystal with $N$ unit cells. (a) Closed system described by \cref{eq:H0:bosonic} illustrating the coupling between the modes. (b) Illustration of the dissipation channels in the crystal considered in the electromagnetic simulations and the input-output formalism.}
    \label{fig:schematic:H0}
\end{figure}

In principle, the magnitude of the photon-photon hopping is controlled by the geometry of the iris: 
the wider it is, the bigger the overlap between the two cavity modes, and hence the stronger the coupling. 
For the rectangular iris presented here, a stronger coupling could be obtained by increasing the length $l$ of the flat surface on either side of the cavity. 
Due to the close proximity between the two cavities, the coupling is ``in-phase", which can be verified through the phase of the transmission coefficient $S_{21}$ (see the supplementary materials). 
%A phase offset could be obtained by connecting the two cavities through a waveguide, of identical dimensions, but whose length will control the phase offset.

%For a single cavity, an eigenmode analysis leads to the cavity resonance $\omega_{a}/2\pi =10.081$ GHz, which is further verified by simulating the reflection and transmission amplitudes shown in \cref{fig:coupled-cavities:s-param:one}. 
%Then, we considered the two-cavity setup of \cref{fig:coupled-cavities:design}, and this time the eigenmode analysis revealed two resonances at $\omega_{a}-\kappa = 2\pi \times 10.031$ GHz and $\omega_{a}+\kappa = 2\pi \times 10.070$ GHz, giving $\kappa / 2\pi = 39$ MHz. The frequency splitting is also shown in reflection and transmission in \cref{fig:coupled-cavities:s-param:two}. 

We have shown that the inter-cavity coupling Hamiltonian of \cref{eq:hamiltonian:photonic-tight-binding} properly accounts for the iris-coupling of the cavities. Therefore, we can use this method to couple $N$ copies of the unit-cell Hamiltonian introduced in \cref{subsec:cavity-magnonics}, hence realising a one-dimensional photo-magnonic crystal.

To write the Hamiltonian of this synthetic crystal, we note $m_{j}$ the annihilation operator of the Kittel mode of the YIG sphere located in cavity $j$. We further assume that all magnon modes have the same frequency $\omega_{m} /2\pi$, and note $g / 2\pi$ the coupling strength. Thus, the Hamiltonian of the crystal reads
%\begin{equation}
    \begin{align}
    {\mathcal H} &= \sum_{j} \qty[ \hbar\omega_{a} a_{j}^\dagger a_{j} +  \hbar\omega_{m} m_{j}^\dagger m_{j} 
        - \hbar(t a_{j}a_{j+1}^\dagger + \hc) ]\nonumber\\ 
        \label{eq:H0:bosonic}
        &\quad +\sum_{j}\hbar (g m_{j} + g^* m_{j}^\dagger)(a_{j}+a_{j}^\dagger).
    \end{align}
%\end{equation}
Note that for simplicity, we can choose $g / 2\pi$ real-valued, without loss of generality. Indeed, the coupling phase can be eliminated through a unitary transformation as discussed in \cref{subsec:cavity-magnonics}, and hence only the magnitude $\abs{g}/2\pi$ is physically relevant.

\begin{figure}[t]
    \centering
    \includegraphics[width=\columnwidth]{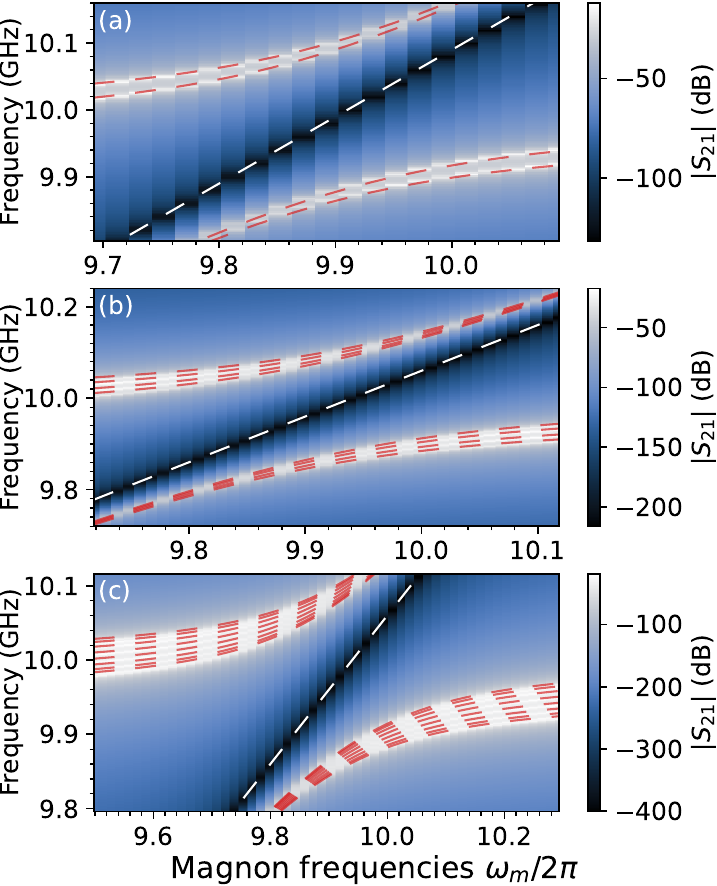}
    \phantomsubfloat{\label{fig:S21-H0-fits:N2}}
    \phantomsubfloat{\label{fig:S21-H0-fits:N4}}
    \phantomsubfloat{\label{fig:S21-H0-fits:N8}}
    \vspace{-3em}
    \caption{Simulated transmission amplitude through the photo-magnonic crystal using COMSOL for (a) $N=2$, (b) $N=4$, and (c) $N=8$ lattice sites. 
    The red dashed lines correspond to the numerical diagonalisation of \cref{eq:H0:bosonic} with the parameters $\omega_a/2\pi=9.9783$ GHz, $\abs{g}/2\pi=112.5$ MHz, $\abs{t}/2\pi=12.7$ MHz and a magnon linewidth $\kappa_m/2\pi=10$ MHz.
    For (a), the magnon frequency is shifted by 30 MHz, see the supplementary materials for a discussion. 
    The diagonal dashed white line corresponds to $\omega=\omega_m$.}
    \label{fig:S21-H0-fits}
\end{figure}

In principle, the Hamiltonian of \cref{eq:H0:bosonic} describes the one-dimensional photo-magnonic crystal shown in~\cref{fig:schematic:H0:closed}. Using the parameters $\omega_a/2\pi=9.9783$ GHz, $\abs{g}/2\pi=112.5$ MHz, and $\abs{t}/2\pi=12.7$ MHz, which correspond to our experimental implementation, the spectrum of this Hamiltonian should allow fitting the transmission amplitude through the chain, just like it did for a unit cell (for details on how the spectrum is computed for the entire crystal, we refer the reader to the supplementary materials). 
To verify this, we have performed COMSOL simulations for chains of length $N \in \qty{2,4,8}$, and obtained the results of \cref{fig:S21-H0-fits}, where we set the dissipation rate of the magnon modes to $\kappa_m/2\pi=10$ MHz. We first notice that in all cases, we still observe level repulsion, signalling the strong coupling between the cavity and magnon modes. Second, the number of resonances matches with the length of the chain, e.g. for $N=2$ we observe two pairs of mode splitting (\cref{fig:S21-H0-fits:N2}), for $N=4$ we have four pairs (\cref{fig:S21-H0-fits:N4}, etc. Furthermore, as expected, the spectrum of \cref{eq:H0:bosonic} agrees with the observed resonances, confirming that this quantum Hamiltonian model captures the physics of this classical electromagnetic simulation.

\paragraph{Input-output formalism.}
The electromagnetic modelling has two drawbacks. 
First, it is computationally expensive for long chains, due to the size of the mesh. 
Second, it does not easily allow probing the individual components of the hybridised modes, or in other words, the composition of the eigenmodes of the crystal. While the Hamiltonian formalism allows examining the eigenmodes numerically, it does not include the effect of dissipation, and does not give access to experimental observables such as microwave reflection and transmission through the crystal. A solution is to use the input-output formalism (whose derivation is detailed in the supplementary materials), which couples each mode of \cref{eq:H0:bosonic} to a heat bath. This enables us to take into account dissipation within a Hamiltonian formalism, and also provides us with an alternative method to compute the reflection and transmission through the crystal. 

\begin{figure}[t]
    \centering
    \includegraphics[width=\columnwidth]{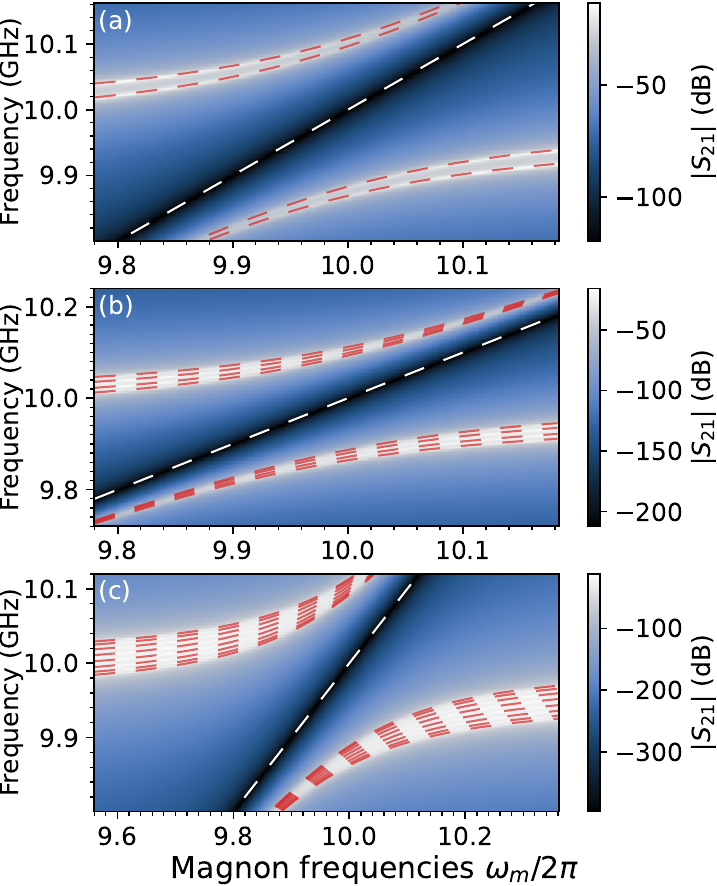}
    \phantomsubfloat{\label{fig:S21-H0-fits:iot:N2}}
    \phantomsubfloat{\label{fig:S21-H0-fits:iot:N4}}
    \phantomsubfloat{\label{fig:S21-H0-fits:iot:N8}}
    \vspace{-3em}
    \caption{Numerical calculation of the transmission amplitude through the photo-magnonic crystal using the input-output formalism for (a) $N=2$, (b) $N=4$, and (c) $N=8$ lattice sites. The dissipation rate of the cavity modes at the boundaries of the crystal are equally set to $\kappa_c/2\pi=0.5$ MHz, while the dissipation rates for the cavity mode in the bulk are set to 0. The legend and the other parameters are identical to those of \cref{fig:S21-H0-fits}. A strong agreement with finite-element results is obtained.}
    \label{fig:S21-H0-fits:input-output}
\end{figure}

For calculations based on the input-output formalism, we have used the same parameters as those obtained from the COMSOL simulations. In the latter, the cavity walls are modelled as perfect electric conductors, and hence there is no intrinsic dissipation for the cavity modes. Therefore, the only source of dissipation for the photons is the coupling of the cavity modes of the first and last lattice sites to the magnetic antenna used to probe the crystal, as per \cref{fig:schematic:H0:open}. Based on the resonance of the cavity mode of the unit cell, see \cref{fig:coupled-cavities:s-param:one}, we deduce a total line-width of 1 MHz. Because the two magnetic antennas are symmetric, we deduce that their coupling of the cavity mode is $\kappa_c/2\pi=0.5$ MHz. Hence, we set $\kappa_c/2\pi = 0.5$ MHz only for the cavity modes at the boundary of the chain, and else $\kappa_c/2\pi = 0$ for the cavity modes in the bulk. For the magnon modes, we assume that they all have a magnetic damping rate $\kappa_m/2\pi=10$ MHz. With these parameters, we obtain the results shown in \cref{fig:S21-H0-fits:input-output}, which agree well with the electromagnetic simulations. As before, because the system is symmetric, we have $S_{21}=S_{12}$ and $S_{11}=S_{22}$. Furthermore, and as expected, the observed resonances agree with the spectrum of \cref{eq:H0:bosonic}, shown in red dashed lines.

To summarise, in this section, we have introduced a theoretical model for a photo-magnonic crystal. 
We have shown that using the input-output formalism or finite-element modelling, we can model the electromagnetic response of the crystal and dissipative effects. In particular, it is worth noting that finite-element modelling of these systems have been shown to agree remarkably well with true experiments, see for instance \cite{2020Bourhill,Gardin2023,2023Gardin}. 
Thus, the microwave responses in \cref{fig:S21-H0-fits,fig:S21-H0-fits:input-output} are expected to occur in experiments, albeit with different linewidths due to additional dissipation.
We have also proposed a concrete experimental platform to realise this crystal. 
Fully electromagnetic simulations based on COMSOL show that the proposed design agrees very well with the theoretical model. 
We have now validated several methods at our disposal to analyse more specific properties of the crystal.

\section{\label{sec:topological-crystal}A topological photo-magnonic chain}
\iffalse
The most delicate point regarding symmetry-protected topological (SPT) physics is that of symmetry protection. The early literature on the integer quantum Hall effect created a lasting but incorrect impression that ``topological physics," generically speaking, is robust somehow simply because it is topologically mandated. However, the presence of \textit{topology} in a quantum system does not always warrant the presence of \textit{protection}. A better statement is that for SPT systems, what is and what is not protected is determined by the classifying symmetry operations. These classifying symmetry operations also inform whether a symmetry class is topologically trivial or not, and if it supports a bulk-boundary correspondence or not. These outcomes can differ when the same symmetry class is considered in different dimensions.
\fi

In this section, we consider a specific instance of the photo-magnonic crystal
%, for $N=4$ lattice sites, 
and provide evidence for it being topologically non-trivial. 
We will show that, for suitable, experimentally implementable boundary conditions, the crystal hosts a zero boundary mode, and that it is topologically mandated. What is not clear at this point is whether the zero edge mode is protected by some physical symmetry. To answer this question, we shall develop in~\cref{sec:topological-classification} a topological classification of free boson Hamiltonians based on basic physical bosonic many-body symmetries, and the basic associated bulk-boundary correspondences in 1-d. Its implications for the 1-d photo-magnonic crystal are investigated in \cref{sec:dicke-chain-revisited}.

\subsection{Electromagnetic finite-elements simulation}

\begin{figure}[!t]
    \centering
    \includegraphics[width=0.85\columnwidth]{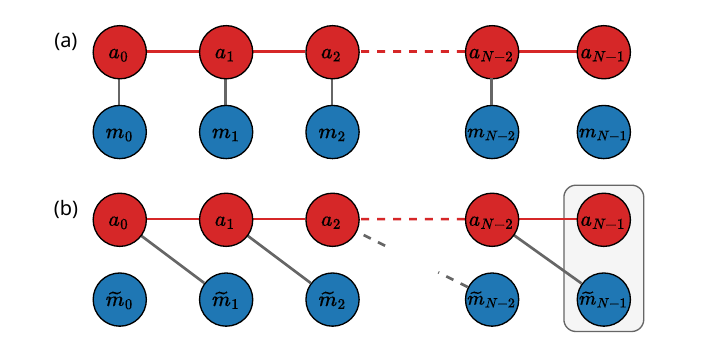}
    \phantomsubfloat{\label{fig:schematic:H1:a}}
    \phantomsubfloat{\label{fig:schematic:H1:b}}
    \vspace{-2em}
    \caption{(a) Photo-magnonic crystal where the magnon mode in the last unit cell is disconnected. (b) Equivalent system after relabelling the magnon modes, described by \cref{eq:Hn:real-space} for $n=1$. An edge mode appears on the last lattice site, indicated by the rectangle.}
    \label{fig:schematic:H1}
\end{figure}

%\paragraph{Finite-element results.}
We have shown in the preceding section an implementation of the photo-magnonic crystal depicted in~\cref{fig:schematic:H0}. To reveal a zero edge modes, here we will examine the same system subjected to different boundary conditions; see \cref{fig:schematic:H1}. They amount to disconnecting the magnon mode for the last lattice site $j=N-1$. How should one achieve something like this in an experimental realisation of the meta-material? A natural possibility is to simply remove the YIG sphere from a boundary cavity, say, the last cavity. We will show below by way of a full numerical simulation of the resulting chain of magnonic cavities yields the desired outcome. 

%\begin{figure}[t]
%    \centering
%    \includegraphics[width=\columnwidth]{S21-H1-fits.pdf}
%    \phantomsubfloat{\label{fig:S21-H1-fits:N4}}
%    \phantomsubfloat{\label{fig:S21-H1-fits:N8}}
%    \vspace{-3em}
%    \caption{Simulated transmission amplitude through the photo-magnonic crystal using COMSOL for (a) $N=2$, (b) $N=4$, and (c) $N=8$ lattice sites, when the YIG sphere in the first cavity is removed. The legend and parameters are identical to those of \cref{fig:S21-H0-fits}.}
%    \label{fig:S21-H1-fits}
%\end{figure}

Electromagnetic simulations for a crystal of $N=4$ sites are shown in \cref{fig:H1-N4:Sparams-fits}, and similar results for $N=8$ lattice sites are available in the supplementary material. The fact that the system is not symmetric -- due to the modified boundary conditions -- is reflected in the asymmetry of the S-parameters: the transmission through the crystal is still symmetric ($S_{12}$, not shown, is identical to $S_{21}$), but the reflection is not. Indeed, the reflection amplitude $\abs{S_{11}}$ at the left edge of the crystal exhibits level repulsion between three resonances. Intuitively, this is interpreted as resulting from the magnon-photon hybridisation in the first three lattice sites only, as per \cref{fig:schematic:H1:a}, since the last lattice site does not have a magnon mode. The behaviour of $\abs{S_{22}}$, corresponding to the right edge, is more peculiar: we observe only a dip at what seems to be the cavity modes' frequency $\omega_a/2\pi$. At first sight, it is not surprising to observe such a dip near $\omega_a/2\pi$, since the cavity mode in the last lattice site is only coupled to other cavity modes at a similar frequency. This suggest that this mode is purely photonic, due to its frequency at $\omega_a/2\pi$. In reality, as we will show below, this mode hybridises, albeit weakly, with the magnon mode in the $j=N-2$ unit cell.
%The reason why it is not appearing in the results of \cref{fig:Sparams-H1-N4-fits} is that the length of the crystal is small, and the photon-photon hoppings $\abs{t}/2\pi=12.7$ MHz is much smaller than the magnon-photon hopping $g/2\pi=112.5$ MHz. 

\begin{figure}[t]
    \centering
    \includegraphics[width=\columnwidth]{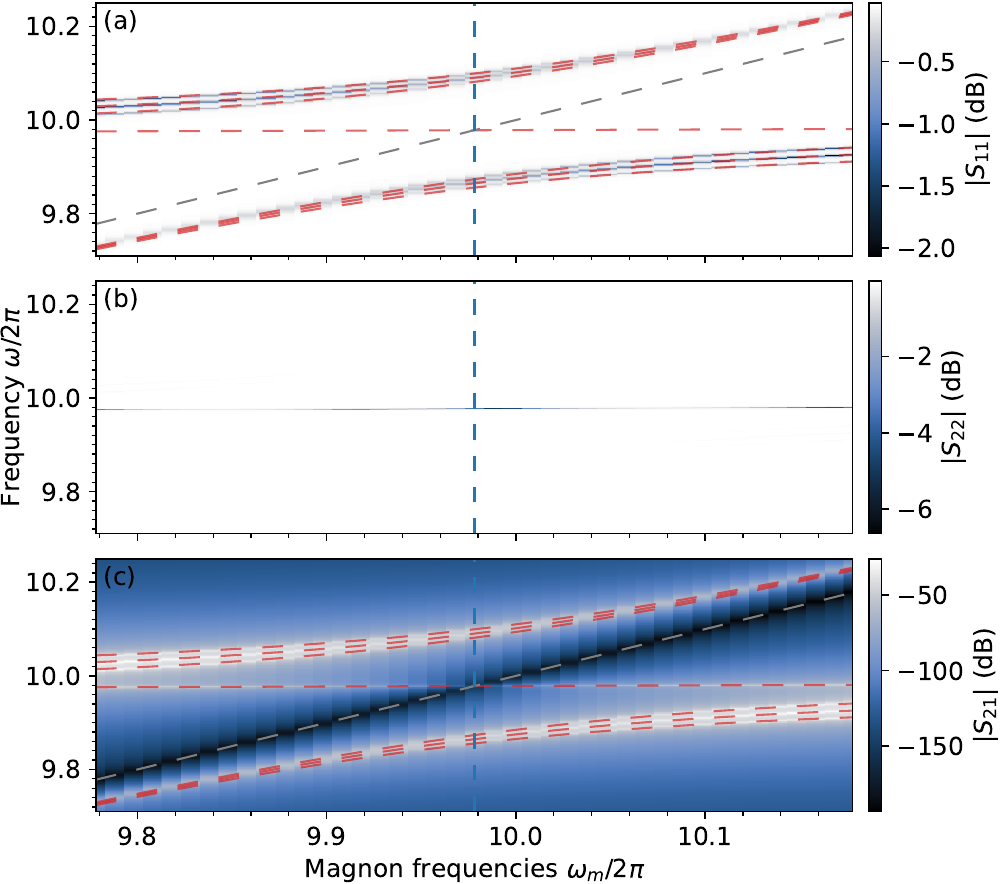}
    \phantomsubfloat{\label{fig:H1-fits-N4:S11}}
    \phantomsubfloat{\label{fig:H1-fits-N4:S22}}
    \phantomsubfloat{\label{fig:H1-fits-N4:S21}}
    \vspace{-3em}
    \caption{Simulated S-parameters amplitude through the photo-magnonic crystal using COMSOL for $N=4$ lattice sites, when the YIG sphere in the last cavity is removed. The legend and parameters are identical to those of \cref{fig:S21-H0-fits} The spectrum is not shown in (b) for readability, and $S_{12}=S_{21}$. The vertical dashed blue line indicates the resonance condition $\omega_m=\omega_a$.}
    \label{fig:H1-N4:Sparams-fits}
\end{figure}

\subsection{\label{subsec:topological-crystal:effective-hamiltonian}Effective Hamiltonian description}
%\paragraph{Effective Hamiltonian.}
Let us investigate next the mode structure for the new boundary conditions starting from the model \cref{eq:H0:bosonic}. To investigate, in addition, topological aspects more broadly, we consider a Hamiltonian which allows us to model various boundary conditions, labelled by an integer $n$. We will refer to this integer as a \emph{coupling offset}. In experimental practice, it
amounts to adding to the photo-magnonic crystal $n$ empty cavities at one end.
To model this situation, we need to relabel the magnon modes; see~\cref{fig:schematic:H1:b}. We thus define shifted magnon operators $\widetilde{m}_{j+n}=m_j$. After switching the notation, the effective Hamiltonian \cref{eq:H0:bosonic} of the cavity array becomes 
%use the input-output formalism with the Hamiltonian
\begin{equation}
\label{eq:Hn:real-space}
\begin{aligned}
    \frac{{\cal H}^{(n)}}{\hbar}
    &=  \sum_{j=0}^{N-1} \qty(\omega_a a_j^\dagger a_j + \omega_m \widetilde{m}_j^\dagger \widetilde{m}_j)
    -\sum_{j=0}^{N-1}(t a_{j} a_{j+1}^\dagger +\hc)\\
    &\quad+ \sum_{j=0}^{N-1} (g a_{j}\widetilde{m}_{j+n}^\dagger+ g a_j \widetilde{m}_{j+n} + \hc).
\end{aligned}
\end{equation}

Given the parameters of the photo-magnonic crystal, we have $g \gg \omega_a,\omega_m$. Hence, we can perform a rotating wave approximation (RWA) and drop the counter-rotating terms $a_j m_{j+n} + \hc$ from~\cref{eq:Hn:real-space}. Furthermore, to simplify our analysis, we consider the magnon and photon to be on resonance, and set $\omega_a=\omega_m$. Thus, in a rotating frame at the frequency $\omega_a/2\pi=\omega_m/2\pi$, the Hamiltonian reduces to 

\begin{equation}
\label{eq:Hn:real-space:rwa-rotating-frame}
\begin{aligned}
    \frac{{\cal H}^{(n)}}{\hbar}
    &= 
    -\sum_{j=0}^{N-1}(ta_{j} a_{j+1}^\dagger + t^*a_{j}^\dagger a_{j+1})\\
    &\quad+ \sum_{j=0}^{N-1} (g a_{j}\widetilde{m}_{j+n}^\dagger+ g^* a_{j}^\dagger \widetilde{m}_{j+n}).
\end{aligned}
\end{equation}

\paragraph{Zero edge modes without dissipation.}
The case $n=1$ corresponds to the boundary conditions of \cref{fig:schematic:H1}, and thus diagonalising
\begin{equation}
\begin{aligned}
    \frac{{\cal H}^{(1)}}{\hbar}
    &= 
    -\sum_{j=0}^{N-1}(ta_{j} a_{j+1}^\dagger + t^*a_{j}^\dagger a_{j+1})\\
    &\quad+ \sum_{j=0}^{N-1} (g a_{j}\widetilde{m}_{j+1}^\dagger+ g^* a_{j}^\dagger \widetilde{m}_{j+1}),
\end{aligned}
\end{equation}
we find the spectrum plotted as dashed red lines in \cref{fig:H1-N4:Sparams-fits}. 
Therefore, the horizontal line observed in $\abs{S_{22}}$ and $\abs{S_{21}}$ is indeed a mode predicted by the Hamiltonian.

%\textbf{maybe cit R. H. Dicke, Coherence in spontaneous radiation processes, Phys. Rev. 93, 99 (1954).?}

Using this theoretical model, we can now show that this mode corresponds to a mode localised on the right edge of the system. The starting point are the commutators with the Hamiltonian
\begin{align}
\frac{1}{\hbar}[a_{N-1},{\mathcal{H}}^{(1)}] &=  -t^* a_{N-2},  \\
\frac{1}{\hbar}[\widetilde{m}_{N-1}, {\mathcal{H}}^{(1)}] &= g a_{N-2},
\end{align} 
for the operators on the right edge of the chain. Therefore, up to a global phase factor, the linear combination
\begin{equation}
    \label{eq:edge-mode:shifted}
    p_R \equiv \frac{ga_{N-1} + t^* \widetilde{m}_{N-1}}{\sqrt{\abs{g}^2 + \abs{t}^2}}
\end{equation}
satisfies $[p_R,p_R^\dagger]=1$ and $\qty[p_R, {\mathcal{H}}^{(1)}]=0$. The first commutator indicates that $p_R$ represents a bosonic mode, while the second tells us that it has zero energy. Let us now return to the magnon operators corresponding the physical system under consideration. Undoing the shift in the magnon operators, the annihilation operator for the edge mode is
\begin{equation}
    \label{eq:edge-mode}
    p_R \equiv \frac{ga_{N-1} + t^* {m}_{N-2}}{\sqrt{\abs{g}^2 + \abs{t}^2}}
\end{equation}
Since the operators $p_R$ and $p_R^\dagger$ only involve modes located on lattice sites $N-1$ and $N-2$, we deduce that it is a bosonic zero-energy mode localised on the right edge of the chain. In the laboratory frame, this edge mode has frequency $\omega_a=\omega_m$. 
Using experimental parameters of the proposed crystal, $\abs{g} \simeq 9\abs{t}$, and thus the edge mode is mostly made of photons. However, this can be tuned by adjusting the magnon-photon coupling. For instance, it is easy to reduce $\abs{g}$ so that $\abs{g}\simeq \abs{t}$ by either moving the YIG spheres away from the maxima of magnetic field, or by reducing the size of the spheres. When the photon-photon and magnon-photon coupling are matched, the edge mode can thus be made to be approximately half magnon and half photon.

\paragraph{Impact on the S-parameters.}
We can now re-interpret the S-parameters of \cref{fig:H1-N4:Sparams-fits}. The horizontal line observed in $\abs{S_{22}}$ and $\abs{S_{21}}$ simply corresponds to the edge mode of \cref{eq:edge-mode}. The reflection and transmission amplitudes are only sensitive to the photon amplitudes in the cavities that couple to the magnetic antennas. Therefore, exciting the edge mode when driving the system from the left edge is impossible, which explains the absence of the horizontal line in $\abs{S_{11}}$ (\cref{fig:H1-fits-N4:S11}). On the other hand, driving the system from the right edge directly couples to the edge mode, leading to the horizontal dip observed in $S_{22}$. However, by the same argument, the transmission amplitudes $\abs{S_{21}}$ and $\abs{S_{12}}$ should vanish on resonance. Due to the limited resolution of the finite-element modelling, it is hard to assess whether this is verified in our simulations. 

\begin{figure}[t]
    \centering
    \includegraphics[width=\columnwidth]{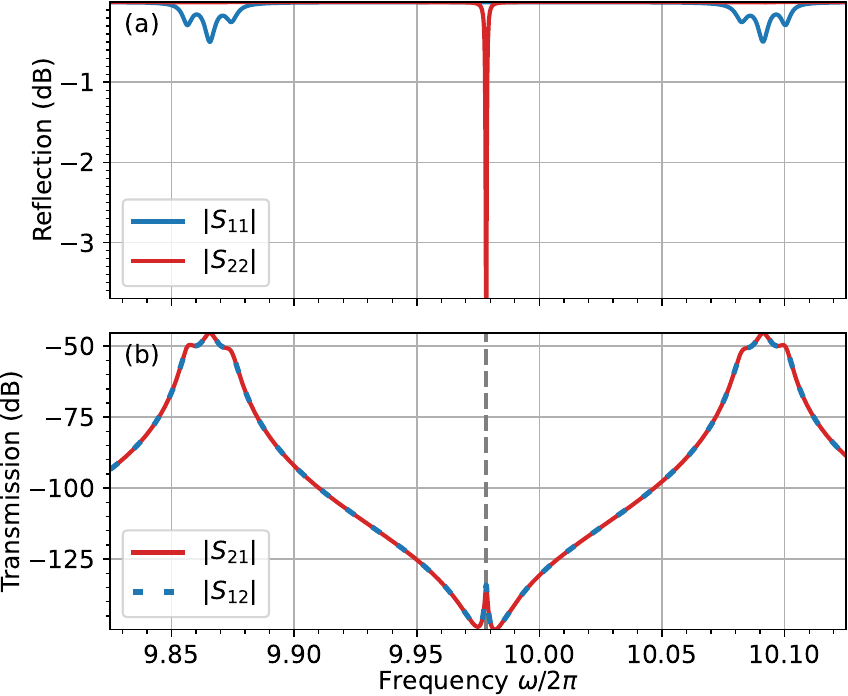}
    \phantomsubfloat{\label{fig:N4-H1:iot:reflection-cuts}}
    \phantomsubfloat{\label{fig:N4-H1:iot:transmission-cuts}}
    \vspace{-2em}
    \caption{Numerical calculation of the S-parameter amplitudes using the input-output formalism for a crystal of $N=4$ lattice sites, where the YIG sphere has been removed from the last lattice site, for $\omega_m=\omega_a$, corresponding to the vertical dashed blue line in \cref{fig:H1-N4:Sparams-fits}. The parameters used are identical to those used for the fit of \cref{fig:S21-H0-fits:input-output}, but the rotating wave approximation has been employed. An extended data range is available in the supplementary materials.}
    \label{fig:N4-H1:iot:Sparams-cut}
\end{figure}

\paragraph{Edge mode excitation.} Instead, we can employ the input-output formalism to calculate numerically the transmission amplitudes. In \cref{fig:N4-H1:iot:reflection-cuts}, we plot the S-parameters at the resonance condition $\omega_m=\omega_a$. We recover the observation of \cref{fig:H1-N4:Sparams-fits}, namely mode splitting in $\abs{S_{11}},\abs{S_{21}}$ and $\abs{S_{12}}$, but only a single resonance for $\abs{S_{22}}$. Interestingly, while the transmission amplitude is expected to vanish when the crystal is driven on resonance, we observe a small resonance instead, indicated by the vertical dashed grey line. This suggests that the edge mode is excited, with the same amplitude (since $\abs{S_{21}}=\abs{S_{12}}$), regardless of which edge is driven.

\begin{figure}[t]
    \centering
    \includegraphics[width=\columnwidth]{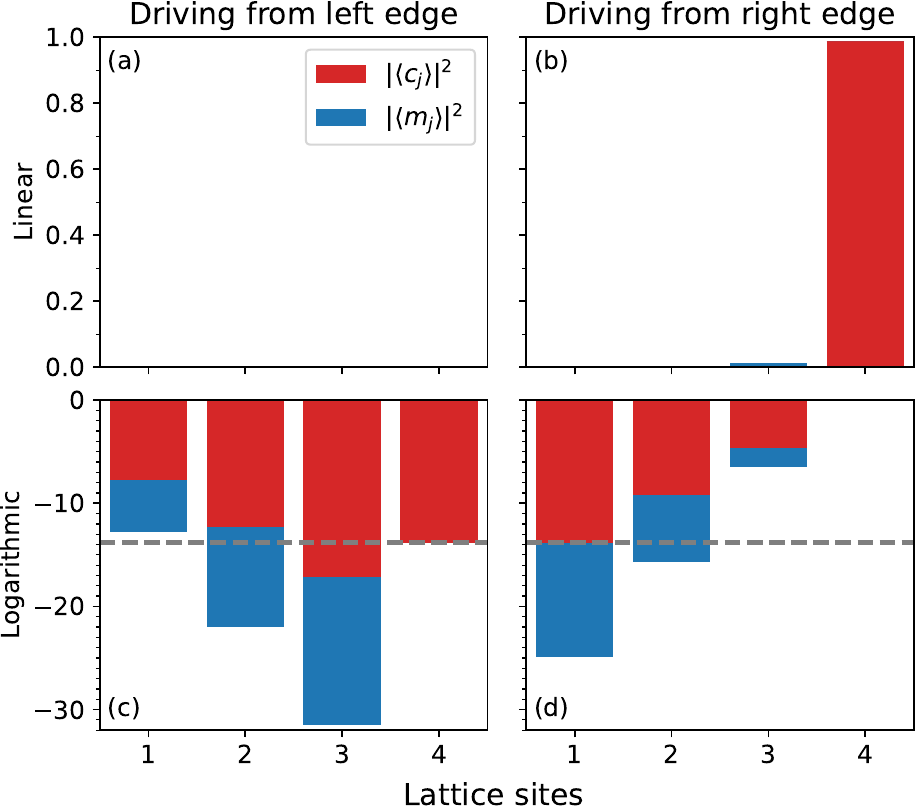}
    \phantomsubfloat{\label{fig:N4-H1:iot:linear:left}}
    \phantomsubfloat{\label{fig:N4-H1:iot:linear:right}}
    \phantomsubfloat{\label{fig:N4-H1:iot:log:left}}
    \phantomsubfloat{\label{fig:N4-H1:iot:log:right}}
    \vspace{-3em}
    \caption{Expectation values of each operator corresponding when driving the crystal at resonance $\omega=\omega_a=\omega_m$. The first column is associated with the $S_{11}$ and $S_{21}$, while the second with the $S_{22}$ and $S_{12}$. The linear values are normalised to the maximum amplitude between 0 and 1. The second row plots the same quantities as in the first row on a logarithmic scale. The horizontal dashed grey line indicates the level of photons at the detection site when measuring the transmission. The parameters are identical to those of \cref{fig:N4-H1:iot:Sparams-cut}. }
    \label{fig:N4-H1:iot:modes}
\end{figure}

To understand this behaviour, we can compute the expectation values of each mode in the crystal when it is driven from either direction. We obtain the results of \cref{fig:N4-H1:iot:modes} when the crystal is driven on resonance. 
\Cref{fig:N4-H1:iot:linear:right} shows a distribution reminiscent of \cref{eq:edge-mode}. Indeed, the system is mostly in a superposition of $a_{N-1}=a_4$ and $m_{N-2}=m_3$ as predicted by \cref{eq:edge-mode}. Additionally, the weight of $a_4$ is expected to be greater than that of $m_3$ since $\abs{g}>\abs{t}.$ 
On the other hand, when the crystal is driven from the left edge, the expectation values suggest that the edge mode of \cref{eq:edge-mode} is not excited as one would expect. 
Instead, we observe a response suggesting that all the energy is absorbed by the magnon mode in the first cavity. Yet, the transmission amplitudes are symmetrical. As shown in the supplementary materials, $S_{21} \propto \expval{c_{N-1}}$, while $S_{12} \propto \expval{c_0}$. In \cref{fig:N4-H1:iot:log:left,fig:N4-H1:iot:log:right} we plot the expectation values in logarithmic scale, and observe that $\expval{c_{N-1}}$ when driving from the left is identical to $\expval{c_{0}}$ when driving from the right. In other words, despite a different repartition of the expectation values in the crystal, the photon amplitude at the detection location remains identical.
%This signature is typical of electromagnetically induced transparency.

The small resonance in the transmission in~\cref{fig:N4-H1:iot:transmission-cuts} is reminiscent of a signature found in electromagnetically-induced transparency, where typically a sharp dip (transparency window) is observed in a larger peak in absorption. In cavity magnonics, this behaviour has been observed and called magnon-induced transparency \cite{2023Wangd}. In particular, ref \cite{2023Wangd} has studied theoretically a system similar to the proposed photo-magnonic crystal, but with only $N=2$ lattice sites. Therein, the magnon-induced transparency was explained in the language of interference and dark mode physics. They recovered the edge mode of \cref{eq:edge-mode}, but referred to it as a dark mode instead, because they only considered the reflection from the left edge of the system. Thus, the readout mechanism did not couple to this eigenmode, rendering it ``dark" for their experimental setup. Here, we have shown that this dark mode exists for an arbitrary long chain, and we will ground its existence in topology.

To summarise, we have shown that an edge mode occurs in the photo-magnonic crystal with an altered boundary condition, and its experimental signature has been examined. 
While we initially expected this edge mode to be trivial, associated with the cavity mode in the last cavity only, we have seen that it strongly hybridises with the magnon mode in the preceding cavity. 
We have further observed an analog of electromagnetically-induced transparency physics on this edge mode. 
One application of such physics is the generation of slow light, and thus if the edge mode is topologically-mandated, this opens the prospect of the topologically-robust generation of slow light.
Hence, we now discuss the topological properties, if any, associated with this edge mode. 
%\textbf{I am thinking here there should be a sentence to outline the important results of the observation of electromagnetically-induced transparency and possibly of some relevant application?}

\subsection{The zero edge modes are topologically mandated}
\paragraph{Reciprocal-space Hamiltonian.}
The photo-magnonic crystal with the standard boundary conditions of~\cref{fig:schematic:H0}) does not possess an edge mode. 
Hence, $\mathcal{H}^{(0)}$ (corresponding to \cref{eq:Hn:real-space:rwa-rotating-frame} with $n=0$) does not host an edge mode, but $\mathcal{H}^{(1)}$ does. This suggests that these two systems represent distinct topological phases. 
We can test this idea by computing a topological invariant, the Berry phase of the bundle of states associated with the low-energy band.

\begin{figure}[t]
    \centering
    \includegraphics[width=0.65\columnwidth]{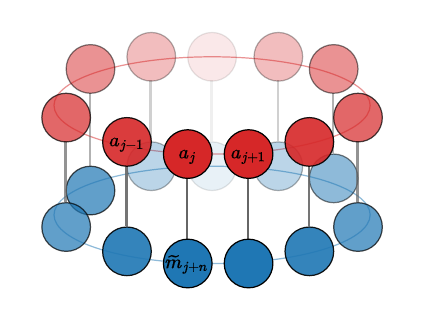}
    \caption{Photo-magnonic crystal with periodic boundary conditions for general $n$. 
    In momentum space, the eigenvalues have no dependence on $n$, but the eigenmodes do.}
    \label{fig:schematic:Hn}
\end{figure}

To examine the band structure, we investigate the Hamiltonian of~\cref{eq:Hn:real-space:rwa-rotating-frame} subjected to periodic boundary conditions. The crystal momentum space operators $a_{q}, \widetilde{m}_{q}$ are the Fourier-transformed operators
\begin{align}
    \label{eq:definition-momentum-c}
    a_{q} &= \frac{1}{\sqrt{ N }} \sum_{j=0}^{N-1} a_{j}e^{2i \pi \frac{jq}{N} },\\
    \widetilde{m}_{q} &= \frac{1}{\sqrt{ N }} \sum_{j=0}^{N-1} \widetilde{m}_{j}e^{2i \pi \frac{jq}{N} }.
\end{align}
Note that here the index $q \in \qty{0, \dots, N-1}$ indexes the momentum $k_q = 2\pi \frac{q}{N}$. For readability, let us rewrite \cref{eq:definition-momentum-c} as
\begin{equation}
    \label{eq:def-momentum-operator}
    a_k = \frac{1}{\sqrt{ N }} \sum_{k \in BZ} a_{j}e^{ikj}
\end{equation}
in what follows (with a similar definition for $\widetilde{m}_k$), where $BZ = [-\pi,\pi)$. Then, in momentum space, the Hamiltonian of~\cref{eq:Hn:real-space:rwa-rotating-frame} becomes ${\cal H}^{(n)} = \sum_{k \in BZ} {\cal H}^{(n)}(k)$ where
\begin{equation}
    \label{eq:hamiltonian:reciprocal-space}
    \begin{aligned}
        {\cal H}^{(n)}(k)
        &= \hbar \Omega(k)a_{k}^\dagger a_{k} + \hbar \omega_m \widetilde{m}_k^\dagger \widetilde{m}_k\\
        &\quad+ \hbar\qty(ge^{ikn} a_{k}\widetilde{m}_{k}^{\dagger} + g^* e^{-ikn} a_{k}^\dagger\widetilde{m}_{k})
    \end{aligned}
\end{equation}
with 
\begin{equation}
    \Omega(k)=-2\abs{t} \cos\qty(k + \phi)
\end{equation}
 for $t=|t|e^{i\phi}$.  
%\textbf{I guess all these calculation have been checked and re-checked?} \Cref{eq:hamiltonian:reciprocal-space} 
%corresponds to a Dicke model labelled by the momentum $k \in BZ$, which can 
%be diagonalised for each $k$ to obtain the band-structure. 
%Following, say, Ref.~\cite{Blaizot1986}, and generalising our previous analysis of a single magnonic cavity, we can introduce the bosonic dynamical matrix \added[id=al]{Cite appendix}

Despite being in momentum space, the Hamiltonian of \cref{eq:hamiltonian:reciprocal-space} can be diagonalised using a similar technique as that used in \cref{subsec:cavity-magnonics}, see the supplementary materials for details. The bosonic dynamical matrix of \cref{eq:hamiltonian:reciprocal-space} is
\begin{align}
    G^{(n)}(k) &=
    \begin{bmatrix}
        K^{(n)}(k) & 0\\
        0 & K^{(n)}(-k)^*
    \end{bmatrix},\\
    %\begin{bmatrix}
    %\Omega(k) & g^* e^{-ikn} & 0 & 0\\ 
    %g e^{ikn} & 0 & 0 & 0\\
    %0 & 0 & \Omega(-k) & ge^{-ikn}\\
    %0 & 0 & g^* e^{ikn} & 0
    %\end{bmatrix}
    \label{eq:Gn:rwa}
    K^{(n)}(k)&\equiv \begin{bmatrix}
    \Omega(k) & g^* e^{-ikn} \\ 
    g e^{ikn} & 0 
    \end{bmatrix}.
\end{align}
The dynamical matrix is block-diagonal because particle number is conserved (recall that, as discussed in \cref{subsec:topological-crystal:effective-hamiltonian}, we have dropped the counter-rotating terms and justified this step by invoking the RWA). The matrix $K(k)$ is the single-particle Hamiltonian, and it is all that is needed to diagonalise the model when particle number is conserved.  Notice how the lower diagonal entry of $K(k)$ vanishes. This is a reflection of the fact that the magnons are completely localised quasi-particles unless they interact with the 
mobile photons.

The eigenvalues of $K^{(n)}(k)$ are 
\begin{align}
    \label{eq:adc:no-pairing:dispersion}
    \omega_\pm(k) = \frac{\Omega(k) \pm \sqrt{\Omega(k)^2 + 4 \abs{g}^2}}{2};
\end{align}
they determine the energy carried by a single ``cavity magnon polariton" 
of crystal momentum $k$. 
%In \cref{fig:spectra-interpolation-H0-H1:bandstructure:0,fig:spectra-interpolation-H0-H1:bandstructure:100} we plot the bandstructure for $n=1$ and $n=0$, respectively. We observe an identical dispersion relation, and we checked that the results agree with the analytical expression of \cref{eq:adc:no-pairing:dispersion}. Indeed, 
Although the integer $n$ parametrises the model, it does not play a role in the band-structure described by \cref{eq:adc:no-pairing:dispersion}. Intuitively, recall that the integer $n$ only serves to impose various boundary conditions \emph{for a finite chain}. Thus, the periodic boundary conditions that are assumed in going to momentum space effectively eliminate the effect of $n$, which indeed amounts to re-labelling the magnon operators, see \cref{fig:schematic:Hn}. Formally, the only dependence on $n$ in \cref{eq:hamiltonian:reciprocal-space} is located in the phase present in the couplings between $c_k$ and $\widetilde{m}_k$. This phase can be removed through the gauge-transformation $U{\cal H}^{(n)}U^\dagger$ with $U=e^{-i kn \widetilde{m}_k^\dagger \widetilde{m}_k}$, since it maps $\widetilde{m}_k \mapsto e^{ikn} \widetilde{m}_k$, and hence the gauge-invariant spectrum, $\omega_\pm(k)$, cannot have a dependence on $n$.
By contrast, the (normalised) eigenvectors 
\begin{align}
\label{eq:adc:no-pairing:eigenvecs}
    \ket{u_{\pm}^{(n)}(k)} 
    &= \frac{1}{{\cal N}_k} \begin{bmatrix} 
        \pm g^* e^{-ikn} \\
        \mp \omega_\mp(k)
    \end{bmatrix} %\begin{bmatrix} 
        %\pm g^* e^{-ikn} \\
        %\mp \frac{\Omega(k)}{2} + \frac{\sqrt{\Omega(k)^2 + 4 \abs{g}^2}}{2}
    %\end{bmatrix}
    \\
    %\label{eq:adc:no-pairing:eigenvecs}
    \nonumber
    &=\frac{1}{{\cal N}_k} 
    \begin{bmatrix} 
        \pm g^* e^{-ikn} \\
        \pm \abs{t}\cos(k) + \frac{\sqrt{\abs{t}^2 \cos(k)^2 + 4 \abs{g}^2}}{2}
    \end{bmatrix}
\end{align}
of $K^{(n)}$ are indeed gauge-dependent, and $N_k$ is a normalisation constant. They determine the quasi-particle creation operators. The destruction operators follow by Hermitian conjugation. The following analysis is inspired by ref.~\cite{2017Alase}. To extract some gauge-invariant information out of the wave functions, let us zoom in on the Berry connection
$$
A_{\pm}(k)=\bra{u_{\pm}^{(n)}(k)}\frac{d}{dk}\ket{u_{\pm}^{(n)}(k)};
$$
see, for example, Ref.~\cite{2013Bernevig} for a textbook discussion.
The integrals
$$
\gamma_{\pm}=\frac{1}{i\pi }\int_{-\pi}^{\pi} A_\pm(k)\,dk 
$$
are integer valued, gauge invariant, and equal and opposite, that is 
$\gamma_+=-\gamma_-$. By  direct calculation, one finds 
that $\gamma_+=n$ provided $g\ne0$. As a particular easy check, 
the reader can confirm, on the back of an envelope, that this is 
the case for $t=0$. We conclude that the zero
boundary modes are topologically mandated. 
\section{\label{sec:context-strategy}From topologically mandated to symmetry protected: 
context and strategy}
We have demonstrated that the photo-magnonic crystal in one dimension happens to
host topologically mandated edge modes. But a question remains: \textit{are they robust?} Topological edge modes can be robust against many perturbations and, at the same time, fragile against many others. In that sense, the case of free fermion topological states is special because one can predict the impact of any given weak perturbation on the topologically mandated boundary modes by a simple symmetry analysis based on physical, many-body fermionic symmetries: perturbations that preserve the classifying symmetries are irrelevant, and perturbations that break them are relevant~\cite{2023Alase}. 
Hence, the sensible question to ask is whether the 1D photo-magnonic crystal hosts \underline{symmetry-protected} topological boundary modes, with the understanding that the symmetries in question should be physical bosonic many-body symmetries. 
In order to answer this question in a principled manner, we will develop in the next section our own topological classification of free boson systems. 

There exist a number of such classifications for free bosons already. Three characteristic 
papers are \cite{PhysRevX.9.041015, PhysRevB.102.125127, Chaudhary2021}. 
While this list is not exhaustive, it is representative of existing strategies. 
There are, in addition, a number of bulk-boundary correspondences for free boson Hamiltonians that are not embedded in some symmetry or topological classification; \cite{Shindou2013, Peano2018} are two characteristic examples. 
Hence, in this section, before tackling the mathematical details in \cref{sec:topological-classification}, we would like to explain our reasons for pursuing yet another topological classification of free boson systems. 
The explanation will also provide the physical intuition behind our theory in the next section.

\subsection{The strength of the fermionic tendold way}
Gapped free fermion Hamiltonians have been classified with symmetry arguments and topological concepts~\cite{Kitaev_2009, Ryu_2010}. To ground these ideas on physical symmetry operations, that is, transformations of the many-body states, one has to focus on spin $j=1/2$ fermions~\cite{2016Anderson, 2023Alase}. The relevant many-body symmetry operations are total fermion number, time reversal (the physical one that properly accounts for spin), spin (internal) rotations, and an anti-unitary transformation that exchanges fermionic creation and annihilation operators. %\textbf{GCT this is not equivalent to chirality? correct?}. 
  
\begin{table*}[]
    \centering
    \begin{tabular}{|c|c|l|}
    \hline
                    &\textbf{Symmetry class} & \textbf{Many-body symmetries}  \\
     Matrix space &    Cartan label       & \\
         \hline
         \hline
          Hermitian & \textbf{A}   &  total fermion number \\
         \hline
         complex    &\textbf{AIII} & total fermion number \\
                    &              & antiunitary exchange $c\leftrightarrow c^\dagger$\\
               \hline
               \hline 
real antisymmetric   &\textbf{D}     & none\\
               \hline
complex antisymmetric &\textbf{DIII} & spin $j=1/2$ time reversal\\
               \hline
               quaternionic Hermitian&\textbf{AII} &total fermion number \\
                & & spin $j=1/2$ time reversal\\
                \hline
                quaternionic & \textbf{CII} & total fermion number\\
                &&antiunitary exchange $c\leftrightarrow c^\dagger$\\
                && spin $j=1/2$ time reversal\\
                                \hline
                quaternionic anti-Hermitian &\textbf{C} & spin $j=1/2$ rotations\\
                \hline
                complex symmetric &\textbf{CI} & spin $j=1/2$ time reversal \\
                && spin $j=1/2$ rotations \\
                \hline
                real symmetric & \textbf{AI} & total fermion number \\
                 && spin $j=1/2$ time reversal\\
                 && spin $j=1/2$ rotations \\
                 \hline
                 real &\textbf{BDI} & total fermion number\\
                 && antiunitary exchange $c\leftrightarrow c^\dagger$\\
                 && spin $j=1/2$ time reversal \\
                 && spin $j=1/2$ rotations \\
                 \hline                
    \end{tabular}
    \caption{The ten Altland-Zirnbauer symmetry classes of free fermion Hamiltonians~\cite{PhysRevB.55.1142} characterized in terms of physical, many-body symmetry operations of spin $j=1/2$ fermions~\cite{2016Kennedy, 2023Alase}. The missing six symmetry classes are not describable in terms of the mathematical theories to which the Cartan labels are native and     do not support interesting mathematical structures. Here, $c\leftrightarrow c^\dagger$ indicates schematically the exchange of fermionic creation and annihilation operators by way of an anti-unitary transformation of the underlying Fock space. The six classes that conserve fermion particle number are classes of insulators; that is, it is implicitly understood that the Fermi level lies on a band gap. The four classes that do not conserve fermion particle number capture the mean-field models of fully gapped superconductors. Thanks to the many-body symmetries, every fermionic Bogoliubov-deGennes (BdG) Hamiltonian in a symmetry class can be put in normal form by way of a fixed (same for every BdG Hamiltonian in the symmetry class) similarity transformation. The blocks of the normal form are matrices in the matrix space listed on the leftmost column of the table. The space dimensionality determines, according to a periodically shifting pattern~\cite{Kitaev_2009, Ryu_2010}, which classes are topologically trivial and which classes are topologically non-trivial. 
    %The topologically non-trivial classes in one dimension are the classes D, DIII, CII, AIII, and
    %BDI~\cite{2023Alase}. 
    Within this many-body framework, the Majorana chain of Kitaev belongs to the class D and the fermionic Su-Schrieffer-Heeger (SSH) model belongs to the class BDI. After \cite{2017Alase}.}
    \label{tab:many_body_AZ}
\end{table*}

Following this line of thought, a \textit{symmetry class} represents a group of second-quantised Hamiltonians that is left invariant by none, one, two, three, or all of these symmetry operations. There are sixteen symmetry classes possible for spin $j=1/2$ fermions, however, only ten of them have interesting properties associated to several major connected topics of algebra and geometry; Table~\ref{tab:many_body_AZ} describes these ten symmetry classes first identified in \cite{PhysRevB.55.1142} in the context of free fermion systems, or the ten Altland-Zirnbauer (AZ) classes for short.

For any fixed space dimensionality, five out of the ten AZ classes are topologically non-trivial, and the subset of non-trivial classes cycles periodically with the dimension of space~\cite{Kitaev_2009, Ryu_2010}.  In a topologically trivial class, any two gapped Hamiltonians can be continuously (\textit{i.e. ``adiabatically"}) deformed into each other without breaking the symmetries and without closing the many-body energy gap. In a topologically non-trivial class, gapped Hamiltonians are further sub-classified by way of a topological invariant. Any two Hamiltonians that produce the same value of this invariant can be continuously deformed into each other; and any two Hamiltonians with a different value of the topological invariant cannot be continuously deformed into each other.%\textbf{GCT without closing the many-body energy gap and undergoing a quantum phase transition}. 

There are two main key consequences of these features of free fermion Hamiltonians. First, if, as a function of some parameter, a topological invariant changes its value without the Hamiltonian ever leaving its symmetry class, then it must be the case that there is a closing of the many-body energy gap of the system, that is, a quantum phase transition somewhere along the way. 
Second, there is the celebrated bulk-boundary correspondence: a non-vanishing value of the bulk topological invariant mandates boundary-localized modes inside some gap between the bulk bands of the system. 
These boundary modes are protected in the following sense. Consider a perturbation small enough to not close the many-body gap. If it does not change the symmetry class of the model Hamiltonian, then it cannot remove the boundary modes either.  More importantly for this story, if the perturbation does change the symmetry class of the Hamiltonian, then there is no way to know, a priori, what it will do to the boundary modes. In short, \textit{boundary modes are topologically mandated}, but the protection mechanism \textit{is linked to the classifying symmetries}. This is the reason why grounding the tenfold way in straightforward many-body symmetries is so powerful.

\subsection{Towards a bosonic tenfold way}
Complementary to the solid state physics community, mostly focused on the electronic degrees of freedom of condensed matter systems, there is a large community focused on systems of weakly-interacting, for the most part
part, bosons: quantum optics and descendants like cavity QED, optomechanics, cavity magnonics, and some systems of cold atoms. The foundation of their theories are the free boson Hamiltonians, albeit quickly updated in one way or another to accommodate dissipation and/or incoherent quantum processes. But one question remains open;~\textit{is there a topological theory of free boson Hamiltonians comparable in analytic and predictive power to the fermionic tenfold way?} If the answer is affirmative, what are its experimental and technological implications? There is at present no universally favoured answer to these questions. There are instead several different proposals, all of them valid, interesting, with distinct technological implications, and more or less tested experimentally. This suggests that either the class of free boson Hamiltonians is not structured enough to support an edifice as neat, comprehensive, and canonical as the fermionic tenfold way, or that we have not yet uncovered the right bosonic foundation for it. There are many confounders and defeaters, all hinting at some important missing piece of this puzzle.

1) \textit{Quantum statistics (defeater).} 
Due to the quantum statistical difference between fermions and bosons, breaking boson number symmetry is a necessary condition for a true many-body gap at zero temperature in free boson systems. 
Put differently, one can export the Hamiltonian of a fermionic topological insulator to bosons by replacing fermionic operators with bosonic operators, and the resulting free boson Hamiltonian will display topologically-mandated bosonic boundary modes. 
What one cannot export is the Fermi sea and its peculiar symmetries, and so this approach misses two physical features of the tenfold way: 1) it is a classification of gapped phases of matter, 2) there is a tight link between topological transitions and quantum phase transitions. 

What is perhaps worse, the fermionic protection mechanism is also lost in this translation. A good example for this is the Su-Schrieffer-Heeger (SSH) model, a fermionic model of the electronic properties of dimerized polyacetylene. 
The fermionic SSH model belongs to the class BDI of \cref{tab:many_body_AZ} \cite{2023Alase}. 
Most of its fermionic many-body symmetries (spin $j=1/2$ rotations, time reversal, and the antiunitary exchange $c\leftrightarrow c^\dagger$) have no bosonic analogue whatsoever. 
It takes all of these fermionic many-body symmetries working in concert, plus fermionic particle number which does have a bosonic analogue,
to endow the fermionic SSH model with the celebrated chiral symmetry of its single-particle Hamiltonian.
The bosonic SSH model, obtained by replacing fermionic operators by bosonic ones (see \cite{Ozawa2019} for a review-level summary of theory and experiments in photonics), inherits this single-particle symmetry of the fermionic SSH model by construction. 
What it does not inherit is the protecting many-body symmetries. 
Hence, it is not clear what, if any, are the generic many-body bosonic symmetries that protect the topologically-mandated edge modes of the bosonic SSH model. We answer this question in \cref{subsec:bosonicSSH}, and a find a topological invariant different from that of the fermionic SSH model.

2) \textit{Semiclassical states (confounder).} When working with free boson Hamiltonians, there is always the nagging suspicion that maybe one's work is not quantum enough. The confusion arises because the bosonic Fock space supports both semiclassical and deeply quantum states. There are no semiclassical states for fermions, or else this confounder would affect the tenfold way as well. The point to remember is that, for bosons, it makes little sense to try to draw a line between quantum and classical by focusing on the dynamics only. The distinction only makes sense if one investigates both the dynamics and the initial quantum states together. See  
 \cite{Peano2016_PhysRevX.6.041026} for a case study.

3) \textit{The loss of Hermiticity (confounder).} For free boson Hamiltonians that break the particle number symmetry and may support a true many-body gap, the equivalent of the Bogoliubov-deGennes Hamiltonian of free fermions is a non-Hermitian bosonic dynamical matrix, see \cite{Flynn2020} for a detailed analysis. For this reason, some of the early topological classifications of free boson Hamiltonians were a direct application of some larger topological classification of non-Hermitian Hamiltonians; see for example \cite{PhysRevX.9.041015} and the recent comprehensive review \cite{Ashida02072020}.

4) \textit{Stability (defeater).} Free fermion Hamiltonians are all dynamically and thermodynamically stable. Neither statement is true of free boson Hamiltonians and this turns out to be a serious complication. See \cite{BARTON1986322, SUBRAMANYAN2021168470, Flynn2020} for illuminating physical discussions of unstable free boson Hamiltonians.

5) \textit{No-go theorem 1: no SPTs for free bosons (defeater).}  
Let us zoom in on gapped, hence stable, free boson Hamiltonians and their quantum phase transitions. 
Then one is immediately defeated by two no-go theorems. 
The first one shows that, for any bosonic symmetry class whatsoever, when restricted to gapped Hamiltonians, it is necessarily topologically trivial directly because of the stability requirement \cite{PhysRevB.102.125127}.
     
6) \textit{No-go theorem 2: no zero modes or surface bands crossing zero for open boundary conditions (defeater).} 
A gapped, and hence stable, free boson Hamiltonian cannot support zero boundary modes, either isolated or as part of a band of boundary modes, for open boundary conditions \cite{PhysRevB.102.125127}. 

7) \textit{The impact of dropping thermodynamic stability (confounder).} 
The no-go theorems force one to give up the constraint of thermodynamic stability. 
The problem though is that the normal frequencies of a thermodynamically unstable free boson Hamiltonian are generically complex numbers (``parametric instabilities"). 
As a consequence, a generic linear observable grows or decays exponentially with time. 
On the bright side, this is the reason why it is possible to model some quantum amplifiers with a free boson Hamiltonian.

8) \textit{Dynamical stability cannot be diagnosed efficiently (defeater).} 
The class of dynamically stable free boson Hamiltonians consists of all the Hamiltonians that can be  fully diagonalised in terms of bosonic
quasi-particle creation and annihilation operators. 
Its normal frequencies are necessarily real. 
This class is interesting because it is free of parametric instabilities and can host topologically-mandated 
zero modes. 
Unfortunately, it is not possible to build a mathematical theory around this class of free boson Hamiltonians because there is no good mathematical way to characterise it~\cite{PhysRevB.102.125127, Flynn2020}.
With the exception of particle number conserving or thermodynamically stable models, dynamical stability seems to be a more or less erratic, accidental property of free boson Hamiltonians \cite{Flynn2020}.
We expect that its topologically-mandated zero modes will have to be explained by way of a framework grounded on a larger class of free boson Hamiltonians.

10) \textit{Several diverse import-export strategies (confounder).} 
An import-export strategy is a mapping of free fermion systems to free boson systems in such a way that one can partially export ideas and tools from the fermionic world of the tenfold way to the world of free boson systems. 
We already mentioned one in 1) above. 
They all necessarily loose in translation the symmetry protection mechanism of free fermion systems because the generic many-body symmetry operations of fermions and bosons do not match in any useful way (recall the example of the SSH model). 
One noteworthy import-export strategy led to the discovery of the bosonic Kitaev chain \cite{2018McDonald}. 
Other intriguing and useful import-export strategies were presented in \cite{PhysRevB.102.125127} (free fermions to free bosons) and \cite{PhysRevLett.122.143901, PhysRevA.103.033513} (quadratic bosonic Lindbladians to free fermions). 
We avoid import-export strategies in this paper, and instead look for a native framework.

11) \textit{The non-Hermitian skin effect (confounder).} 
If we drop thermodynamic stability, we cannot count on the topologically-mandated boundary modes to be the only boundary modes. 
The reason is that the non-Hermitian skin effect is a generic feature of unstable 
free boson Hamiltonians, see, for example, \cite{Brunelli2023} for a good discussion of this issue. 

12) \textit{Too few interesting many-body symmetries (defeater).}
The idea of refocusing the tenfold way on spin \(j=1/2\) fermions so that it can be fully grounded on many-body symmetry transformations is a recent development~\cite{2016Kennedy, Alldridge2020, 2023Alase}. 
It does not qualitatively change the tenfold way but makes it easier to test experimentally. 
In contrast, generic many body-symmetries of stable, gapped free boson Hamiltonians seem too simple to support anything like the tenfold way. 
However, after dropping thermodynamic stability, squeezing transformations can play the role of symmetries of free boson Hamiltonians. 
As shown in the next section, this indeed partly circumvent this defeater. 

13) \textit{Too many interesting single-particle symmetries (confounder).}
The standard analysis that leads to the fermionic tenfold way is built on the single-particle or the Bogoliubov-de Gennes Hamiltonians subjected to some single-particle ``symmetry" constraints, as opposed to many-body symmetry constraints~\cite{Ryu_2010}. 
This has not been a problem in practice because the link between single-particle ``symmetries" and many-body symmetries is tight for fermion \cite{PhysRevB.55.1142, 2016Kennedy}.
For bosons, building a tenfold-way-like theory starting from the effective single-particle bosonic dynamical matrix 
is somewhat self-defeating because there are too many interesting ``symmetries" of the bosonic dynamical 
matrix and most of them have no many-body significance. 
Without a many-body interpretation of a symmetry of the dynamical matrix, claims of symmetry protection become weaker from a physical perspective.

\subsection{Strategy}
The bottom line is that there seems to be no tight bosonic analogue of the tenfold way. Something must give, and the different proposals in the literature reflect the choices of  the authors. In this paper we
explore the theoretical and experimental implications of the following choices:
\begin{itemize}
    \item Dropping all stability constraints, and thus considering all free boson Hamiltonians on equal footing.
    \item Breaking the universe of free boson Hamiltonians into symmetry classes characterised by generic, for bosons, local many-body symmetry operations. 
    \item Matching symmetry classes to index theorems in order to 
    \begin{enumerate}
        \item identify a suitably generalised notion of band gap,
        \item identify the topologically non-trivial bosonic symmetry classes, and 
        \item extract the associated bulk-boundary correspondences.
    \end{enumerate}  
\end{itemize}

\iffalse
The interesting stability constraints are thermodynamic stability and dynamical stability. Thermodynamic stability excludes the possibility of having a tight link between topologically-mandated boundary modes and the 
quantum phase diagram of the system. Dynamic stability without thermodynamic stability is an erratic, fine-tuned property of some free boson Hamiltonians and cannot be used to usefully identify Hamiltonian classes
\fi

By ``generic, for bosons, local many-body symmetry operations" we mean unitary or anti-unitary transformations 
of the bosonic Fock space that have a clear meaning without reference to any particular Hamiltonians, 
and act locally with respect to the underlying spatial lattice. 
Total particle number is a good example. Bosonic time reversal ${\cal T}_B$ can be more detail-dependent in implementation, but because of the general relation between time reversal and spin rotations, one can expect ${\cal T}_B^2=1$ for boson~\cite{ballentine2014quantum} (while by contrast, for fermions, ${\cal T}_F^2=-1$). 
The exchange $c\leftrightarrow c^\dagger$ operation of fermionic creation
and annihilation operators is an important ingredient of the tenfold way, see Table~\ref{tab:many_body_AZ}. Unfortunately, for bosons, it is strictly forbidden: there is no unitary or anti-unitary
transformation of the bosonic Fock space that exchanges bosonic creation and annihilation 
operators~\cite{PhysRevB.102.125127, Zirnbauer2021}. 

\iffalse
Regarding the notion of many-body gap, the class of thermodynamically stable free boson 
Hamiltonians is precisely the class that supports the standard notion of many-body gap to
a ground state and no 
notion of symmetry-protected topological quantum order \added[id=g]{citation?}. 
For dynamically stable Hamiltonians, it has recently been discovered 
that a new kind of gap, called the Krein gap in Ref.~\cite{Mariam2025}, is the appropriate tool 
for detecting and characterising critical points in parameter space where the correlation length 
diverges in the quasi-particle vacuum state. 
Because we drop stability conditions altogether, we can't use this notion of many-body gap. Instead, we work with a notion of gap that matches well the mathematical framework of this paper. Let us call it the Fredholm gap for definiteness, since it signifies whether the 
Bosonic dynamical matrix is a Fredholm operator \cite{lax2014functional}. 
The Fredholm gap is a sharply defined quantity that is easy to compute for translation-invariant systems. 
If the system happens to be dynamically stable close to a point
in parameter space where the Fredholm gap closes, then the closing of the Fredholm gap is associated to 
a divergent correlation length in the bosonic quasi-particle vacuum. In this way we recover a 
fairly tight connection between topology and a suitably generalised notion of quantum phase transition.
\fi

Finally, since we have dropped all stability conditions, the non-Hermitian
skin effect is a definite possibility for the systems we consider~\cite{Ashida02072020}.
Within our framework, the distinction between topologically-mandated boundary modes and other ``skin modes" is 
in the robustness of the former. To tell the difference, one idea
is to investigate the response to the system to perturbations. 
Indeed, the non-Hermitian skin effect is generic and responds erratically to perturbations. 
By contrast, the  topologically-mandated boundary modes respond to perturbations, both symmetry-preserving
and symmetry-breaking, in very definite and predictable ways. 
For example, their localisation length decreases with a definite scaling on approaching a closing of the (generalised) band gap. 
See \cite{Brunelli2023, bomantara2025floquetbosonickitaevchain} for additional discussion and ideas.

%%%%%%%%%%%%%%%%%%%%%%%%%%%%%%%%%%%%%%%%%%%%%%%%%%%%%%%%%%%%%%%%%%%%%%%%%%%%%%%%%%%%%
% Section V
%%%%%%%%%%%%%%%%%%%%%%%%%%%%%%%%%%%%%%%%%%%%%%%%%%%%%%%%%%%%%%%%%%%%%%%%%%%%%%%%%%%%%
\section{\label{sec:topological-classification}Many-body symmetry 
classes, topological classifications, 
and bulk-boundary correspondences}
%protected topology in free boson systems} %(TSP)}

To continue with our development of our theory, in this section we introduce a self-contained framework, based directly on elementary bosonic many-body symmetries and basic index theorems~\cite{lax2014functional, bleecker2013index}, for building a theory of bosonic symmetry classes, topological classifications, and bulk-boundary correspondences. To focus on the concepts and minimise the mathematical complexity of the exposition, the core of this work will be mainly towards one-dimensional systems. Higher dimensions require more sophisticated index theory. We will extend our framework to higher dimensions in a forthcoming publication to address the key issue of periodicity, namely 
identifying the  periodic pattern of our classification as a function of space dimension. 
 %Next, our choice of classifying bosonic symmetries is discussed in details with the uncovering of a set of topologically non-trivial symmetry classes in one space dimension, and with the unfolding of bulk-boundary correspondences appropriate for
%each non-trivial class. 

\subsection{Background}

\subsubsection{
The structure of free boson
Hamiltonians}
\label{subsec:formalism-qbh}
This section introduces a formalism to analyse quadratic bosonic Hamiltonians (QBHs) in terms of matrices. 
The tools are old~\cite{Blaizot1986}; our modern style is that of Refs.~\cite{Alldridge2020} and \cite{2023Alase} suitably adapted to bosons~\cite{Flynn2020}. 
This formalism has already been used in \cref{sec:photo-magnonic-crystals,sec:topological-crystal} to diagonalise Hamiltonians and find the normal modes of the system.
%The methods introduced should be familiar to most readers working with bosonic systems, and has already been discussed in \cref{subsec:qbh-diagonalization}. 
In the present section the level of formality will be higher, as required to establish our theory.  
We refer the reader to the supplementary materials for illustrative applications of this formalism.

To illustrate the main points within a minimal setting, let us focus on a system of independent bosons with abstract single-particle state labels \(j=1,\dots,N\). The operators $a_j^\dagger$ that create bosons in these states and their Hermitian conjugates, the destruction operators, satisfy the canonical commutation relations
\begin{align*}
    [a_i,a_j^\dagger]=\delta_{ij}1_B, \quad
    [a_i^\dagger,a_j^\dagger]~=~[a_i,a_j]=~0.
\end{align*}
Here, $1_B$ is the identity operator of the 
many-body bosonic Fock space. 
This space is built out of a normalizable vacuum state $|\Omega\rangle$ annihilated by all the destruction operators. The single-particle states,
or ``orbitals," are the states 
$|\psi_i\rangle\equiv a^\dagger_i|\Omega\rangle$
for $i=1,\dots,N$.

The Hamiltonian for a closed system of independent bosons is some Hermitian 
\emph{quadratic bosonic form}, conventionally presented in second quantisation ~\cite{Blaizot1986} as 
\begin{equation}
\label{eq:tracefulQBH}
{\cal H}=\sum_{i,j}\left(K_{ij}a_i^\dagger a_j + \frac{1}{2}\Delta_{ij}a_i^\dagger a_j^\dagger +\frac{1}{2}\Delta_{ij}^*a_ja_i\right).
\end{equation}
To get a physical picture of this object,
consider first the case $\Delta=0$. Then,
$$
K_{ij}=\langle \psi_i|{\cal H}|\psi_j\rangle
\quad (\Delta =0)
$$
is the expectation value of the many-body Hamiltonian ${\cal H}$ with respect to
the single-particle states $|\psi_i\rangle =
a_i|\Omega\rangle$. This is the reason
why the matrix the $N\times N$ matrix $K$ is
Hermitian and is often referred to as the 
``single-particle Hamiltonian." The $N\times N$ matrix $\Delta$ is symmetric and its physical
meaning is context dependent. For example,
quantized magnetic waves, magnons,
described within the quadratic approximation, display a non-zero $\Delta$ as a rule because 
magnon number need not be conserved. 
In any case, the matrices $K$ and $\Delta$
are arbitrary at this point. 

\paragraph{The transition from many-body bosonic to matrix algebra.} 
A \emph{linear bosonic form} is a linear combination of creation and destruction operators. That is, it is an \emph{operator} of the form 
$$
\hat{v}\equiv 
\sum_{i=1}^Nv_{0i}a_i^\dagger+v_{1i}a_i
%v_{01}a_1^\dagger +\dots+ v_{0N}a_N^\dagger+v_{11}a_1+\dots+v_{1N}a_N.
$$
In this linear combination of creation
and annihilation operators, the complex 
numbers $v_{0j}$ are the coefficients
for the bosonic creation operators and the $v_{1j}$ are the coefficients for the destruction 
operators.
The most important examples of linear bosonic
forms are the quasi-particle creation and annihilation operators. These
are the linear bosonic
forms that create and destroy the 
quasi-particles of the system described by 
${\cal H}$. 

In the expression for $\hat{v}$, let us collect the complex coefficients $v_{\tau i}$, $\tau=0, 1$, $i=1,\dots,N$ into a column vector 
\begin{align}
    v\equiv\begin{bmatrix}
        v_{01}\\
        \vdots\\
        v_{0N}\\
        v_{1N}\\
        \vdots\\
        v_{1N}
    \end{bmatrix}\in \Sigma.
\end{align}
Here, $\Sigma=\mathds{C}^2\otimes \mathds{C}^N$ is the auxiliary space of numerical column vectors.
Let us similarly arrange 
the bosonic operators into a row array
\begin{equation}
\label{eq:qbh:nambu-vector}
\Phi^\dagger=\begin{bmatrix}
a_1^\dagger &  \dots & a_N^\dagger & a_{1}& \dots&a_N
\end{bmatrix}.
\end{equation}
The associated array $\Phi$ is the column array
\begin{equation} 
\Phi=\begin{bmatrix}
a_1 \\ \vdots \\ a_N^\dagger.
\end{bmatrix}
\end{equation}
Both $v$ and $\Phi$ are column vectors with $2N$ elements, but $v$ contains complex scalars while $\Phi$ contains operators.
With these definitions in hand, we can compactly describe an arbitrary linear bosonic form as
the matrix product of the row array of operators $\Phi^\dagger$ and the column array
of complex numbers $v$, that is,
\begin{align}
%\label{eq:oneform}
\nonumber
\hat{v}=&\Phi^\dagger v = 
\begin{bmatrix}
a_1^\dagger &  \dots & a_N^\dagger & a_{1}& \dots&a_N
\end{bmatrix}
\begin{bmatrix}
        v_{01}\\
        \vdots\\
        v_{0N}\\
        v_{1N}\\
        \vdots\\
        v_{1N}
    \end{bmatrix}\\
    \label{eq:oneform}
    =&v_{01}a_1^\dagger +\dots+ v_{0N}a_N^\dagger+v_{11}a_1+\dots+v_{1N}a_N
%v_{01}a_1^\dagger +\dots+v_{1N}a_N
\end{align}
in terms of the usual algebra of arrays.
From this point on, we will use the shorthand $\tau_s$, $s=1, 2, 3$, in replacement of \(\sigma_s\otimes I_N\) with $\sigma_s$ the Pauli matrices and $I_N$ the identity matrix of size $N$. With this notations in place, one can check
that 
\begin{equation}
    \label{eq:oneform2}
    \Phi^\dagger v= v^T \tau_1 \Phi.
\end{equation}

After some rearrangement, the general quadratic bosonic Hamiltonian (QBH) of \cref{eq:tracefulQBH} can be rewritten as \cite{Flynn2020}
\begin{equation}
\label{eq:qbh:G-hat}
{\cal H}=\widehat{G}-\frac{1}{2}\tr(K)1_{B} 
\end{equation}
in terms of 
\begin{equation}
\label{eq:hatH}
\widehat{G}=\frac{1}{2}\Phi^\dagger G\tau_3\Phi=\widehat{G}^\dagger, \quad G=\begin{bmatrix} K & -\Delta\\ \Delta^* & 
-K^*\end{bmatrix}.
\end{equation}
The $2N\times 2N$ \emph{complex-valued matrix} $G$ is called the (bosonic) dynamical matrix.
In general, for a $2N \times 2N$ matrix $M$, \(M\tau_3\) has the effect of switching the sign of the $N \times N$ blocks in the second column of $M$. The awkward difference between ${\cal H}$
and $\widehat{G}$,  a shift by a constant multiple
of the identity operator, see Eq.~\eqref{eq:qbh:G-hat}, is peculiar to the
conventions of second quantisation. On one hand, it is important for modelling the phenomenon of Bose-Einstein condensation~\cite{10.1093/acprof:oso/9780198758884.001.0001}. On the other hand, the shift vanishes automatically for systems described more naturally in terms of position and momentum
quadratures 
\begin{align}
    \label{eq:quadx}
    x_j= \frac{a_j+a_j^\dagger}{\sqrt 2},\\
    \label{eq:quadp}
    %p_j=(a^\dagger_j-a_j)/i\sqrt{2}, EMILIO version
    p_j=i \frac{a_j^\dagger -a_j}{\sqrt 2},
\end{align}
like photons and phonons. Whatever its provenance,
for us, the central object of investigation is $\widehat{G}$.

\paragraph{Three fundamental formulas} The next steps in combining the bosonic and matrix algebra is to notice that %\added[id=al]{I find this confusing, since until now $\Phi^\dagger$ is a row vector, and $\Phi$ a coliumn vector. I understand the commutator is taken element-wise but this is not straightforward in a first read.}
\begin{equation}
    [\Phi,\Phi^\dagger] \equiv
\begin{bmatrix}[a_1, a_1^\dagger]& \cdots & [a_1,a_N]\\
\vdots & \ddots& \vdots\\
[a_N^\dagger, a_1^\dagger]&\cdots & [a_N^\dagger, a_N]
\end{bmatrix}= \tau_3 1_B.
\end{equation}
By combining this observation with  Eqs.~\eqref{eq:oneform}, \eqref{eq:oneform2},
and \eqref{eq:hatH}, one obtains the three fundamental formulas 
%\added[id=g]{Comments on three fundamental boson relations: bit of explanation about the many-body space, and matrix multiplication}
\begin{align}
\label{eq:conj}
\hat{v}^\dagger &= \widehat{{\cal C}v}, 
\quad {\cal C}v\equiv\tau_1 v^*,\\
\label{eq:indef}
[\hat{v}^\dagger, \hat{w}]&=\langle v|\tau_3|w\rangle 1_B,\quad \langle v|w\rangle \equiv v^\dagger w,\\
\label{eq:dynamicalm}
[\widehat{G}, \hat{v}]&=\Phi^\dagger Gv=\widehat{Gv}.
\end{align}
%\added[id=g]{is C defined somewhere?}
They provide a dictionary between many-body
operators and numerical vectors and matrices. 
The Eq.~\eqref{eq:dynamicalm}
indicates that one can compute the 
commutator $[\widehat{G}, \hat{v}]$ simply
by multiplying the numerical vector $v$ by
the dynamical matrix $G$. Similarly, \cref{eq:indef} indicates that one can
compute the commutator of two bosonic linear 
forms by taking the $\tau_3$ inner product
of two numerical vectors. 
Finally, \cref{eq:conj} indicates that one
can take the Hermitian conjugate of a bosonic
linear form $\hat{v}$ by replacing the numerical
vector $v$ with ${\cal C}v=\tau_1v^*$. 
%The usual procedure for diagonalising QBHs is built on these formulas, starting with Eq.~\eqref{eq:dynamicalm}, see for example Ref.~\cite{Blaizot1986}. 

\paragraph{Quantum dynamics}
The matrix $G$ is called the (bosonic) dynamical matrix because one can solve the Heisenberg equations of motion of the system of free
bosons by diagonalising $G$. One can see
that this is the case by calculating  
\begin{align}
    \dot{\Phi}^\dagger = \frac{i}{\hbar}[{\cal H}, \Phi^\dagger]=\Phi^\dagger G,
\end{align}
where the last equality follows from  Eq.~\eqref{eq:dynamicalm}.
Thus, the eigenvalues of $G$ are the normal 
frequencies of the system,
and the eigenvectors of $G$ determine the quasi-particle creation and annihilation operators
associated to ${\cal H}$. We already saw a concrete example in \cref{sec:photo-magnonic-crystals}.   

A bosonic dynamical matrix satisfies two and only two structural constraints \cite{Blaizot1986},
\begin{eqnarray}
\label{eq:pseudo}
\tau_3G^\dagger \tau_3&=&G,\\
\label{eq:ph}
\tau_1G^*\tau_1&=&-G. 
\end{eqnarray}
One can check directly that the block matrix $G$ of Eq.~\eqref{eq:hatH} satisfies both conditions. 
They capture
aspects of the many-body operator algebra. It
is illuminating to see exactly how. 
 First, since
\begin{equation*}
    0=[\widehat{G}, [\hat{v}^\dagger, \hat{w}]]=
-[[\widehat{G},\hat{v}]^\dagger, \hat{w}]+[\hat{v}^\dagger,
[\widehat{G}, \hat{w}]]
\end{equation*}
(the first equal sign follows from Eq.~\eqref{eq:indef} and the second one from
the Jacobi identity
for commutators), 
it follows that 
\begin{equation}
\label{eq:pseudoH}
0=-\langle Gv|\tau_3|w\rangle + \langle v|\tau_3|Gw\rangle.
\end{equation} 
That is, the bosonic dynamical matrix is self-adjoint with respect to the indefinite inner product derived from the bosonic commutation relations. Since $v$ and $w$ are arbitrary vectors, Eq.~\eqref{eq:pseudoH} boils down to Eq.~\eqref{eq:pseudo}. Following the convention in the literature, we say that $G$ is pseudo-Hermitian \cite{Flynn2020}. Second, 
\begin{equation}
    \widehat{{\cal C}Gv}=\widehat{Gv}^\dagger=[\widehat{G},\hat{v}]^\dagger=
-[\widehat{G},\widehat{{\cal C}v}]=-\widehat{G{\cal C}v}.
\end{equation}
That is,
\begin{equation}
\label{eq:partholesym}
{\cal C}G=-G{\cal C}.
\end{equation}
This last property is sometimes called the particle-hole symmetry of the bosonic 
dynamical matrix. Referring back to the definition of 
${\cal C}$ in Eq.~\eqref{eq:conj}, one
can check that Eq.~\eqref{eq:partholesym} 
boils down to the Eq.~\eqref{eq:pseudo}.

To summarize,
a) according to Eq.~\eqref{eq:dynamicalm},
a complex $2N\times 2N$ bosonic dynamical matrix $G$ acts on numerical column vectors $v$ from of a $2N$ dimensional auxiliary space 
$
\Sigma \equiv \mathds{C}^2\otimes \mathds{C}^N $
by matrix-vector multiplication. 
b) Along the same lines, by way of 
Eq.~\eqref{eq:indef}, the space 
$\Sigma$ acquires an indefinite inner product $\langle v|\tau_3|w\rangle=v^\dagger \tau_3 w$. It is important for
generalising these ideas to infinitely many bosonic creation
and destruction operators ($N=\infty$) that 
the indefinite inner product is related in
a simple way to the basic Hermitian inner
product $\langle v|w\rangle =v^\dagger w$.
And, c) there is a distinguished map of auxiliary vectors ${\cal C}v=\tau_1v^*$ 
that entered the picture with Eq.~\eqref{eq:conj}. The map
${\cal C}$ is anti-linear but it is not
related to any time reversal symmetry operation. 
The auxiliary space $\Sigma$ inherits these 
three features a), b), and c) 
from the bosonic many-body algebra by way
of the ``hat maps" $v\mapsto \hat{v}$ and $G\mapsto \widehat{G}$
of numerical vectors and bosonic dynamical matrices to bosonic linear and Hermitian quadratic forms respectively. The auxiliary
space $\Sigma$ is sometimes called the Nambu
space in the physics literature \cite{Ryu_2010}.
Mathematically, $\Sigma$ it is an example of a Krein space obtained as the complexification of 
a canonical symplectic space. The reader
interested in these technical terms 
should consult Ref.~\cite{GohbergIndefiniteLA}.

\paragraph{The physical meaning of $\Sigma$} 
%What is the physical meaning of 
%the complex numbers $v_{\tau i}$, that is,
%the entries of the auxiliary vector $v\in \Sigma$?
The technique of second quantisation
links the auxiliary space $\Sigma$ to the 
Hilbert space ${\cal H}_{sp}$ of states of 
a single 
boson. In second quantisation, the labels of 
the creation and destruction operators are 
also, by construction, the 
labels of some orthonormal basis of ${\cal H}_{sp}$. In fact, more is true: the creation operators produce a copy of ${\cal H}_{sp}$ 
inside the infinite-dimensional Fock space 
${\cal H}_{F}$ of bosonic many-body states.
Let us recall briefly how this works.
By construction, the Fock space includes
a normalisable state $|\Omega\rangle\in{\cal H}_{Fock}$ with no particles in it, that is,
$a_i|\Omega\rangle=0$ for any label $i$.
The states 
$$|\psi_i\rangle\equiv a_i^\dagger |\Omega\rangle
\in {\cal H}_{Fock},\quad i=1,\dots,N,
$$ 
are orthonormal and span the Hilbert space ${\cal H}_{sp}$. 

The bosonic linear forms $\hat{v}$ and
$\hat{v}^\dagger$ map the Fock vacuum to the states 
\begin{align}
    \widehat{v}|\Omega\rangle
    &= \sum_{i=1}^N  v_{0i}|\psi_i\rangle
%\in {\cal H}_{Fock}
,\\
    \widehat{v}^\dagger |\Omega\rangle &=
    \sum_{i=1}^N v_{1i}^*|\psi_i\rangle
    %\in {\cal H}_{Fock}
    ,
\end{align}
for one boson. Since one can think of $\hat{v}$
(there is no typo, we do mean $\hat{v}$)
as the creation operator of some quasi-particle,
we can also think of $\widehat{v}|\Omega\rangle$
as a ``particle state" and $$\widehat{v}^\dagger |\Omega\rangle$$ as the corresponding 
``hole state."
Thus, the Krein space $\Sigma$ is
an augmentation of the single-particle Hilbert space ${\cal H}_{sp}$ that accommodates both
particles and holes.  From now on, we will describe the auxiliary Krein space as 
$$
\Sigma = \mathds{C}^2\otimes {\cal H}_{sp}.
$$

\subsubsection{Topological invariants in 1D}
\label{subsubsec:topinvs}

In this section we will briefly summarise 
topological properties of some translation invariant operators that act on the auxiliary
Krein space $\Sigma$ or its factor ${\cal H}_{sp}$. In addition, we will
consider translation symmetry but in one direction only for simplicity. Let us sharpen 
the notation
accordingly by decomposing the single-particle
states according to a lattice site label
and and a label for an internal state. 
The quantum state for being at
site $r$ of an infinite, one-dimensional lattice is denoted by $|r\rangle$. Within the tight-binding approximation, different lattice sites are associated to orthonormal states. 
%\added[id=g]{may be stupid question why is this the case?}.
%They span the lattice Hilbert space $l^2(\mathds{Z})$. 
The state $|\phi\rangle$ for being in some 
internal state at site $r$ belongs to the
Hilbert space ${\cal H}_{int}$ of
internal states. Again, within the tight
binding approximation, it is assumed that 
${\cal H}_{internal}$ is finite-dimensional.
A generic state in the single-particle 
state space 
$$
{\cal H}_{sp}\equiv {\cal H}_{lattice} \otimes {\cal H}_{internal}
$$ 
is
\begin{equation}
    |\psi\rangle=\sum_{r\in\mathds{Z}}|r\rangle|\phi_r\rangle, \quad \langle \psi|\psi\rangle=\sum_{r\in\mathds{Z}}\langle\phi_r|\phi_r\rangle < \infty.
\end{equation}
In the following we will ignore the difference between $\Sigma$ and ${\cal H}_{sp}$ as a 
detail encoded together with the internal
states if appropriate. The focus is on the lattice.
%Let us go ahead and point out that, in this section,
%non-normalisable states $(\langle \Psi|\Psi\rangle = \infty)$ are not allowed into the picture. Eigenvector of operators in particular refer to normalisable eigenvectors only. 

With these notations in place, one can describe
the left and right translation (or ``shift") operators 
of the infinite lattice. The left-shift operator is
\begin{equation}
    \mathbf{T}=\sum_{r\in\mathds{Z}} |r\rangle\langle r+1|\otimes I
\end{equation}
($I$ is the identity matrix for $\mathds{C}^n$.)
and, as one can check, $\mathbf{T}^\dagger$ is
the right-shift operator. A 
translation invariant operator can be 
conveniently expressed in terms of shift 
operators as 
\begin{equation}
    \label{eq:boldO}
    \mathbf{B}=\sum_{n\in\mathds{Z}} \mathbf{T}^{\dagger\, n} \otimes B_n.
\end{equation}
The Fourier transform of the lattice
degrees of freedom, by way of a change of 
basis to eigenstates of momentum
\begin{equation}
    |k\rangle=\sum_{r\in\mathds{Z}}\frac{e^{irk}}{\sqrt{2\pi}}|r\rangle,
\end{equation}
yields the periodic, matrix-valued function of $k$ 
\begin{align}
    \label{eq:FCO}
    \nonumber
    \mathbf{B}&=\sum_{k\in[-\pi,\pi)}|k\rangle\langle k|\otimes B(k),\\
    B(k)&=\sum_{n\in\mathds{Z}}e^{ink}B_n.
\end{align}
What we have described is the usual workflow 
in physics. For the sake of bringing topology into the picture,
we need to reverse it. Let us take as the starting 
point the space of \underline{continuous}, periodic of period $2\pi$, 
complex matrix-valued functions $B(k)$. Continuity is enough to guarantee 
that the coefficients $B_r$ in Eq.~\eqref{eq:FCO} are uniquely
determined by the function. Hence, we can plug them back
in Eq.~\ref{eq:boldO} to build, out of any continuous function
and without ambiguity, a corresponding translation invariant operator $\mathbf{B}$. 
To remind ourselves of this inverted workflow, we call  $B(k)$ the symbol of the 
translation invariant operator $\mathbf{B}$, following established language~\cite{bleecker2013index}. 
%These fine points do not pose any computational issues if one works with short range models.

\paragraph{The winding invariant.} Being continuous functions, we can investigate the class of symbols with tools from 
topology and ask what that tells us about the associated translation
invariant operators. For starters, as it turns out, if the symbol is algebraically 
invertible, that is, if 
\begin{align}
    \label{eq:fredholmgap}
    \det\,B(k)\neq 0\quad \mbox{for all}\ k\in [-\pi,\pi),
\end{align}
so that $B(k)^{-1}$ exists for each value of the crystal momentum, 
then  $\mathbf{B}$ is an invertible operator. 
Let us focus on the subset of algebraically invertible 
symbols. For them, the map $k\mapsto \det B(k)$ 
is a map from the Brillouin zone, topologically,
a one-dimensional torus, to the complex plane
that misses the origin. Such maps are topologically
classified by the integer-valued winding number
\begin{align}
    &\mbox{winding invariant}(\det B,0) = \nonumber \\
    &\frac{1}{2\pi i}\int_{-\pi}^\pi
    \frac{d}{dk}\log\det B(k)\, dk.
\end{align}
The statement that the winding invariant takes the
integer value $n$ means
that $\det B(k)$ circles the origin of the complex plane $|n|$ 
times. A positive (negative) integer $n$ indicates 
counterclockwise (clockwise) overall circulation. 
For example, for the symbol $B(k)=e^{ink}I$, the winding 
number is $n$. 

We can think of the situation as follows:
the set of invertible symbols consists of a disjoint
collection of path-connected components 
labelled by the winding number. There is no continuous path 
connecting two different components. This picture about 
continuous functions, the symbols, repeats
itself for the invertible translation invariant operators 
in Hilbert space because the mapping of symbols into 
operators is continuous.
\iffalse
This observation can be physically interesting because of the
following implication. Say one is looking for
an adiabatic deformation $\textbf{O}_\tau,$ $\tau\in[0,1]$, of
the invertible operator $\textbf{O}_0$ into the invertible operator
$\textbf{O}_1$. If the winding numbers of the 
corresponding symbols $O_1(k)$, $O_2(k)$ are different,
then one can be certain that there is no such adiabatic
deformation.
\fi

\paragraph{The Pfaffian invariant.} For a subset of symbols with vanishing winding number,
and their associated invertible operators, there exists 
a refinement of this picture \cite{PMIHES_1969__37__5_0, schulzbaldes2015z2indicesfactorizationproperties}. 
The starting point is to focus on even-dimensional internal 
spaces and invertible symbols $A(k)$ such that
\begin{eqnarray}
    A(k)^*=A(-k),\quad A(k)^T=-A(-k).
\end{eqnarray}
Taken together, these conditions imply that the operator
$\textbf{A}=\sum_{r\in\mathds{Z}}\mathbf{T}^\dagger\otimes A_r$
is real and antisymmetric.
They also imply that the winding invariant of $\det A(k)$ is necessarily zero. Nonetheless, this subset of symbols
consists of two different components that cannot be connected
by any continuous path. This feature transfers to the 
associated set of operators. One can assign a label $\pm 1$ to 
the two components and relate it to elements by way of the topological invariant  
\begin{eqnarray}
    \mbox{Pfaffian invariant}(A)=\mbox{sign}\,
    \frac{\mbox{Pf}\,A(0)}{\mbox{Pf}\,A(-\pi)}.
\end{eqnarray}
The Pfaffian $\mbox{Pf}\,A$ of a real and antisymmetric matrix $A=[a_{ij}]$ of even 
order is calculated by way of the formula
\begin{equation}
    \mbox{Pf}(A)= 
\frac{1}{2^{n}n!}\sum _{\sigma \in S_{2n}}\mbox{sgn}(\sigma)\prod_{i=1}^{n}a_{\sigma(2i-1),\sigma(2i)}
\end{equation}
in terms of the permutations $\sigma$ of $2n$ letters. 
In particular, for a $2 \times 2$ matrix, 
\begin{equation}
    \label{eq:pfaffian:2}
    A = \mqty[0 & a \\ -a & 0]
\end{equation}
the Pfaffian is $\mathrm{Pf}(A)=a$ and for a $4 \times 4$ matrix
\begin{equation}
    \label{eq:pfaffian:4}
    A = \mqty[0 & a & b & c \\ -a & 0 & d & e \\ -b & -d & 0 & f \\ -c & -e & -f & 0]
\end{equation}
$\mathrm{Pf}(A)=af-be+cd$.
The symbol $A(k)$ is real and antisymmetric precisely for
the two values of crystal momentum $k=0$ and $k=-\pi$. 
%\added[id=g]{we may miss a sentense to conclude this section}

\subsubsection{Translation invariant 
bosonic dynamical matrices} 
If the many-body Hamiltonian $\widehat{G}$ is translation invariant, then 
same is true of the dynamical matrix $G$~\cite{Blaizot1986};
see the supplementary materials for a worked
out example. Let's consider calculating 
the invariants of this section for $G(k)$.
The Pfaffian invariant is simply not relevant for a generic $G(k)$: most bosonic dynamical 
matrices are neither real nor anti-symmetric.  
The winding number invariant is naturally applicable, but it vanishes automatically 
because a bosonic dynamical matrix is necessarily 
pseudo-Hermitian (see Eq.~\eqref{eq:ph}). 
We are failing to achieve interesting results
because we are missing a key ingredient: physically meaningful symmetry conditions.

\subsection{Classifying symmetry operations}

\subsubsection{Continuous symmetries: ${\cal N}$, ${\cal S}$}
\label{subsubsec:ContinuousSymmetries}

Three of our four fundamental bosonic symmetries that we will consider are continuous symmetries generated by one of the following Hermitian quadratic bosonic forms.
\begin{itemize}
\item A squeezing transformation
\begin{equation}
    {\cal S}_1=\frac{1}{2}\widehat{\beta_1}=\frac{i}{4}\sum_j 
(a_j^{\dagger\, 2}- a_j^{2}).
\end{equation}
%\item bosonic time reversal ${\cal T}$.
\item Another squeezing transformation 
\begin{equation}
    \label{eq:beta_2}
    {\cal S}_2=\frac{1}{2}\widehat{\beta_2}=\frac{1}{4}\sum_j
    (a_j^{\dagger\, 2}+ a_j^{2}).
\end{equation}
\item Particle number
\begin{equation}
    {\cal N} = \frac{1}{2}\widehat{\beta_3}=\frac{1}{4}\sum_j 
(a_j^\dagger a_j + a_ja_j^\dagger).
\end{equation}
\end{itemize}
As in the background section, $j$ is an abstract, catch-it-all label. 
For example, it could include a spatial, lattice site label and some other labels for internal degrees of freedom.
The bosonic dynamical matrices of the symmetry generators are 
\begin{equation}
\beta_1=-i\tau_1,\ \beta_2=-i\tau_2,\ \beta_3=\tau_3.
\end{equation}

 One can check that 
\begin{align}
    [{\cal S}_1, {\cal S}_2]&= - i {\cal N},\\
    [{\cal N}, {\cal S}_1]&=i{\cal S}_2,\\
    [{\cal N}, {\cal S}_2]&=-i{\cal S}_1.
\end{align}
There is a shortcut for checking these relations; see Eqs.~\eqref{eq:BracketQBFs} and \eqref{eq:BracketBDMs}.
A neat physical interpretation of this Lie algebra follows from noticing that the quadratic Casimir operator is
$$
{\cal Q}^2= {\cal N}^2-{\cal S}_1^2-{\cal S}_2^2,
$$
meaning that ${\cal Q}^2$ commutes with all three symmetry generators.
Hence, we are dealing with the Lie algebra of the Lorenz group in $2+1$ space-time dimensions as have been noticed every now and then over the years; see Ref.~\cite{SUBRAMANYAN2021168470} for
a recent discussion. 

The fundamental difference between particle number and squeezing symmetry operations
is that the particle number operator generates compact phase rotations while 
the squeezing transformations generate non-compact hyperbolic rotations. 
With hindsight, the best way to handle the pair of squeezing transformations is to investigate 
classes of Hamiltonians that commute 
with some fixed but arbitrary normalised combination
\begin{align}
    {\cal S}=n_1{\cal S}_1+n_2{\cal S}_2,\quad n_1^2+n_2^2=1.
\end{align}
No information is lost (with respect to keeping both ${\cal S}_1$ and 
${\cal S}_2$ in the picture) because a class of Hamiltonians that 
commute with both squeezing transformations can be equivalently 
characterised as being a class that commutes with a single suitable 
${\cal S}$ and ${\cal N}$.

\subsubsection{Time reversal: ${\cal T}$}
We will assume for simplicity that time reversal is the antilinear
transformation of the bosonic Fock space that acts on the basic 
creation and destruction operators as 
$
{\cal T}a_i{\cal T}=a_{i}
$, 
$
{\cal T}a_i^\dagger {\cal T}=a_{i}^\dagger.
$ 
For lattice systems for example, this means that the creation
and destruction operators are labelled by lattice sites. 
Then, the effect of time reversal is to
map the dynamical matrix to its complex conjugate,
$$
{\cal T} \widehat{G} {\cal T} = \widehat{G^*}.
$$
We do not expect that more complicated implementations of bosonic 
time reversal (e.g., by including an integer spin degree of freedom) will yield any more interesting results. The key issue at stake is 
that for any system of bosons, ${\cal T}^2=1_B$ due to the general
relationship between rotations and time reversal \cite{ballentine2014quantum} and the spin-statistics theorem.

\subsection{Symmetry classes}

\begin{table*}[]
\begin{center}
\begin{tabular}{|c | c| c| c | c |} 
 \hline
Many-body symmetry class&  Parametrisation of $G$ in& Possible & Topological invariant & Analytical index \\
(independent of space dimension) & terms of square matrices & descriptor & (one dimension) &(one dimension) \\[0.5ex] 
 \hline\hline
 $\{\}$ & one Hermitian matrix and & symplectic & trivial & 0 \\
         & one complex symmetric matrix&         &        & \\
 \hline
 $ \{{\cal T}\} $ & two real symmetric matrices &chiral symmetric & trivial & 0 \\
\hline
 $\{{\cal N}\}$ & one Hermitian matrix & A & trivial& 0 \\
 \hline
 $\{{\cal S}\}$& one real matrix & BDI & \textbf{winding invariant} & $\mathds{Z}$ \\
 \hline 
 %$\{{\cal S}_2\}$& real symmetric & AI & \textbf{winding number} & $\mathds{Z}$ &\textbf{YES} \\
 %\hline 
  $\{{\cal T}, {\cal N}\}$ & one real symmetric matrix & AI & trivial & 0 \\
 \hline  
 $\{{\cal T}, {\cal S}\}$ & one real symmetric matrix & AI & trivial & 0 \\
 \hline 
$\{{\cal N}, {\cal S}\}$
& one real antisymmetric matrix & D & \textbf{Pfaffian invariant} & $\mathds{Z}_2 $ \\
 %& of \textbf{even} order &  &  & &  \\
\hline
 %& not DAZ, & & & & \\
 %$\{{\cal N}, {\cal S}\}$
%& real antisymmetric &  & trivial & 0  & N/A\\
 %& of odd order &  &  & &  \\
 %\hline
$\{{\cal T}, {\cal N}, {\cal S}\}$ &  $G=0$ &  & trivial & 0 \\
\hline 
\end{tabular}
\end{center}
\caption{The bosonic many-body symmetry classes of this paper and some of their properties. The symmetries are ${\cal T}$, many-body 
time reversal with ${\cal T}^2=1$, particle number ${\cal N}$, 
which generates a compact group of unitary transformations of 
the Fock space, and a squeezing transformation ${\cal S}$, 
which generates a non-compact group of unitary transformations 
of the many-body Fock space. The bosonic class D is topologically non-trivial if and only if the systems feature an even number of 
bosonic degrees of freedom per lattice site. Since this is not 
a  symmetry condition, our scheme goes slightly beyond pure symmetry protection. 
}
\label{table:symmclasses}
\end{table*} 

A symmetry class is a class of QBHs such that every member is left invariant by some fixed set of symmetry operations. Space dimensionality plays no role at this basic level of classification. It is sensible to include in this framework a class for which the set of symmetry operations is empty. Every QBH that one can define with the prefixed set of single-particle labels belongs to this class. The opposite situation is the class containing only the zero QBH with dynamical matrix $G=0$. In between these two extreme cases, the symmetry conditions induce some general structure on the dynamical matrices of the members of a symmetry class. Zooming in on our specific scenario, we will consider time reversal which, according to our comments above, forces the dynamical matrix to be real, and symmetry operations generated by the three quadratic bosonic forms 
$\hat{\beta}_1$, $\hat{\beta}_2$, and $\hat{\beta}_3$. To understand the impact of these symmetries on the members of a symmetry class, we need to combine two observations.

First, one can expand a general dynamical matrix as 
\begin{widetext}
\begin{align}
\label{eq:gen_bdm}
G=\begin{bmatrix} 
K& -\Delta\\
\Delta^* & -K^* \\
\end{bmatrix}
= \beta_1\otimes \Delta_{im} + \beta_2\otimes\Delta_{re} + \beta_3\otimes K_{re} + i I_2\otimes K_{im}
\end{align}
\end{widetext}
where $K$ and $\Delta$ are the hopping and pairing matrices introduced in \cref{subsec:formalism-qbh}.
Notice that $K_{im}$, the imaginary part of the $N\times N$ Hermitian matrix $K$, is a real antisymmetric matrix, and 
$K_{re}$, the real part of $K$, is a real symmetric matrix.
Since $\Delta$ is some complex symmetric matrix, 
$\Delta_{re}$ and $\Delta_{im}$ are real symmetric matrices. 

Second, a symmetry condition, if it is associated to a continuous symmetry generated by a quadratic bosonic form, induces a corresponding symmetry condition on the dynamical matrix. We need to make this statement explicit for characterizing symmetry classes. The starting point is to notice that the commutator of two Hermitian quadratic bosonic forms is another Hermitian quadratic bosonic form, up to a factor of $i$. That is,
\begin{align}
\label{eq:BracketQBFs}
    [\widehat{G_1},\widehat{G_2}]= i \widehat{G_3}.
\end{align}
If $G_1$ and $G_2$ are given as input, what is $G_3$? Building on the formalism of the background section,
one finds that  
\begin{align}
   i\Phi^\dagger G_3v&= i[\widehat{G_3}, \widehat{v}]= \nonumber\\
   [[\widehat{G_1},\widehat{G_2}], \hat{v}]&=\widehat{G_1G_2v}  -\widehat{G_2G_1v}=\Phi^\dagger[G_1,G_2]v
   \nonumber
\end{align}
for an arbitrary linear bosonic form $\hat{v}$. Hence,
\begin{align}
\label{eq:BracketBDMs}
    [G_1,G_2]=iG_3.
\end{align}
In particular, any two quadratic bosonic forms commute if and only if their respective bosonic dynamical matrices commute.

The main results of this section of the paper are summarised in \cref{table:symmclasses}. The figure \cref{fig:topo-classes} provides a more
visual presentation of some of the same information. 
%Here, $AZ$ stands Altland and Zirnbauer after their classical paper Ref.~\cite{PhysRevB.55.1142}. 
It is convenient to develop some notation for the symmetry classes.  The notation \(\{\cdots\}\)
means ``the set of all one-dimensional QBHs invariant
under every operation in the list \(\cdots\)." Hence, \(\{\}\) denotes the set of all one-dimensional QBHs, as there are no symmetry constraints, and so we call the associated class of dynamical matrices symplectic. 
The Cartan labels are borrowed from Table~\ref{tab:many_body_AZ}; they reflect the type of matrix parametrisation
of the bosonic dynamical matrix in the cases where it matches most closely some a fermionic symmetry class.

\subsubsection{Class $\{\}$}
This is the class of generic bosonic dynamical matrices 
subject to no symmetry conditions. It contains every other 
symmetry class. The dynamical matrices are parameterised 
by four independent real matrices, three symmetric and one 
antisymmetric, of arbitrary order; see Eq.~\eqref{eq:gen_bdm}
and the surrounding explanation. This comes relatively
close to the class C of quaternionic anti-Hermitian matrices
to the extent that these matrices are also parameterised by
four real matrices, three symmetric and one antisymmetric,
\underline{of even order}. However, even if one were to zoom 
in on bosonic dynamical matrices of even order, it is still 
the case that the constitutive real matrices are not fit together 
in the same way for the symplectic and anti-Hermitian quaternionic
case. Hence, not surprisingly, this bosonic symmetry class is not related
to any of the Altland-Zirnbauer (Cartan inspired) labels. 
As it stands, this class is topologically trivial in one dimension. 

\subsubsection{Class $\{{\cal T}\}$} 
The dynamical matrix of a QBH in this symmetry class is of the form
$$
G_{\cal T} = \beta_2\otimes \Delta_{re}+\beta_3\otimes K_{re}.
$$
Hence, this symmetry class can be characterised, at the ``single-particle" 
level, as the class of bosonic dynamical matrices possessing $\tau_1=i\beta_1$ as a chiral symmetry, that is, 
$$
\{G_{\cal T}, \tau_1\}=0, \quad \tau_1\tau_1^\dagger=\tau_1^2=I_2.
$$
Unfortunately, nothing useful follows from this elegant 
characterisation.  

Let us transform the bosonic dynamical matrix
$G_{\cal T}$ to the off-diagonal basis determined by
the chiral symmetry $\tau_1$. In this basis, the same 
for all members of the symmetry class, it takes the 
auxiliary form 
$$
A_{\cal T}= MG_{\cal T}M^{-1} = 
\begin{bmatrix}
0 & K_{re} - \Delta_{re}\\
K_{re} + \Delta_{re} & 0
\end{bmatrix}.
$$ 
We conclude that the $G_{\cal T}$ are parameterised 
by two independent real symmetric matrices. The 
class CI of complex symmetric matrices is also parameterised
by two real symmetric matrices. However, these two real
symmetric matrices are not fit together in the same way 
for the complex symmetric case on one hand 
and the time-reversal invariant symplectic case on the other.  
As far as we can 
see, this class is topologically trivial in one dimension.

\iffalse
A chiral symmetry is a very powerful constraint for fermionic 
single-particle, that is, Bogoliubov-deGennes Hamiltonians because these 
Fermionic dynamical matrices are Hermitian. Hence, in the off-diagonal
basis associated to the chiral symmetry, a fermionic single-particle
Hamiltonian in the chiral symmetry class takes the form 
$H_{BdG}=\begin{bmatrix}0 & C\\ C^\dagger &0\end{bmatrix}$ with $C$ a 
matrix in one of the AZ classes depending on the additional classifying symmetries.
This structure is ripe for a bulk-boundary correspondence, as we will illustrate
for bosons later.
\fi

\subsubsection{Class $\{ {\cal N}\}$} 
This is the class of systems characterised by the conservation of particle number. 
The dynamical matrix takes the form
$$
G_{\cal N} =   \beta_3\otimes K_{re} + iI_2\otimes K_{im}
$$
and so it is parameterised by the Hermitian matrices. Hence,
this bosonic symmetry classes realises class A. 
It is topologically trivial in one dimension.

\subsubsection{Class $\{{\cal S}\}$: winding invariant}
If $G$ is in the symmetry class associated to 
$\mathcal{S}=n_1\widehat{\beta}_1+n_2\widehat{\beta}_2$, 
then $[G,\beta]=0$ for $\beta_1'\equiv n_1\beta_1+n_2\beta_2$. Let us introduce 
$\beta_2'\equiv \pm(-n_2\beta_1+n_1\beta_2)$ and $\beta_3'\equiv\beta_3$. 
The overall sign of $\beta_2'$ should be 
chosen so that 
$$
\beta_1\beta_2\beta_3=\beta_1'\beta_2'\beta_3'.
$$ 
Let's assume
for concreteness that the overall positive sign is the appropriate
choice. The other case yields to the same analysis modulo some sign
flips here and there. The matrices 
$\beta_1'$, $\beta_2'$, and 
$\beta_3'$ satisfy, by construction, 
the same algebraic relations as the original 
$\beta_i$, $i=1, 2, 3$, matrices. It follows that there is 
a similarity transformation mapping 
$\beta_1,\ \beta_2,\ \beta_3$ to $\beta,\beta', \beta''$, 
respectively. 
 
In terms of the new beta matrices adapted to the symmetry operation 
${\cal S}$, the decomposition of the bosonic dynamical matrix is 
\begin{widetext}
\begin{align}
G
= \beta_1'\otimes \frac{n_1\Delta_{re}+n_2\Delta_{im}}{2n_1n_2} + \beta_2'\otimes\frac{n_2\Delta_{re}-n_1\Delta_{im}}{2n_1n_2} 
+ \beta_3'\otimes K_{re} + i I_2\otimes K_{im}
\end{align}
\end{widetext}
Hence, the dynamical matrix of a QBH in this symmetry class is of the form  
\begin{align*}
G_{{\cal S}}=\beta_1'\otimes \frac{n_1\Delta_{re}+n_2\Delta_{im}}{2n_1n_2} +i I_2\otimes K_{im}.
\end{align*}
Since $\beta_1'$ is equivalent to $-i\tau_1$ up to a change of basis, 
one can block-diagonalize 
any bosonic dynamical matrix in this class by way of some
fixed (for the whole symmetry class) change of basis $M$. The 
resulting auxiliary matrix is
\begin{eqnarray}
\label{eq:aux_b1}
A_{{\cal S}} &=& MG_{{\cal S}}M^{-1} = i \begin{bmatrix}N & 0\\
0 & -N^T\end{bmatrix}
\end{eqnarray}
in terms of the real matrix
\begin{eqnarray}
N&\equiv& K_{im}-\frac{n_1\Delta_{re}+n_2\Delta_{im}}{2n_1n_2}.
\end{eqnarray}
Any real matrix can be obtained from some QBH in this symmetry class. 
Hence, this class is a bosonic realisation of the class BDI, see 
Table~\ref{tab:many_body_AZ}. Having noted this, it is convenient to 
absorb the factor of $i$ into the definitions so that 
\begin{eqnarray}
A_{{\cal S}} =   \begin{bmatrix}B & 0\\
0 & B^\dagger \end{bmatrix}, \quad B\equiv iN.
\end{eqnarray}

Let us focus next on translation invariant
systems and investigate the symbol $B(k)$ associated
to an operator $\mathbf{B}$ derived 
from a translation invariant bosonic dynamical matrix
$\mathbf{G}_{{\cal S}}$. Since this symbol is in one-to-one
correspondence with the dynamical matrices of this
symmetry class, we can classify topologically the translation
invariant dynamical matrices in terms of the winding invariant
\begin{align}
    \label{eq:frakn}
    {\mathfrak n}(\mathbf{G}_{{\cal S}}) = 
    \mbox{winding invariant}(\det B(k), 0) 
    %=\frac{1}{2\pi i} \int_{-\pi}^\pi \frac{d\log\det B(k)}{dk}\,dk 
\end{align}
provided that the symbol $B(k)$ is algebraically invertible. 
We can check whether this is the case directly 
from the dynamical matrix since 
$$
\det(G_{{\cal S}}(k))=|\det(B(k))|^2.
$$
Hence, we can describe the situation directly in
terms of the band structure of the dynamical matrix:
a translation invariant QBH in the symmetry class 
$\{{\cal S}\}$ can be 
topologically classified by a winding invariant 
if and only if $0$ is in a band gap. Generically, the
bands are complex-valued.

\subsubsection{Class 
$\{{\cal T}, {\cal N}\}$}
Since
$$
G_{{\cal T},\, {\cal N}}= \beta_3\otimes K_{re},
$$
is parametrised by the real symmetric matrices, 
this bosonic symmetry class realises the class AI. 
It is topologically trivial in one space dimension.

\subsubsection{
Class $\{{\cal T}, {\cal S}\}$}
Since only ${\cal S}_2$ is left invariant by ${\cal T}$,
the only available option is the case ${\cal S}={\cal S}_2$. 
Then, 
$$
\quad G_{{\cal T}, {\cal S}_2}= \beta_2\otimes \Delta_{re}.
$$
This is another bosonic realization of the class AI.

\subsubsection{Class $\{{\cal N}, {\cal S}\}$: Pfaffian invariant}
The structure of the general bosonic dynamical matrix 
in this symmetry class is
$$
G_{{\cal N}, {\cal S}}= iI_2\otimes K_{im}.
$$
At this point, the analysis branches out.  If the number 
of degrees of freedom 
per lattice site is odd, then this symmetry class is 
topologically trivial. If the number of 
Bosonic degrees of freedom per lattice site is even, then 
this symmetry class is a bosonic realisation of 
class D and it is topologically non-trivial in one dimension.
The additional constraint on the number of local degrees of 
freedom is not a symmetry 
condition. However, it play a crucial role as a protecting 
condition together with the symmetry conditions. For this reason,
this bosonic symmetry class is protected by a little
bit more than symmetries. 

Let's zoom in on the even case and add translation
invariance to the picture. Then, 
one can classify the bosonic dynamical matrices in
this symmetry class by way of the $\mathds{Z}_2$-valued
quantity
\begin{equation}
    \label{eq:fraks}
    {\mathfrak s}(\mathbf{G}_{{\cal N}, {\cal S}}) = 
    \mbox{Pfaffian invariant}(K_{im}(k)).
\end{equation}
That is, provided the symbol $K_{im}(k)$ is algebraically invertible
or, equivalently, provided that the bands of the system, necessarily real-valued in this case due to the number
symmetry, do not touch $0$.

\subsubsection{
Class 
$\{{\cal T}, {\cal N}, {\cal S}\}$}
In this symmetry class,
\begin{equation}
    G_{{\cal T}, {\cal N}, {\cal S}}=0.
\end{equation}
That is, this symmetry class contains only the zero QBH.  

\iffalse
\subsubsection{Remark: A physical comparison of 
$\{{\cal T}, {\cal N}\}$ and $\{{\cal T}, {\cal S}\}$}
These classes are structurally closely related. In the first class,
the general dynamical matrix is of the form $\beta_3\otimes S$
for some real symmetric matrix, and in the second class it is of the
form $\beta_2\otimes S$. In this sense, they are both realizations
of the AZ class AI, according to our convention for exporting
AZ labels to bosons. However, these two classes are are physically 
very different. While the free boson systems associated to $G_{{\cal T}, N}$ 
are dynamically stable, the systems associated to $G_{{\cal T}, {\cal S}}$ are 
dynamically unstable (they can be used to model quantum parametric amplifiers). 
This physical distinction is a manifestation
of the mathematical fact that there is no symplectic transformation
that can rotate $\beta_2$ into $\beta_3$.
\fi

\subsection{Bulk-boundary correspondences}
The topologically non-trivial, in one dimension, 
symmetry classes $\{\cal S\}$ and $\{\cal N, S\}$ 
support a bulk-boundary correspondence. To avoid 
mathematical complications, we 
demonstrate this point in the simplest possible way, 
building on
the two basic index theorems for Fredholm operators and 
no additional formalism \cite{bleecker2013index}. 

\subsubsection{Basic index theory in 1D}

\begin{figure*}
    \centering
    \includegraphics[width=0.75\linewidth]{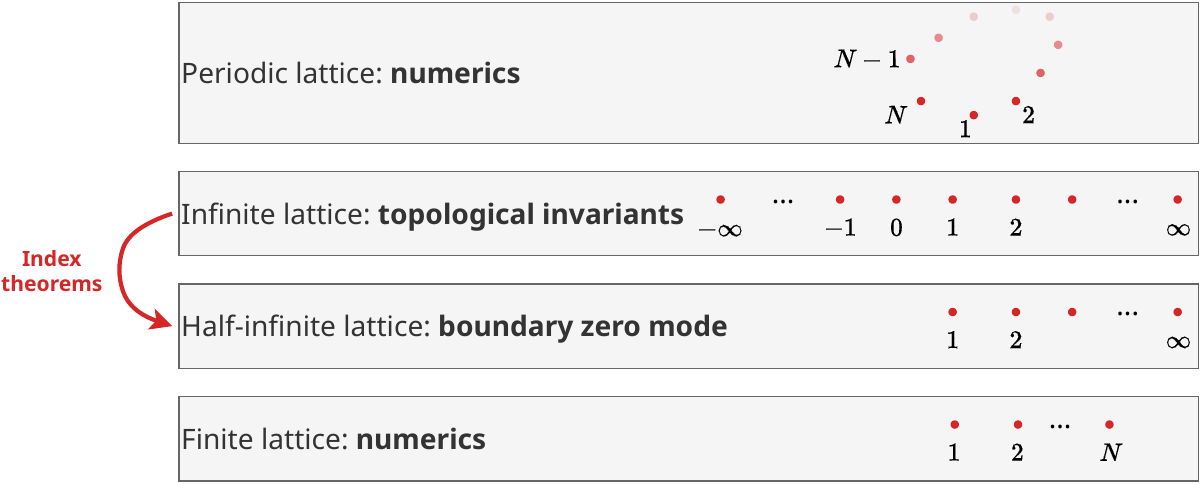}
    \caption{A bulk-boundary correspondence establishes a relationship between the bulk topological properties of a translation-invariant system, supported on an infinite lattice and the exponentially decaying zero modes that the same system might develop in a half-infinite lattice. As such, the predictions of a bulk-boundary correspondence are idealised versions of what a numerical simulation might show. On one hand, there are satisfactory methods for computing topological invariants from finite systems subjected to periodic boundary conditions. On the other hand, a finite system subjected to open boundary conditions necessarily has two terminations. If the terminations are sufficiently far apart, then one can expect 
    modes with energy very close to zero and localised on the components of the boundary (the left or right termination for example). 
    The energy of a topologically mandated boundary mode scales to 
    zero exponentially fast with the number of sites $N$. Generically,
    other skin-effect modes, if any, will not display any particular scaling with system size.}
    \label{fig:IndexThrms}
\end{figure*}

We are picking up the story where we left it at the end 
of Sec.~\ref{subsubsec:topinvs}. 
The next step is to map the translation invariant operator $\mathbf{B}$
associated to the symbol $B(k)$ to the corresponding operator for a half-infinite
lattice with one termination on the left, see \cref{fig:IndexThrms}. 
The lattice Hilbert space becomes 
$l^2(\mathds{N})$ spanned by the lattice states $|r\rangle$, $r\in\mathds{N}$. We visualise
this chain as growing towards the right from its termination on the left.
One can embed these states of the half-infinite lattice into the state space 
of the infinite lattice by thinking of them as states of the latter with vanishing amplitude for $r<0$. 
Let's call this mapping $E$. Similarly, we can
take a state of the infinite lattice and project out
half of the lattice to obtain a state of the half-infinite lattice. Let's call
this projector $P$. With the help of these mappings, we can transfer the bulk operator
$\textbf{B}$ to the half-infinite lattice by
way of the formula
$$
B \equiv P\mathbf{B}E.
$$
Schematically,  
$$
B:l^2(\mathds{N})\stackrel{E}{\longrightarrow}
l^2(\mathds{Z})\stackrel{\mathbf{B}}{\longrightarrow}
l^2(\mathds{Z})\stackrel{P}{\longrightarrow}l^2(\mathds{N})
$$
where we have suppressed any explicit reference
to the internal state space $\mathds{C}^n$. For example,
$$
T=P\mathbf{T}E=\sum_{r\in \mathds{N}} |r\rangle\langle r+1|\otimes I_n.
$$

Let us now consider the following question. 
We know that $\mathbf{B}$ is 
an invertible operator precisely when its 
symbol is invertible in an algebraic sense.  
Does it follow that $B=P\mathbf{B}E$ is invertible?
Clearly it does not, since $\mathbf{T}$ (with symbol $T(k)=e^{ik}I$) is invertible
and $T$ is not: $T|r=0\rangle|\psi\rangle=0$. However, it is true that
$B$ is ``almost invertible" or ``Fredholm" \cite{lax2014functional} 
(an idea that only makes sense in infinite dimensions) if $\mathbf{B}$ is
invertible. What this means is that the
dimension of the kernel of $B$, is finite and so
is the dimension of the complement of the range
of $B$. This last quantity is called the dimension
of the cokernel, and can be calculated as the dimension 
of the kernel of $B^\dagger$. The 
analytical (as in functional analytical, meaning
linear algebraic) index of $B$ is defined as
\begin{equation}
\mbox{index}(B)=\dim\,\ker\,B-\dim\,\ker\,B^\dagger. 
\end{equation}
Now we can state the result
we are after: there is an index theorem that
has as a corollary that
\begin{equation}
    \label{ToeplitzITh}
    \mbox{index}(B) = 
    \mbox{winding invariant}(\det B(k), 0). 
\end{equation}
For more details, we recommend consulting first
Ref.~\cite{lax2014functional} and follow with Ref.~\cite{bleecker2013index}.

Physicist often work with ``state" vectors of infinite norm like the eigenstates of translations. It is important 
to keep in mind that, in the mathematical context of the index theorem, non-normalisable vectors
\underline{do not count}. Only normalisable vectors count as 
kernel or co-kernel vectors. This is how boundary localisation enters the picture in connection to 
topologically mandated zero modes.

There is a second index theorem for the real antisymmetric
Fredholm operators $\mathbf{A}$. In this case, the appropriate analytical
index is~\cite{PMIHES_1969__37__5_0}
\begin{equation}
    \mbox{index}\, A\equiv(-1)^{\dim\ker A},
\end{equation}
with as a corollary
\begin{eqnarray}
    \label{PfaffianIth}
    \mbox{index}(A)=\mbox{Pfaffian invariant}(A(k)).
\end{eqnarray}

Deploying index theorems for making physical
predictions requires ingenuity. Symmetries 
play a crucial role, as we will illustrate below
(see also Ref.~\cite{2023Alase} for a similar
approach to the fermionic bulk-boundary correspondence). 
Before we do that, let's us conclude with some further 
remarks about the ideas of this section.

First, index theorems connect ideal infinite systems
to ideal half-infinite systems. Neither one is,
strictly speaking, amenable to numerical simulation.
As a proxy for the former, one can simulate a finite
system subjected to periodic boundary conditions, and
there are numerical recipes for computing topological
invariants that yield exact (integer valued) results. As 
a proxy for the latter, one is forced
to simulate a chain with two terminations subjected
to open boundary conditions, and the comparison to
the predictions of the index theorems can be
less straightforward, specially for non-Hermitian operators.
A basic rule of thumb is that the 
chain should be ``long enough", but
there is no way to know a priori how long is long enough. 
In addition, the topologically 
mandated zero modes (the kernel states)  
will be split away from zero. However, 
the splitting should scale down with system size, exponentially 
fast. This is one way to tell
these boundary modes apart from skin-effect modes.

Second, while our discussion so far puts
all the emphasis on open boundary conditions, 
even basic index theorems accommodate arbitrary boundary conditions
from the outset, in the following sense. Consider 
a finite block matrix
$$
M=\sum_{r,r'=0}^{m<\infty} |r\rangle\langle r'|\otimes M_{r r'} 
$$
for some $n\times n$ matrices $M_{rr'}$.
The way we have described this matrix makes
it clear that we can have it act on the 
state space of the half-infinite lattice.
By way of this trick, we can rethink the set
of all finite-dimensional matrices of any order 
as a set of operators acting on $l^2(\mathds{N})\otimes \mathds{C}^n$. 
The defining characteristic of this set of operators
is that their range is finite dimensional. The 
closure of the set of operators of finite range, 
in the operator norm topology, is the algebra 
of  compact operators. 

The range of a compact operator is no longer necessarily finite dimensional. 
However, it is still true that a compact operator decays into the bulk 
so that its action becomes 
indistinguishable from that of the zero 
operator far enough from the boundary, up to some tolerance. The smaller the 
tolerance, the further away from the boundary one must move. In this
sense, if $O=P\mathbf{O}E$ is some lattice operator,
translation invariant in the bulk, and $C$ is
a compact operator, then one can think of $O'=O+C$ as the same operator up to a 
change of boundary conditions, in the sense that
$O$ and $O'$ become indistinguishable far away from the boundary. Alternatively, 
one can think of $C$ as an impurity on the boundary. This observation is useful 
because an analytical index, any analytical index, is unaffected by such a modification, 
that is, $\mbox{index}\, O = \mbox{index}\, O'$. 
Similarly, if $C_a$ is a real and antisymmetric compact operator and
$A'=A+C_a$, then $\mbox{index}\, A = \mbox{index}\, A'$. This kind of robustness (against boundary perturbations), can also be
tested numerically with proper attention to finite-size scaling. The
bigger the impurity, the longer the chain has to be.

\subsubsection{Class $\{{\cal S}\}$}
In this section, the matrix $G_{{\cal S}}$ describes specifically
a clean (no bulk disorder), half-infinite wire in the symmetry class 
$\{{\cal S}\}$ and subjected to open 
boundary conditions. As before, and the symbol $G_{\cal S}(k)$ describes 
the associated translation invariant system. 

The number of boundary zero modes of the half-infinite chain is precisely 
the dimension of the
kernel of $G_{{\cal S}}$. Looking back at the 
auxiliary matrix $A_{\cal S}$ of Eq.~\ref{eq:aux_b1}, one concludes that
\begin{equation}
    \label{eq:dim-ker-G_S}
    \mbox{dim}\,\ker\,G_{\cal S} = n+\bar{n},
\end{equation}
in terms of 
\begin{align*}
n&\equiv \dim\,\ker\, B,\quad 
\bar{n} \equiv \dim\,\ker\, B^\dagger.
\end{align*}
Now, referring back to the topological classification, Eq.~\eqref{eq:frakn} and the index theorem~\eqref{ToeplitzITh}, we obtain the additional
relationship 
\begin{eqnarray}
    \label{eq:useful}
    n-\bar{n}=\mathfrak{n}(\mathbf{G}_{\cal S}).
\end{eqnarray}
Assuming, that is, that $G_{\cal S}$ is a Fredholm operator and $B$ with
it. Combining Eqs.~\eqref{eq:dim-ker-G_S} and \eqref{eq:useful}, we conclude that
\begin{align}
|\mathfrak{n}(G_{\cal S})|\leq \dim\, \ker\,G_{\cal S} 
%&=2\bar{n} + {\mathfrak n}(G_{\cal S})
%\label{eq:bbb1}
%\geq |{\mathfrak n}(G_{\cal S})|.
\end{align}
This is our bulk-boundary correspondence for 1D QBHs in 
the symmetry class $\{{\cal S}\}$: it connects a bulk property
of a translation invariant system, the topological integer 
$\mathfrak{n}(G_{\cal S})$, to the number of boundary zero
modes of the same system with one termination only, on the left.
In this symmetry class and in 1D, the 
bulk topological invariant mandates bosonic boundary zero modes. 
Moreover, the existence of bosonic boundary zero modes is a prediction 
robust against small perturbations that do not break the classifying 
symmetries, the squeezing symmetry ${\cal S}$ in this case.

\subsubsection{Class $\{{\cal N}, {\cal S}\}$}
In this symmetry class, 
$$
G_{\cal N, S}=I_2\otimes K_{\cal S},\quad K_{\cal S}=iK_{im}. 
$$
Since particle number is conserved, one 
can focus on the single-particle Hamiltonian $K_{\cal S}$. Moreover, due to the 
squeezing symmetry $K_{\cal S}=iK_{im}$ for some real and antisymmetric matrix 
$K_{im}$. For the half-infinite chain configuration, the number of boundary zero
modes is 
\begin{eqnarray}
\dim\,\ker\, K_{\cal S} = \dim\,\ker\,K_{im}. 
\end{eqnarray}
This is as far as symmetry cant take us. To achieve a bulk-boundary 
correspondence, we need to assume further that the number of bosonic degrees of 
freedom per lattice site is even. So suppose this is the case,
and suppose that $0$ is in a band gap of $K_{\cal S}$ so that $K_{im}$ is a Fredholm operator.
Then we can use the index theorem~\ref{PfaffianIth} and the 
bulk invariant of Eq.~\eqref{eq:fraks} to conclude that 
$$
\mathfrak{s}(\textbf{G}_{\cal N, S}) = (-1)^{\dim\,\ker\,K_{\cal S}}=
(-1)^{\frac{1}{2}\dim\,\ker\,G_{\cal N, S}}. 
$$
This is our bulk-boundary correspondence for one-dimensional QBHs in 
the symmetry class $\{\cal S, N\}$: in this symmetry class, a non-trivial 
bulk topological invariant $\mathfrak{s}(\textbf{G}_{\cal N, S})=-1$ 
mandates an odd number of kernel vectors for $K_{\cal S}$. Notice that 
each kernel vector of $K_{\cal S}$ corresponds to one creation
and one destruction bosonic operator. Moreover, the existence of an odd 
number of zero boundary modes is a prediction 
robust against small perturbations that do not break the classifying symmetries 
(the squeezing and particle number symmetries in this case).

\subsection{\label{subsec:topological-classification-summary}
Section Summary and Case Studies}

\subsubsection{The take-home message}
\begin{figure}[t]
    \centering
    \includegraphics[width=\columnwidth]{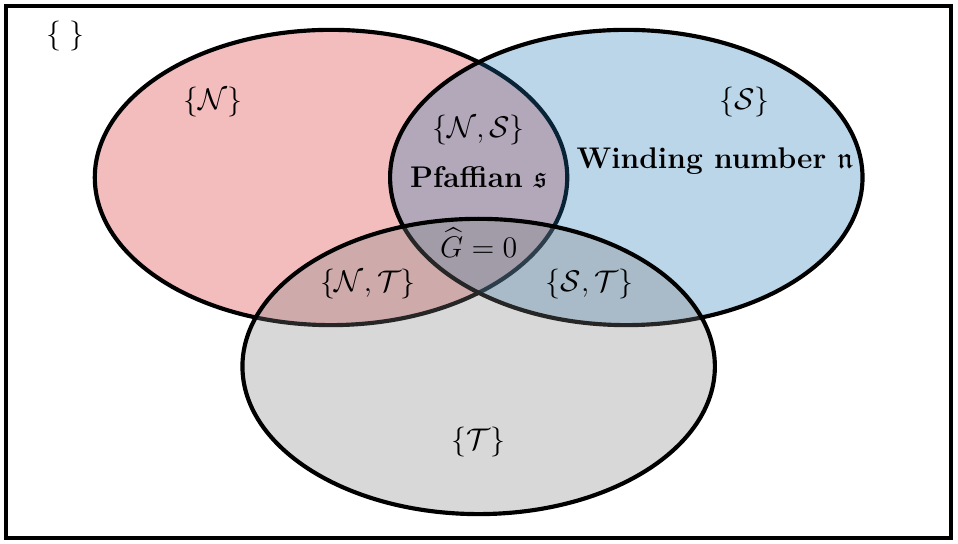}
    \caption{Overview of the different topological classes associated with time-reversal $\mathcal T$, squeezing $\mathcal{S}=n_1{\cal S}_1+n_2{\cal S}_2$, and particle number $\mathcal{N}$ symmetries. Only the classes $\qty{\mathcal{S}}$ and $\qty{\mathcal{N}, \mathcal{S}}$ are topologically non-trivial in $1D$. They are characterised by a winding number and Pfaffian invariants, respectively. Any transition between components with different invariants occur through a quantum phase transition with a closing of the band gap. There is a link to the theory of quantum phase transitions; see Ref.~\cite{Mariam2025} for details on notions of generalised criticality for free boson Hamiltonians. }
    \label{fig:topo-classes}
\end{figure}

Let us summarise the actionable takeaways. Based on our choice of classifying
many-body symmetries, time reversal, squeezing transformations, and particle number, 
there are, in one dimension, two symmetry classes of QBHs that are topologically nontrivial, see \cref{fig:topo-classes}. 
They are $\qty{\mathcal{S}}$, the class of one-dimensional QBHs that commute with some squeezing transformation, and $\qty{\mathcal{N}, \mathcal{S}}$, the class  that commutes with a squeezing transformation
and particle number. The squeezing 
transformation ${\cal S}$ is, in general, a linear combination of the two basic squeezing 
transformations ${\cal S}_1$ and ${\cal S}_2$
introduced in \cref{subsubsec:ContinuousSymmetries}.
Any combination can play the role of ``the" ${\cal S}$, and each combination
defines its own symmetry class. 
Any two such classes are connected by a rotation with ${\cal N}$ which, physically, changes the phase of the Bosonic pairing potential $\Delta$.

\begin{figure}[t]
    \centering
    \includegraphics[width=\columnwidth]{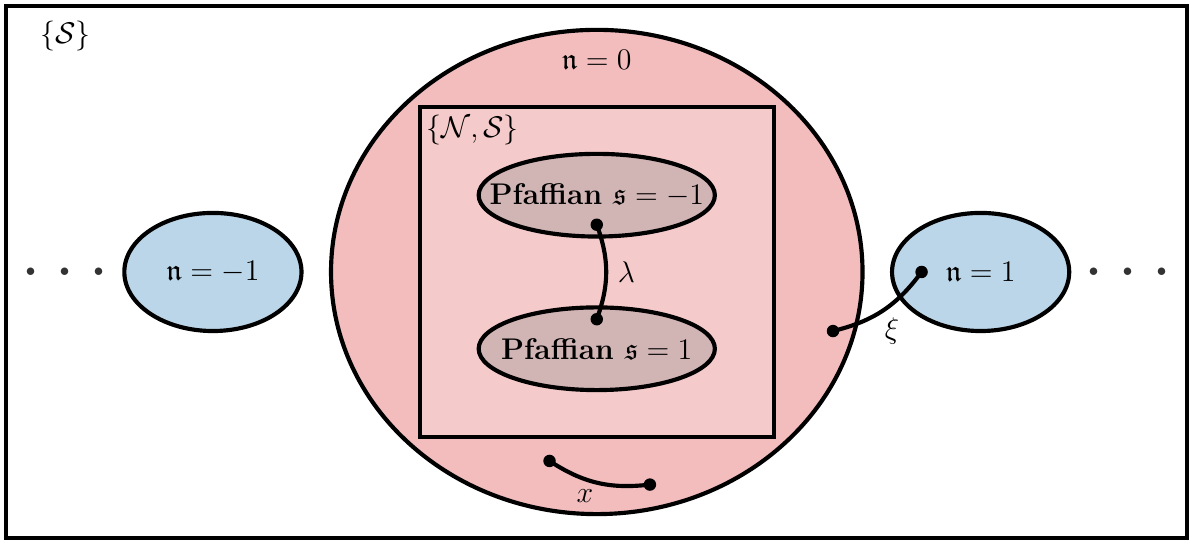}
    \caption{The set of $1D$ and gapped (in the sense discussed in the main text) Hamiltonians in the class $\qty{\mathcal{S}}$ breaks up into disconnected components labelled by the winding number invariants $\mathfrak{n} \in \mathds{Z}$. Any two Hamiltonians in any one component can be adiabatically deformed into each other; see
    e.g. the path labelled by $x$. A continuous path from on component to another one, that is, a deformation that changes the topological invariant, cannot be adiabatic;
    see e.g. the path labelled $\xi$. The $\qty{\mathcal{N},\mathcal{S}}$ class is fully contained in the component with $\mathfrak{n}=0$. It splits into two disconnected sub-components labelled by the  Pfaffian invariant $\mathfrak{s} \in \mathds{Z}_2$.}
    \label{fig:winding-vs-pf}
\end{figure}

Let us focus in the following on ${\cal S}_1$ as the squeezing symmetry operation.
To determine whether a QBH $\mathcal{H}$ has many-body symmetry-protected edge 
modes, or, to engineer an Hamiltonian with such modes, the first step 
is to identify the dynamical matrix
\begin{equation}
    G=\begin{bmatrix} 
    K & -\Delta\\ 
    \Delta^* & -K^*
    \end{bmatrix}
\end{equation}
of $\mathcal{H}$ in terms the hopping and pairing matrices $K$ and $\Delta$, as introduced in \cref{subsec:formalism-qbh} (pedagogical examples of finding such a decomposition are provided in \cref{subsec:cavity-magnonics} and the supplementary materials).
If the real parts of the matrices $K$ and $\Delta$ vanish, that is, if $K_{re}=\Delta_{re}=0$, 
then the QBH is in the $\qty{\mathcal{S}_1}$ symmetry class. %As a reminder,
%${\cal S}_1=\frac{i}{4}\sum_j (a_j^{\dagger\, 2}- a_j^{2})$. 
The second step is 
to compute the auxiliary matrix $B = i \qty(K_{im}-\Delta_{im})$ in position space, 
after which we go to momentum space to compute the complex-valued function of crystal 
momentum $\det B(k)$. As momentum $k$ is swept, $\det B(k) $ traces a closed path in the 
complex plane and, if the system has a point gap at $0$, meaning that $\det B(k) 
\neq 0$ for all $k$, we can compute its winding number around $0$. We call this 
integer $\mathfrak{n}(\mathbf{G})$ 
and we use it as a topological classifier of the gapped, in the above sense, QBHs in 
this symmetry class.
In other words, the class $\qty{\mathcal{S}_1}$ is made up of disconnected components labelled by the winding number $\mathfrak{n}$, see \cref{fig:winding-vs-pf}.
The implication is a bulk-boundary correspondence
\begin{equation}
    \dim\,\ker G\geq |\mathfrak{n}|,
\end{equation}
that is, there are at least $|\mathfrak n|$ 
symmetry-protected and topologically-mandated edge modes on the left edge, immune to 
any small perturbation whose dynamical matrix commutes with ${\cal S}_1$.
Strictly speaking, $G$ here is the dynamical matrix of a half-infinite system
terminated on the left.

As it turns out,
the bosonic Kitaev (BKC) chain of Ref.~\cite{2018McDonald} 
belongs to the symmetry class $\qty{\mathcal{S}_1}$; see the first
example below for details.
It will not surprise readers familiar with
the BKC to find out that its topological
properties are protected by a squeezing symmetry. However, it may surprise them
to observe there is a unitarily equivalent
of the BKC in the class $\qty{\mathcal{S}_2}$
which is just as topologically non-trivial
but does \underline{not} behave as a
directional amplifier for the basic quadratures. Hence, directional amplification is not a fully topological feature. Rather, it overtly coexists with topology in the class $\qty{\mathcal{S}_1}$ due to the peculiar structure of the symmetry ${\cal S}_1$. This said, parametric instabilities
are the rule in any class ${\cal S}$ and
zero boundary modes often coexist with the
non-Hermitian skin effect.

For a 1D QBH in the symmetry class $\qty{\mathcal{N}, \mathcal{S}_1}$, 
1) the real part of $K$ necessarily vanishes due to the squeezing symmetries 
and, 2) the pairing matrix $\Delta$ vanishes due to the particle number symmetry ($K_{re}=\Delta_{re}=\Delta_{im}=0$). 
If, in addition, there is an even number of bosonic modes per lattice site, then we can proceed with the topological analysis by computing $K(k)$ in momentum space.
If $0$ lies in a band gap, then $\det K (k) \neq 0$ and we can proceed with the
topological classification and bulk-boundary correspondence for this symmetry class in 1D. 
First, we compute the Pfaffian (see the formula in \cref{eq:pfaffian:2,eq:pfaffian:4}) of $-iK(k)$ for $k=0$ and $k=-\pi$, and we determine the parity index
$\mathfrak{s}=\mbox{sign}\frac{Pf(-iK(0))}{Pf(-iK(-\pi))}=\pm 1$ as the label of the two topological classes in this symmetry class, see \cref{fig:winding-vs-pf}. 
The bulk-boundary correspondence is, in this case, that the number of left boundary zero modes is constrained by the formula
\begin{equation}
   (-1)^{\dim\,\ker\,K}=(-1)^{\frac{1}{2}\dim\,\ker\,G}=\mathfrak{s}.
\end{equation}
Hence, this time, it is the \emph{parity} of the number of edge zero modes which is 
protected against small perturbations that commute with the classifying symmetries. 
Strictly speaking, $K$ here is the single-particle Hamiltonian of a half-infinite system
terminated on the left.

The central example of a model in this symmetry class is the celebrated bosonic SSH model~\cite{Ozawa2019}; see the second example below for details. 
This is shocking because it identifies the Pfaffian as the appropriate invariant for this model, 
\underline{not} the time-honored winding number which is, moreover, tightly linked to the celebrated chiral symmetry of its single-particle Hamiltonian. 
With hindsight, it was unreasonable to export to the bosonic SSH model \underline{the details} of the analysis of the fermionic SSH model without some physical justification. 
The two models share, by construction, the same zero boundary modes at the single-particle level. 
The underlying physical protecting symmetries are, however, very different. 
Moreover, it cannot be otherwise as the only basic symmetry shared by fermions and bosons is particle number. 
The difference in protecting physical symmetries translates into a difference in invariants and robustness properties.

\subsubsection{\label{subsec:applications-to-bosonic-models} The bosonic Kitaev chain}

The Hamiltonian of the BKC chain, the 
first example of a topological directional amplifier~\cite{2018McDonald}, is
\begin{align}
    {\cal H}_{BKC} = \frac{1}{2}\sum_{j=1}^{N} \qty(ita_j a_{j+1}^\dagger+ i\delta a_j^\dagger a_{j+1}^\dagger + \hc)
\end{align}
with $t,\delta$ real parameters.
The hopping and pairing matrices are (see the supplementary materials for details)
\begin{align}
    K &=\frac{it}{2}(T_N - T_N^\dagger)\\
    \Delta &= \frac{i\delta}{2}(T_N + T_N^\dagger).    
\end{align}
where $T_N$ is the $N \times N$ matrix with ones on the first lower diagonal.
Since they are both pure imaginary, we conclude that 
\begin{equation}
    [{\cal H}_{BKC},\ {\cal S}_1]=0
\end{equation}
and, as long as neither vanish, this
is the only basic symmetry of the BKC.

\begin{figure}[t]
    \centering
    \includegraphics[width=0.9\columnwidth]{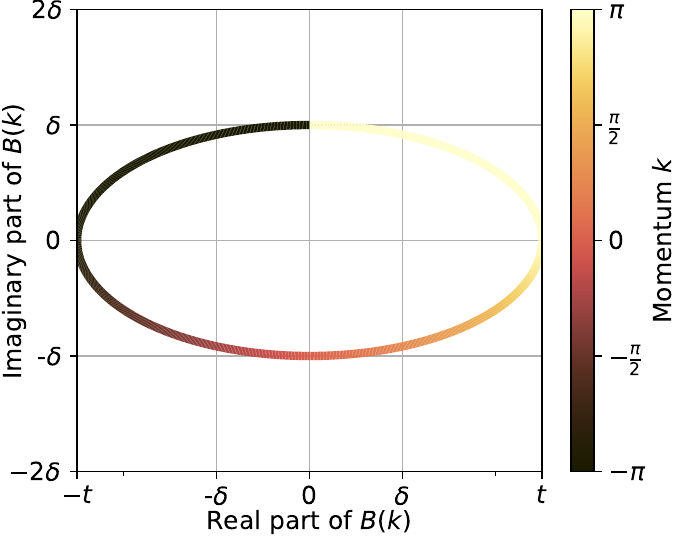}
    \caption{Plot of $B(k)$ given by \cref{eq:bkc:B1} for $\delta = 0.4t$ as the momentum is swept in the Brillouin zone. We see that the winding number is $\mathfrak{n}=1$ suggesting the presence of at least one edge mode.}
    \label{fig:bkc-winding}
\end{figure}

\paragraph{The topological classification of the BKC}
The auxiliary matrix and symbol are 
\begin{align}
    B &= i(K_{im}-\Delta_{im}) = i\frac{t-\delta}{2}T_N^\dagger - i\frac{t+\delta}{2} T,\\
    B(k) &= i\frac{t-\delta}{2}e^{-ik} - i\frac{t+\delta}{2} e^{ik}\\
    \label{eq:bkc:B1}
    &= t \sin(k)-i\delta \cos(k)
\end{align}
By plotting $B(k)$ in the complex plane, see \cref{fig:bkc-winding}, we confirm that the winding number 
$\mathfrak{n}(\det(B(k)), 0)=1$ for $<|\epsilon|<1$ with 
\begin{align}
    \label{eq:BKCloclength}
    \epsilon \equiv\frac{\delta-t}{\delta+t}
\end{align}
The band gap 
(in the sense of this paper) 
closes for $|\epsilon|= 1$ and reopens for 
$|\epsilon|>1$ where the winding number is now $\mathfrak{n}=-1$. In summary, the BKC is indeed an example of a symmetry-protected (by ${\cal S}_1$) and topologically non-trivial free
boson Hamiltonian in one dimension. 

\paragraph{The boundary physics of the BKC} 
The BKC illustrates very well  some differences, rooted in non-Hermitian physics, between topological free fermions 
and free bosons. The model is exactly solvable for open boundary conditions~\cite{2018McDonald} and so one can make some rigorous statements. 
First, the normal frequencies of the model are
complex for periodic boundary conditions and
purely real for open boundary conditions. This 
kind of ``spectral sensitivity to boundary conditions" in the words of Ref.~\cite{2018McDonald} 
is rigorously forbidden for systems governed by Hermitian dynamical
matrices. The mathematical background for understanding this phenomenon is the theory
of the pseudo-spectrum of an operator; see Ref.~\cite{PhysRevA.110.032207} 
and references therein. 

Moreover, for open boundary conditions, every
eigenfunction of $G$ is exponentially localised at the end-points of the chain. Hence, the BKC illustrates the celebrated ``non-Hermitian skin effect"~\cite{Ashida02072020}. The non-Hermitian
skin effect is unrelated to topology but can 
be a confounder from the point of view
of the bulk-boundary correspondence because it
could mean that the topologically-mandated boundary modes are embedded in a much larger set
of boundary-localised modes.   
Nonetheless, the topologically mandated zero modes show characteristic behaviour that is 
easy to spot. 

It is instructive to compute the topological
zero modes of the BKC, which is easier in the quadrature basis of Eqs.~\eqref{eq:quadx} and \eqref{eq:quadp}.  
In this basis, the BKC Hamiltonian becomes
\begin{align}
    \label{eq:BKCquads}
    {\cal H}_{BKC}&=\frac{\delta+t}{2}{\cal H}_\epsilon,\\
    {\cal H}_\epsilon&\equiv \hbar\sum_{j=1}^{N-1}
    \left(x_j p_{j+1} + \epsilon p_j x_{j+1}\right).    
\end{align}
There is an even-odd effect associated to
the length of the chain $N$. The odd
case is more interesting and so we focus 
on $N=2M+1$ odd in the following. The even
case can be handled similarly but the algebra
of the topologically mandated zero modes is
less interesting.

The starting point are the commutators 
\begin{align*}
    \frac{i}{\hbar}[{\cal H}_\epsilon, x_1]&=\epsilon x_2,\\
    \frac{i}{\hbar}[{\cal H}_\epsilon, x_3]&= x_2+\epsilon x_4,\\
    &\vdots\\
    \frac{i}{\hbar}[{\cal H}_\epsilon, x_{N-2}]&=x_{N-3}+\epsilon x_{N-1},\\
    \frac{i}{\hbar}[{\cal H}_\epsilon, x_N]&=x_{N-1}.
\end{align*}
It follows that the Hermitian bosonic
linear form 
\begin{align}
\label{eq:majboleft}
L&\equiv \sum_{j=0}^{M}(-\epsilon)^jx_{2j+1}\\
\nonumber
&=x_1-\epsilon x_3 +\epsilon^2 x_5-\dots
+\epsilon^{N-1} x_N
\end{align}
commutes with the Hamiltonian. This is 
the topologically mandated zero mode of 
the BKC chain. Its finite-size counterpart, localized on the right edge, can be determined
from the commutators 
\begin{align*}
    -\frac{i}{\hbar}[{\cal H}_\epsilon, p_N]&=\epsilon p_{N-1},\\
    -\frac{i}{\hbar}[{\cal H}_\epsilon, p_{N-2}]&= p_{N-1}+\epsilon p_{N-3},\\
    &\vdots\\
    -\frac{i}{\hbar}[{\cal H}_\epsilon, p_{3}]&=p_{4}+\epsilon p_{2},\\
    -\frac{i}{\hbar}[{\cal H}_\epsilon, p_1]&=p_2.
\end{align*}
The corresponding Hermitian zero mode is 
\begin{align}
   \label{eq:majboright}
R&\equiv\sum_{j=0}^M(-\epsilon)^j p_{N-2j}\\
\nonumber
&= p_N-\epsilon p_{N-2}+\epsilon^2 p_{N-4}-\dots +\epsilon^{N-1} p_1. 
\end{align}

The commutator of the topological zero modes
is
$
[L,R]=i(M+1)(-\epsilon)^{M}.
$
Thus, one can define left and right 
localised quadratures
$$
X_l\equiv ((M+1)(-\epsilon)^{M})^{-1/2}L,\quad
P_r\equiv ((M+1)(-\epsilon)^{M})^{-1/2}R
$$
such that
$$
[{\cal H}_\epsilon,X_l]=0,\quad [{\cal H}_\epsilon, P_r]=0,
$$
and satisfy the Heisenberg commutation relations
$$
[X_l,P_r]=i.
$$
In other words, the topologically mandated 
Hermitian bosonic zero edge modes of the BKC can be regarded as an instance of of Majorana bosons, 
that is, tight bosonic analogues of the Majorana
fermions of the fermionic Kitaev~ chain\cite{Kitaev_2001}.
The analogy is perfect at the level of the modes
themselves but breaks down when one
considers the the rest of the mode spectrum. 
In the fermionic case, the Majorana modes are midgap states; in the bosonic case, they are embedded in the rest of the spectrum due to the non-Hermitian skin effect. This could not 
possibly happen in a Hermitian systems~\cite{PhysRevA.110.032207}.
See Refs.~\cite{PRLFlynnCobaneraViola2021, PhysRevB.108.214312}, for a careful discussion
of the concept of Majorana boson.  
Other Hamiltonian models featuring Majorana bosons can be found in 
Refs.\cite{PhysRevB.102.125127, Flynn2020}.
See also Refs.~\cite{bomantara2025nonhermitiantopologicalphaseshermitian, bomantara2025floquetbosonickitaevchain}
for a recent alternative perspective on the topological boundary physics of the BKC.

\iffalse
For $N=2M$ even, one can check that the
Hermitian bosonic linear forms
\begin{align}
    L&=\sum_{j=0}^{M} (-\epsilon)^{j-1}x_{2j-1},\\
    R&=\sum_{j=1}^{M} (-\epsilon)^{j-1}p_{N-2j}
\end{align}
satisfy the commutation relations
\begin{align}
    [{\cal H}_\epsilon, L]&= (-\epsilon)^{N/2}x_N,\\
    [{\cal H}_\epsilon, R]&=-(-\epsilon)^{N/2}p_1,\\
    [L,R]&=0.
\end{align}
Hence, in the topological phase $|\epsilon|<1$, $L$ is an exact zero mode for $N=\infty$; it is the topologically mandated (approximate) zero mode with finite-size counterpart $R$. 
\fi

\subsubsection{
\label{subsec:bosonicSSH}
The bosonic SSH chain}
%\paragraph{A symmetry protected bosonic SSH model}

The Hamiltonian of the bosonic Su-Schrieffer-Heeger (SSH) model is 
\begin{align}
    \label{eq:ssh:hamiltonian}
    {\cal H} =  \sum_{j=1}^N t_1(b_j^\dagger a_j+a_j^\dagger b_j) + \sum_{j=1}^{N-1}t_2(a_{j+1}^\dagger b_j+b_j^\dagger a_{j+1})
\end{align}
where $a_j,b_j$ are bosonic annihilation operators and $t_1>0$ and $t_2>0$. The model is number conserving
and the number of degrees of freedom
per lattice site is even. If, in addition,
the model commutes with some squeezing symmetry, then it belongs to a bosonic
symmetry class $\{\cal N, S\}$. 
As they were characterised earlier in this section,
the squeezing symmetries do not commute with
the SSH model. However, a local gauge transformation places the rotated SSH model
squarely in the class $\{\cal N, S\}$ just
as described above. This
observation points the way forward.
\begin{figure}[t]
    \centering
    \includegraphics[width=\columnwidth]{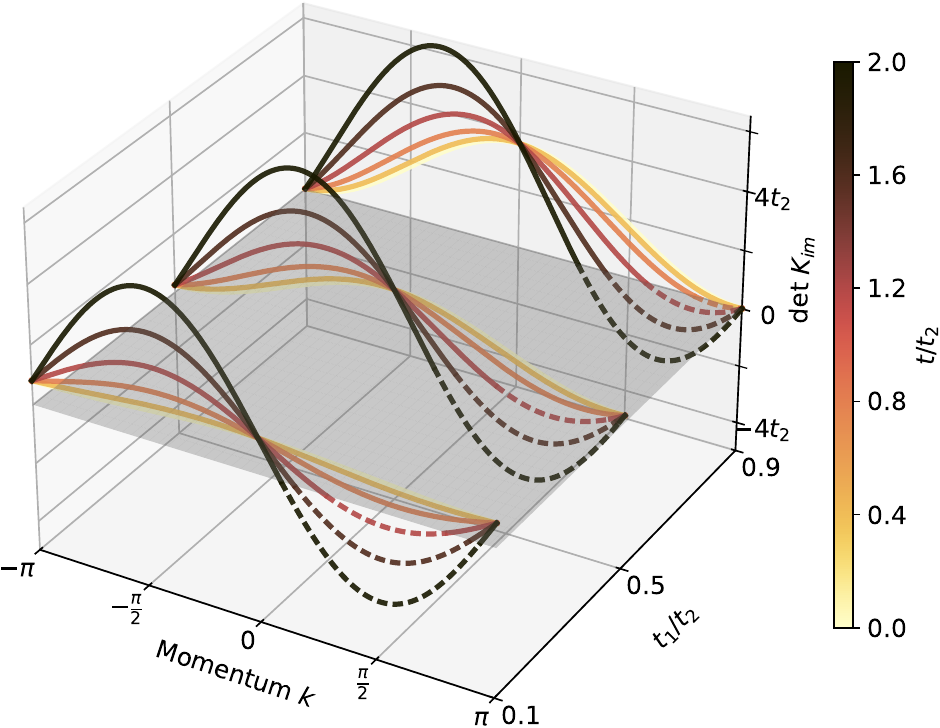}
    \caption{Plot of $\mathrm{det}\, K_{im}$ from \cref{eq:ssh:det-kim} as a function of $t_1$  and the perturbation strength $t$. The solid lines indicate positive values, while the dashed lines indicate negative values. The grey plane indicate the $z=0$ region.}
    \label{fig:ssh:det-kim}
\end{figure}

\begin{figure*}
	\centering
	\includegraphics[width=0.95\linewidth]{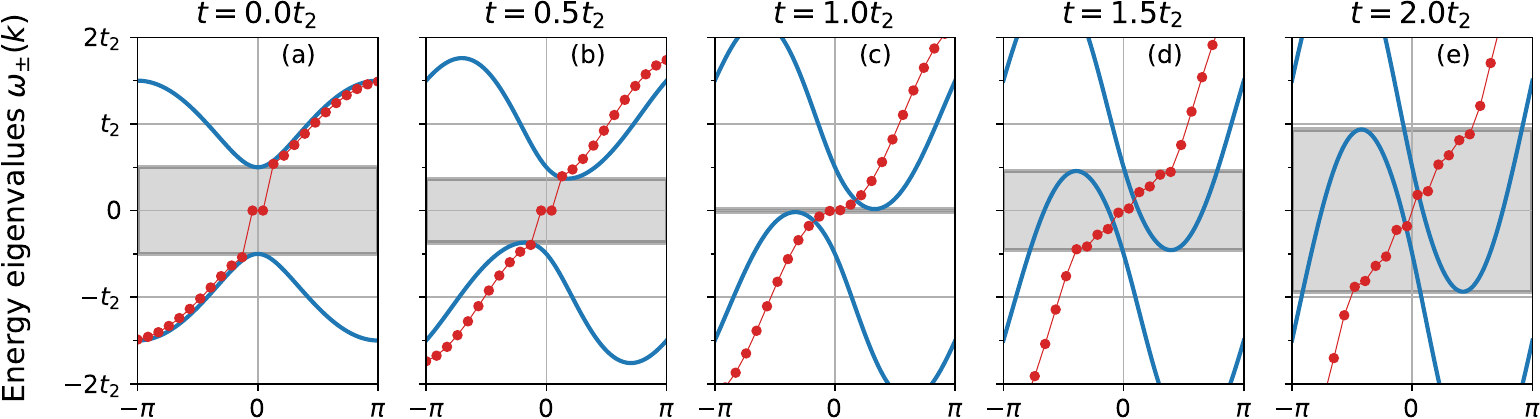}
	\phantomsubfloat{\label{fig:ssh:0}}
	\phantomsubfloat{\label{fig:ssh:0.5}}
	\phantomsubfloat{\label{fig:ssh:1}}
	\phantomsubfloat{\label{fig:ssh:1.5}}
	\phantomsubfloat{\label{fig:ssh:2}}
	\vspace{-4em}
	\caption{Spectra of the symmetry-protected bosonic Su-Schrieffer-Heeger model of \cref{eq:ssh:hamiltonian} perturbed by nearest-neighbour hoppings of \cref{eq:ssh:perturbation}. The blue solid lines correspond to the bandstructure, while the red dots correspond to the spectrum in position space (open boundary conditions) for $N=12$ lattice sites. The parameters used are $t_2=20$ MHz and $t_1 = t_2/2$.}
	\label{fig:ssh}
\end{figure*}

It is instructive to look for the correct
symmetries directly. Since squeezing symmetries
do not commute with the particle number 
operators, we need to work with 
the dynamical matrix 
\begin{align}
 G_{SSH}&= \beta_3\otimes K_{SSH},\\
 K_{SSH}& \equiv  \begin{bmatrix}
    0 & t_1 + t_2 T_N^\dagger\\
    t_1 + t_2T_N & 0
    \end{bmatrix}   
\end{align}
where as before $T_N$ is the $N \times N$ matrix with ones on the first lower diagonal.
We see then by inspection that  $G_{SSH}$ commutes with the dynamical matrices
 \begin{align*}
     \beta_1^{SSH} &= \beta_1\otimes
     \begin{bmatrix}
         I_N & 0\\
         0& -I_N
     \end{bmatrix},\\
     \beta_2^{SSH} &= \beta_2\otimes
     \begin{bmatrix}
         I_N & 0\\
         0& -I_N
     \end{bmatrix},\\
     \beta_3^{SSH} &= \beta_3\otimes
     \begin{bmatrix}
         I_N & 0\\
         0& I_N
     \end{bmatrix}.\\
 \end{align*}
The  $\beta_i^{SSH}$ satisfy the same algebra as the basic $\beta_i$.
It is immediate to work out the corresponding many-body squeezing symmetries and confirm that they are local in real space. 
Hence, the bosonic SSH model belongs to the Pfaffian class $\{{\cal N}^{SSH}, {\cal S}^{SSH}\}$. In other wors, these are
the protecting symmetries of the bosonic SSH model. By contrast, the fermionic SSH model
is protected by four many-body fermionic symmetries (class BDI of \cref{tab:many_body_AZ}).

One can further check, for example, by actually carrying out the
gauge transformation mentioned above, that the Pfaffian invariant $\mathfrak{s}(\textbf{G}_{SSH}) = -1$.
Thus, the zero boundary modes of the bosonic SSH model is topologically mandated by the Pfaffian invariant. However, this means that the robustness properties of the bosonic and fermionic SSH models are quite different.
In particular, if this analysis of the model is correct, we should be able to break the chiral symmetry of the bosonic SSH model
with any small perturbation of the form 
\begin{equation}
    \delta G= i I_2\otimes 
\begin{bmatrix}
    \delta A &0\\
    0 & \delta A
\end{bmatrix}
\end{equation}
without lifting its zero boundary modes.
Here, $\delta A$ is any small real antisymmetric matrix. 
Such perturbations of $K_{SSH}$ break its chiral symmetry without taking it out of its symmetry class.
%$\{{\cal N}^{SSH}, {\cal S}^{SSH}\}$. 

Let's test our predictions. 
Having clarified the point of symmetry protection, we can go back to working with the single-particle Hamiltonian $K_{SSH}$. 
Let us add to the bosonic SSH model the perturbation
\begin{align}
    \label{eq:ssh:perturbation}
    \delta{\cal H}=-\frac{it}{2}\sum_{i}
    (a_{i+1}^\dagger a_i + b_{i+1}^\dagger b_i - \hc) 
\end{align}
of strength $t$, which clearly breaks the sub-lattice/chiral symmetry of the model. 
In momentum space, the symbol of the perturbed SSH model is 
\begin{align}
    K(k)=
    \begin{bmatrix}
        -t\sin(k) & t_1 + t_2 e^{ik}\\
        t_1+t_2 e^{-ik} & -t\sin(k)    
    \end{bmatrix}.
\end{align}
%\begin{align}
%    K_{im}=
%    \begin{bmatrix}
%        -it\sin(k) & t_1 - t_2 e^{ik}\\
%        -t_1+t_2 e^{-ik} & -it\sin(k)    
%    \end{bmatrix}.
%\end{align}
Our expectation is that the zero modes, being mandated by the non-trivial value of the Pfaffian invariant, survive for $t$ small enough not to violate the gap condition even though the sub-lattice/chiral symmetry of the model is now badly broken. 
The fate of the gap can be diagnosed by way of the determinant
\begin{align}
    \label{eq:ssh:det-kim}
    \mathrm{det} \, K(k)= t^2\sin(k) - t_1^2-t_2^2-2t_1t_2\cos(k).
\end{align}
For $t=0$, the gap is $|t_1-t_2|$. As $t$ increases, it diminishes gradually until it finally closes and never reopens. 
\Cref{fig:ssh:det-kim} displays the evolution of the gap with $t$, and \Cref{fig:ssh} confirms that the zero boundary modes survive all the way to the gap closing, nicely pinned at zero even for short chains. 
%That is,
%$$
%i\delta A=  i\frac{t}{2}
%\begin{bmatrix}
%   T - T^\dagger & 0\\
%   0 & T - T^\dagger
%\end{bmatrix}.
%$$
%\onecolumngrid 

%\twocolumngrid 

\iffalse
{\color{blue} FORMERLY 
\begin{align}
    %\label{eq:ssh:det-kim}
    \mathrm{det} \, K_{im}(k)= -t^2\sin(k) + t_1^2+t_2^2-2t_1t_2\cos(k),
\end{align}
please update figure 14 accordingly. Thank you! Also, I thin there's something WRONG here.
\\
\\
which is plotted in \cref{fig:ssh:det-kim}. For $t=0$, \cref{eq:ssh:det-kim} is symmetric in momentum $k$, and the ratio $t_1/t_2$ determines the curvature. Notably, for $t=0$, $\mathrm{det} \, K_{im}(k)>0$ over the whole Brillouin zone. The effect of the perturbation strength $t$ is to negatively bias towards positive momentum.}
\fi

%the gap is $|t_1- t_2|$ at $t=0$ closes at $t=?$ (choose, say, $t_1=0.1$, $t_2=1.0$ if hard to do in general). 

\section{\label{sec:dicke-chain-revisited} 
The photo-magnonic chain is symmetry protected}
%SPT physics in 
%Topology$\mathbf{+}$: a many-body symmetry-protected photo-magnonic crystal

We saw already that the photo-magnonic crystal in 1d (photo-magnonic chain) can be topologically non-trivial as a matter of band structure.  
Let us consider next the problem of symmetry protection: what, if any, are the physical symmetries that protect the topological features of the photo-magnonic chain for example? If we can answer this question, we will know which perturbations are irrelevant from the point of view of the topological
boundary physics.  

We will stick with the rotating wave approximation (RWA) and so focus on the effective Hamiltonian of \cref{eq:Hn:real-space:rwa-rotating-frame}. 
The only chance for symmetry protection then is for the system to belong to the Pfaffian class $\{\cal N, S\}$.
One can see that this is indeed the case by way of a gauge transformation. 
First, one can adjust the phase of the photon hopping amplitude to make it pure imaginary by way of a gauge transformation
$a_j\mapsto e^{ij\theta}a_j$, with $\theta$ 
such that $t=i|t|e^{i\theta}$.
Second, one can adjust the phase of $g$, now partly including the phase introduced by the gauge transformation of the photon operators, to also be pure imaginary ``for free", see \cref{subsubsec:phase}. At
this point, the gauge rotated model 
belongs to the Pfaffian class $\{\cal N, S\}$
as described in \cref{sec:topological-classification}. 
Undoing the gauge transformation reveals the physical protecting symmetries of the actual photo-magnonic chain. We worked out this kind of calculation in detail for the Bosonic SSH model in the previous section. We conclude that the photo-magnonic chain is more than topological: it is also symmetry-protected.

In this section we will continue our analysis in the gauge transformed presentation of 
the system for conciseness in the analysis. 
For $|t|$ small compared to $|g|$, the photo-magnonic chain host $n$ topologically mandated,
by a Berry phase, zero boundary modes.
This suggests that the ${\cal H}^{(n)}$  with $n$ even belongs to the topologically trivial subclass (Pfaffian invariant $=1$), and the rest belong to the topologically non-trivial subclass (Pfaffian invariant $=-1$). 
The bulk-boundary correspondence for
the Pfaffian class implies that,
for $n$ odd, a small perturbation also in the symmetry class cannot remove all of the edge modes: there must be at least one survivor 
per edge until the perturbation is strong enough to close the energy gap.

\begin{figure*}
	\centering
	\includegraphics[width=2\columnwidth]{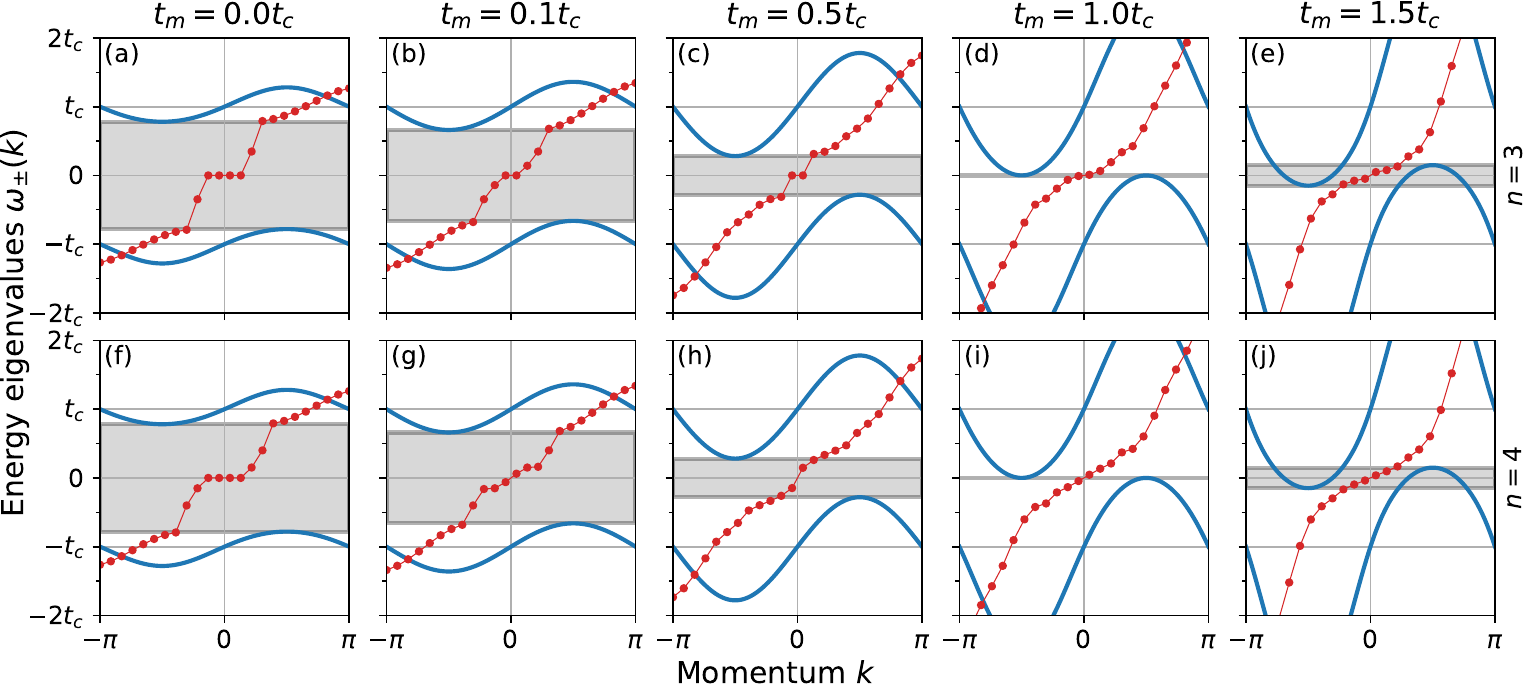}
	%\vspace{-1em}
	\caption{Comparison of the spectra of the photo-magnonic crystal $K^{(3)}_{spt}$ (first row) and $K^{(4)}_{spt}$ (second row) with magnon hoppings. 
    The blue solid lines correspond to the bandstructure, while the red dots correspond to the spectrum in position space (open boundary conditions) for $N=12$ lattice sites. The parameters are $\omega_a=\omega_m=0$, and hoppings $\abs{g}/2\pi=100$ MHz and $\abs{t}=\abs{g}/4$}
	\label{fig:H3-H4-pfaffian}
\end{figure*}
Let us quickly check this picture.
Suitably instantiating \cref{eq:Gn:rwa}, one obtains the single-particle Hamiltonian $K^{(n)}(k)=iK_{im}(k)$ with
% Emilio version, which I think is wrong because of differing conventions:
%$$
%    K_{im}(k) = \mqty(
%        -2it \sin(k) & ge^{ikn}\\
%        -ge^{-ikn} & 0
%    ).
%$$
% My version
\begin{equation}
\label{eq:Kim:adc:pfaffian}
    K_{im}(k) = \mqty[
{-2it\sin\qty(k)} & -ge^{-ikn}\\
ge^{ikn} &0
].
\end{equation}
We find
\begin{align}
    \mathrm{Pf}\qty(K_{im}^{\qty(n)}\qty(k=0))&=-g \\
    \mathrm{Pf}\qty(K_{im}^{\qty(n)}\qty(k=-\pi)) &= -\qty(-1)^{n}g
\end{align}
and thus the Pfaffian invariant evaluates to $\mathfrak{s}=(-1)^n$, as predicted. 

To test our claims about the behavior of the zero edge modes, let us add to the Hamiltonian some generic perturbation that does not break the classifying many-body symmetries. 
As an example, let us add 
\begin{equation}
   \delta{\cal H}= -\sum_{j=0}^{N-1}it_m(m_{j} m_{j+1}^\dagger -m_{j}^\dagger m_{j+1})
\end{equation}
to the basic Hamiltonian, 
%\cref{eq:Hn:real-space:rwa-rotating-frame:spt}, 
where $t_m$ is real and positive. 
Physically, this corresponds to allowing magnon hopping. 
It is not natural to try to implement this kind of interaction in a photo-magnonic crystal. However, it provides a compelling 
test of the Pfaffian bulk-boundary correspondence.
With this perturbation,
% Emilio version below
%\begin{align*}
%    K_{spt}^{(n)}(k, t_m)&=iK_{im}(k, t_m), \\
%    K_{im}(k, t_m) &= \mqty(
%        -i2t \sin(k) & ge^{ikn}\\
%        -ge^{-ikn} & -i2t_m\sin(k)
%    ).
%\end{align*}
\begin{align}
    \nonumber
    K^{(n)}(k, t_m)&=iK_{im}(k, t_m), \\
    K_{im}(k, t_m) &= \mqty[
        -2it \sin(k) & -ge^{-ikn}\\
        -ge^{-ikn} & -2it_m\sin(k)
    ]
\end{align}
where it is understood that $g,t$ are now real parameters. % as in \cref{eq:Kim:adc:pfaffian}. 
The determinant of $K_{spt}^{(n)}(k, t_m)$ is $g^{2}-4tt_{m}\sin^2 k$. 
Hence, the gap closes at $k=\pm \frac{\pi}{2}$ provided that $t_{m}> t_c= \frac{g^{2}}{4t}$, as illustrated by \cref{fig:H3-H4-pfaffian}.

Let us investigate the cases $n=3$ and $n=4$. 
For $n=3$, the Pfaffian invariant is non-trivial and we see numerically that, for $t_m=0$, the unperturbed photo-magnonic crystal hosts two zero boundary modes per termination. The additional third zero mode that exists for $n=3$ and $t=0$ remains localised but has already migrated away from zero energy in response to the non-vanishing photon hopping. We expect that increasing $t_m$ will move another one of the two zero boundary modes away from zero, bringing the system down to one protected zero boundary mode per termination. 
This protected zero boundary mode should survive all the way until $t_m$ is large enough to close the energy gap. The numerics of \cref{fig:H3-H4-pfaffian} confirm these
expectations.

For $n=4$, the Pfaffian invariant is trivial and we see again that, for $t_m=0$, the unperturbed photo-magnonic crystal hosts two zero boundary modes per termination. An additional pair of localised modes
have already migrated away from zero energy where they are originally for $n=4$ and $t=0$. 
Increasing $t_m$ should move all the boundary modes away from zero before the gap closes because none of these zero modes are protected. Again, the numerics of \cref{fig:H3-H4-pfaffian} confirm these expectations. 
In fact, all the boundary modes delocalise 
and join the band modes well before the gap closes. 

\paragraph{Experimental implementation of 
complex photon hopping.} 
It might be desirable to implement directly
the photo-magnonic chain with pure imaginary
parameters. The issue at stake is that
changing
the presentation of the protecting symmetries changes the set of irrelevant perturbations.
Since the phase of the magnon-photon coupling is an adjustable gauge degree of freedom,
the challenge is to modulate the phase of the photon hopping between \emph{unit cells}. 

For three-dimensional cavities, a simple idea to adjust the phase of the photon hopping $t/2\pi$ is to adjust the length of a waveguide between the two cavities. 
However, this assumes that the mode in the waveguide is propagating, i.e. it is a travelling wave of the form $e^{i\beta x}$, with $x$ the propagation direction, and $\beta$ the propagation constant. 
For the cavity design we have adopted, the waveguide would be rectangular, with dimensions \SI{5}{\milli\meter} by \SI{5}{\milli\meter}.
Thus, the lowest-order mode that can propagate are the (degenerate) transverse-electric modes TE$_{01}$ and TE$_{10}$ with frequency \cite{2011Pozar}
\begin{equation}
    \omega_c/2\pi = \frac{1}{2l\sqrt{\epsilon_r}}.
\end{equation}
For an air-filled ($\epsilon_r=1$) square section of length $l =$ \SI{5}{\milli\meter} (corresponding to the size of the iris, see \cref{subsec:implementation-Dicke-unit-cell}), we find $\omega_c/2\pi \simeq 30$ GHz, which is far above the cavity mode's resonance $\omega_a/2\pi \simeq 10$ GHz. 
Thus, the wave is evanescent, with an exponential decay and no phase offset. A potential solution, which keeps the dimensions identical, is to insert a material with a high dielectric constant, thus reducing the cut-off frequency. However, in this case the waveguide section becomes resonant and acts as a new cavity by itself.

Instead, a simple solution is to consider another kind a waveguide: a simple coaxial cable terminated by magnetic loop antennas on either side. Indeed, in a coaxial waveguide the transverse electromagnetic (TEM) mode propagates at all frequencies. 
Thus, in this case, a phase shift can be induced by adjusting the length of the coaxial cable. 
Concerning the loop antennas, they are designed like those used to excite and readout the cavity, see \cref{fig:cavity-magnonics-experiment:cavity} and \cref{fig:coupled-cavities:design} for instance. 
Notably, the area of the loop antenna protruding in the cavity can be used to adjust the magnitude of the coupling, as realised in several experiments \cite{2021Rao,2017Zhang,2020Zhaoa}.

\section{\label{sec:conclusion}Conclusions and outlook}
\paragraph{Summary of the results.}
In this paper we have introduced a new experimental platform for microwave topological photonics, namely the photo-magnonic crystals. 
Photo-magnonic crystals are naturally prone to displaying non-trivial band topology because they mix heavy degrees of freedom, the magnons, with highly-mobile photonic degrees of freedom. 
The magnons provide flat bands that hybridise with the photon bands. This kind of band mixing is a well-know precursor of topologically non-trivial band structures. 
In addition, photo-magnonic crystals also support local synthetic gauge fields, another precursor of topological physics. 
In this paper we have focused on the rich topological physics of the simplest photo-magnonic chain.
We will return to higher dimensions and synthetic gauge fields in future publications. 

Photo-magnonic crystals are
highly tunable systems and can be reliably characterised in terms of various modelling techniques. The most important ones are finite-element modelling of the cavity electromagnetic field, and the tight-binding  Hamiltonian formalism, which can be complemented with quantum Langevin equations. They are deployed progressively, each layer building on the previous one, 
to achieve an increasingly accurate characterisation of the dynamic of the quasi-particles (sometimes called ``cavity magnon polaritons"). 
In this paper, the tight binding modelling layer is the one that reveals the topological physics of photo-magnonic crystals.

Because photo-magnonic crystals are so well controlled engineered systems, they are a natural platform to raise the stakes and seek true bosonic many-body symmetry protection comparable to that available for fermions. 
To tackle this challenge, in this paper we have described a symmetry classification of free boson Hamiltonians based on three basic bosonic many-body symmetries and we have found seven different symmetry classes. 
Of these, two are topologically non-trivial in one dimension and support bulk-boundary correspondences. 
The symmetries that define these classes are the protecting symmetries of the models in these classes. 
In this way, we confirm that the protecting symmetry of the bosonic Kitaev chain \cite{2018McDonald}, a recent model of directional quantum optical amplification, is a squeezing symmetry. 
The bosonic SSH model, a fundamental model of topological photonics, is protected by two squeezing symmetries and particle number. 
As a consequence, the topological invariant for the bosonic SSH model is not the winding number invariant (the correct invariant for the fermionic SSH model), but rather the Pfaffian invariant.  
This is a surprising outcome and so we have justified it extensively. 
With hindsight, it is a physically sensible result. 
 
\paragraph{Integration.} 
The symmetry classification we developed applies to closed quantum systems, where no dissipative processes take place. 
Hence, in order to be able to investigate the topological physics of photo-magnonic crystals within this formalism, it was crucial to suppress the impact of dissipation as much as possible. 
For this reason, in this work we have focused on  three-dimensional microwave cavities. 
One negative aspect of this architecture is that the size of the cavities limits the scalability of the crystals. 
Quantum effects in the microwave frequency range typically require milli-Kelvin temperatures and, while integrating a unit cell in a dilution refrigerator can be achieved \cite{2018GoryachevWattBourhillKostylevTobar}, doing so for a large enough crystal presents a significant challenge. 
On the bright side, well-differentiated topological boundary modes can emerge even for a very small number of lattice units, specially if one takes advantage of the high tunability of the platform. 

Another way to address the challenge of scalability is to move to planar architectures. One recent approach takes advantage of substrate integrated waveguide (SIW) technology, which allows integrating three-dimensional cavity designs on a chip. This approach was recently experimentally demonstrated in ref \cite{2025Ardisson}, where the torus cavity of ref \cite{PhysRevApplied.19.014030} integrated on a chip. 
Alternatively, standard two-dimensional cavity designs, such as split-ring resonators or cross-line cavities can be employed, but at the expense of stronger dissipation rates for the cavity and magnon modes. Both of these have been experimentally demonstrated, see refs \cite{2017Bhoi,2022Kaffash,2022Ma,2024Wagle} for plit-ring resonators, and refs \cite{2019Wang,2021Rao,2022Zhong} for cross-line cavities.
Taken together, these various platforms offer a path towards scalability, especially if the theory can keep up with the experimental capabilities and 
provide a suitable framework for dissipative topological physics protected by physical symmetries. 
We will address this challenge for the theoretical foundation of the subject in a forthcoming publication. 

\paragraph{Topologically-robust control of microwave light.}
Cavity magnonics systems have been integrated with superconducting coplanar resonators in ways that are  compatible with the constraints
imposed by dilution refrigerators. 
These architectures are promising for investigating
finer quantum effects and could open the way for new parametric amplifiers along the lines of Ref.~\cite{2018McDonald} and quantum sensors~\cite{2020McDonald}. 
Other applications of synthetic photo-magnonic crystals, now in the classical regime, revolve around the control of microwave signals. We have seen in \cref{sec:topological-crystal} that the transmission through the photo-magnonic chain is reciprocal despite two structurally different edge modes being excited on each end. Furthermore, while the transmission is reciprocal, the reflection is not, exhibiting starkly different behaviour.
This is a new addition to the list of non-reciprocal effects
documented in cavity magnonics~\cite{2019Wang,2021Wang,2022Zhong,PhysRevApplied.19.014030,2025Ardisson}, including perfect absorption and reflections \cite{2021Rao,2023Qian}. 
Building on our work in this paper, these applications could 
become enhanced with symmetry-protected topological robustness by scaling up the single magnonic cavity to a potentially quite small (in terms of number of unit cells) photo-magnonic crystal.

\paragraph{Towards higher-dimensional crystals.}
Another interesting topic is the extension of our theory to higher dimensions. The three-dimensional cavities presented in this work can easily be organised in a two-dimensional lattice, owing to the fact that the cavity modes's field distribution remain in the $(\vu{x},\vu{y})$ plane. However, different microwave cavity architectures could be used to create a full three-dimensional photo-magnonic crystal. Alternatively, Floquet driving can be used to induce synthetic dimensions. In the context of cavity magnonics, this has been experimentally realised by driving a YIG sphere with a magnetic loop antenna \cite{2020Xub,wc1j-mq69}.

%\added[id=e]{I would keep silent about this. We already have plenty of exciting ideas above IMO.}
%\deleted[id=e]{\emph{Topological superradiance.}
%In this work, we have made two approximations to describe the photon-magnon Hamiltonian, (1) the Holstein-Primakoff transformation (to use a bosonic representation of the macrospins) and (2) the rotating wave approximation (RWA). 
%Both approximations are well justified for the photo-magnonic crystal proposed here, but we note that the photon-magnon ultrastrong coupling regime -- where the RWA loses applicability -- has been realised in both three-dimensional \cite{2014Goryachev,2023Bourcin} and planar cavity architectures \cite{2021Golovchanskiy,2021Golovchanskiya,2023Ghirri}. 
%If one does not employ the RWA, and uses a macrospin representation instead, our proposed photo-magnonic crystal becomes a one-dimensional Dicke chain. 
%This naturally motivates the examination of the interplay between the symmetry-protected topological features of such a crystal and the superradiant phase transition of the standard Dicke model.}

In short, this paper raises exciting experimental
prospects by introducing a new quantum meta-material and provides for the first time a fairly simple and conceptually compelling framework for physical symmetry protection in bosonic systems. 
We expect many more experimental and theoretical discoveries to follow.

\begin{acknowledgments}
    We thank Christophe Fumeaux for his insights concerning the coupling of microwave cavities and Jonathan Marenkovic for proofreading the manuscript. 
    Emilio Cobanera gratefully acknowledges many stimulating discussions with Amit Sangwan on the topic of mesoscopic and nanoscopic arrays of electromagnetic devices. Giuseppe C. Tettamanzi acknowledge funding from the Australia's Economic Accelerator Innovate Program scheme project number IV240100119.
    Giuseppe C. Tettamanzi also acknowledges the generous support of the Organisation of Naval Research Global (ONRG) Prestigious Visiting Scientist Program that has allowed him to visit SUNY Polytechnic Institute, Utica, and has allowed him to start the collaboration that has led to this publication. 
\end{acknowledgments}

\bibliography{references}

\end{document}

\title{Supplementary notes for ``Many-body symmetry-protected zero boundary modes\\ of synthetic photo-magnonic crystals''}
%Synthetic Topological Magnonic Crystals}
%Quantum Phase Transitions and Symmetry-Protected Topological edge modes in a lattice Dicke model}

\author{Alan Gardin}
\affiliation{Quantum and Nano Technology Group (QuaNTeG), School of Chemical Engineering, The University of Adelaide, North Terrace Campus, Adelaide, 5000, South Australia, Australia}

\author{Emilio Cobanera}
\affiliation{Department of Physics, SUNY Polytechnic Institute, 
%100 Seymour Ave, 
Utica, NY 13502, USA}
\affiliation{Department of Physics and Astronomy, Dartmouth College, 6127 Wilder Laboratory, Hanover, New Hampshire 03755, USA}

\author{Giuseppe Carlo Tettamanzi}
\affiliation{Quantum and Nano Technology Group (QuaNTeG), School of Chemical Engineering, The University of Adelaide, North Terrace Campus, Adelaide, 5000, South Australia, Australia}

\maketitle
\tableofcontents
\section{\label{app:fem}
The finite-elements simulations}

%\paragraph{Further details on the simulations.}
In this appendix, we detail the finite-element simulations of the cavity magnonics systems using \comsol. The geometry of the cavity ``unit cell'' was described in section II.A.2. The cavity itself was modelled as a perfect electric conductor. For the magnetic loop antenna, the conductor were modelled as copper, based on the default settings provided by \comsol. The dielectric material for the antenna has a relative permittivity of 2.1, vanishing conductivity, and a magnetic permeability of 1. The YIG material has a relative permittivity of 15, vanishing conductivity, and a magnetic permeability of 1. 

For frequency domain simulations, we have considered the linearisation of the Landau-Lifschitz-Gilbert equation around the equilibrium magnetisation. The details of this method are covered in the textbook by \textcite{2011Pozar}. In our simulations, we have set the magnon's linewidth to be 10 MHz, which is typical of YIG spheres. The effective loss tangent is set to $2 \times 10^{-3}$ and the saturation magnetisation to 0.178 Tesla.

\paragraph{Magnon frequency shifts.} 
For all fits of COMSOL simulations,the magnon resonance frequencies $\omega_m/2\pi$ are shifted by 60 MHz, except for fig 5.a of the main text which is shifted by 90 MHz. The overall shift of 60 MHz most likely originates from the imperfect meshing of the YIG spheres in the simulations, which produce imperfect demagnetisation factors. Indeed, our simulations on a unit cell show that the location of the magnon resonance with the cavity mode depends on the mesh refinement of the YIG sphere. The differing shift of 90 MHz for fig 5.a of the main text may be due to a different mesh refinement for this simulation compared to the others. Note that in practice, small frequency shifts are expected due to anisotropy in YIG spheres. However, they can always be compensated for by adjusting the applied magnetic field strength.

\begin{figure}[t]
    \centering
    \includegraphics[width=\columnwidth]{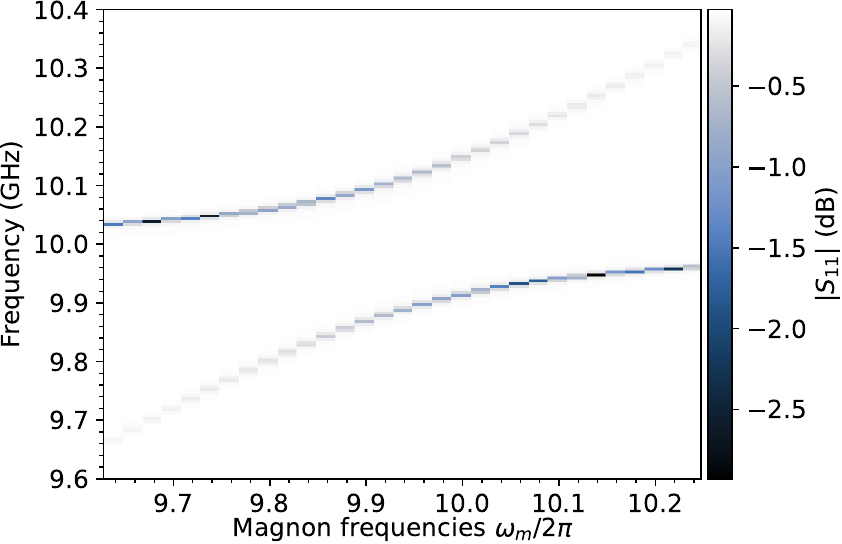}
    \caption{Simulated reflection spectrum for the cavity unit cell in the presence of a YIG sphere.}
    \label{fig:N1-s11}
\end{figure}

\paragraph{Cavity unit cell.}
Since this system is symmetric, the S-parameters are also symmetric, and thus we have $S_{11}=S_{22}$ and $S_{12}=S_{21}$. Therefore, only $S_{11}$ is displayed in \cref{fig:N1-s11}. Similar to the transmission $\abs{S_{21}}$, we observe an anti-crossing signalling the strong coupling with the magnon mode.

\begin{figure}[t]
    \centering
    \includegraphics[width=\columnwidth]{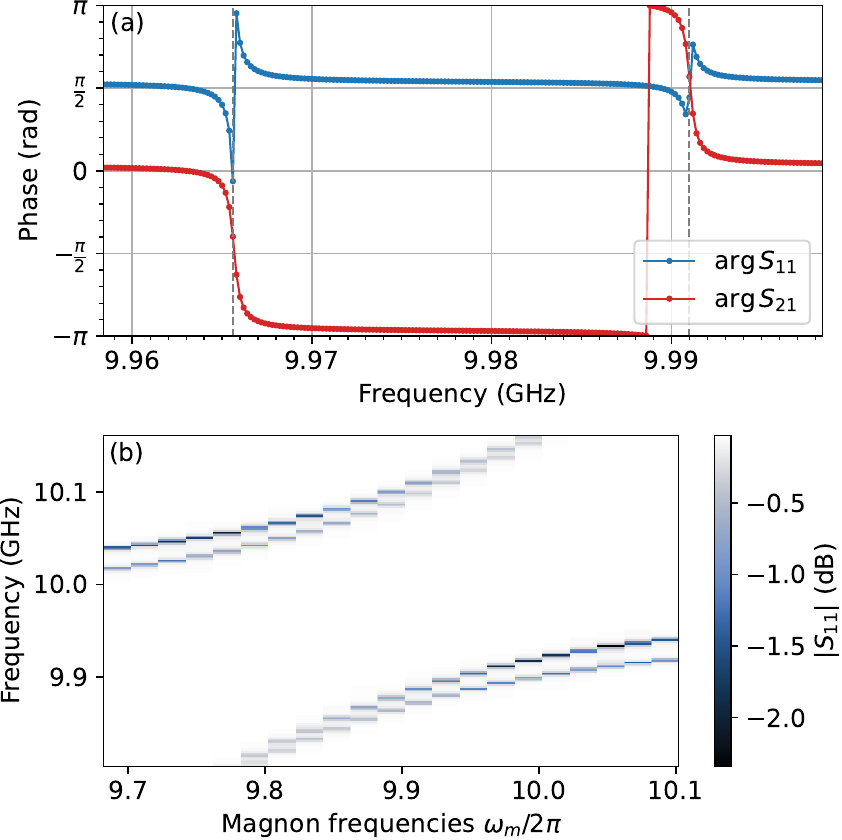}
    \phantomsubfloat{\label{fig:N2:s-params-phase}}
    \phantomsubfloat{\label{fig:N2-s-params:S11}}
    \vspace{-2em}
    \caption{Simulated S-parameters for two coupled unit cell cavities. (a) Phases of the S-parameters with no YIG sphere (b) Amplitude of the reflection coefficient where each cavity is loaded with a YIG sphere. }
    \label{fig:N2-s-params}
\end{figure}

\paragraph{Two coupled cavities.}
\Cref{fig:N2-s-params} compiles additional plots for two coupled cavities.
The phase of the S-parameters for two coupled cavities, without the magnon modes, are given in fig 3.c of the main text. Of interest is the transmission coefficient $S_{21}$, since it carries a signature of the photon-photon hopping coupling $t/2\pi$. We observe an expected $\pi$ phase shift at the location of the two resonances, indicates by the grey dashed lines. We note that a small phase shift exists, signalling that the phase of $t/2\pi$ is non-zero. 

\begin{figure}[t]
    \centering
    \includegraphics[width=\columnwidth]{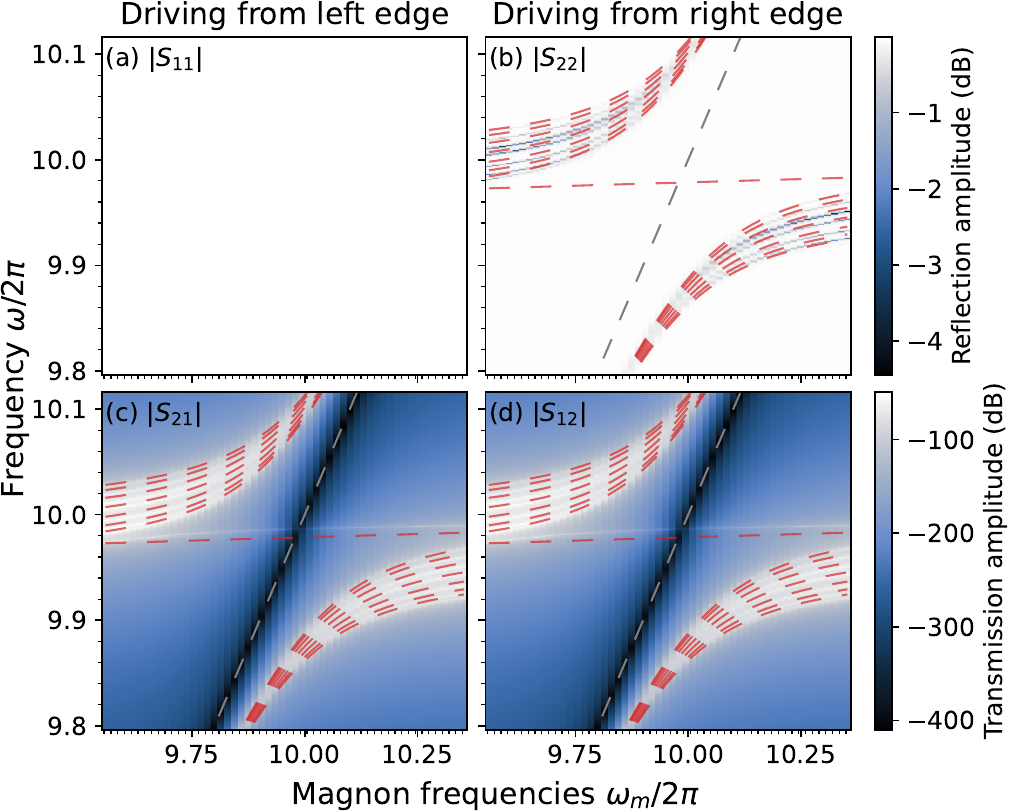}
    \phantomsubfloat{\label{fig:H1-fits-N8:S11}}
    \phantomsubfloat{\label{fig:H1-fits-N8:S22}}
    \phantomsubfloat{\label{fig:H1-fits-N8:S12}}
    \phantomsubfloat{\label{fig:H1-fits-N8:S21}}
    \vspace{-3em}
    \caption{Simulated S-parameters amplitude through the photo-magnonic crystal using COMSOL for $N=8$ lattice sites, when the YIG sphere in the first cavity is removed. The legend and parameters are identical to those of fig 5 of the main text. The spectrum is not shown in (a) for readability.}
    \label{fig:Sparams-H1-N8-fits}
\end{figure}

\paragraph{Eight coupled cavities.}
Finally, we present the results for $N=8$ lattice sites. \Cref{fig:Sparams-H1-N8-fits} presents the fit of the using the usual parameters. We note that contrary to a lattice of size $N=4$, the agreement between the S-parameters and the fits is not as good. However, we observe similar trends to those mentioned in the main text for $N=4$ lattice sites, see fig 8 of the main text. 

\begin{figure}[t]
    \centering
    \includegraphics[width=\columnwidth]{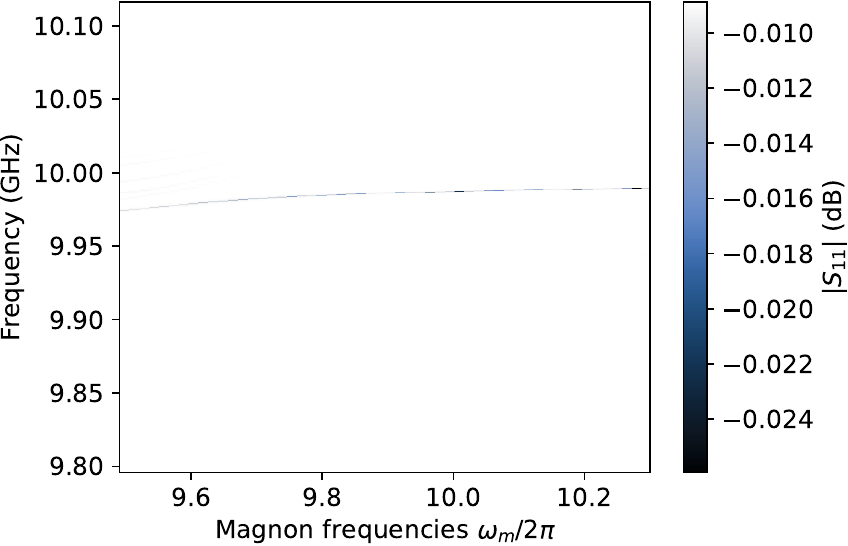}
    \caption{Simulated $S_{11}$ amplitude through the photo-magnonic crystal using COMSOL for $N=8$ lattice sites, when the YIG sphere in the first cavity is removed.}
    \label{fig:N8-H1-S11}
\end{figure}

It seems that the reflection parameter at the left edge, described by $S_{11}$, completely vanishes. In reality, the $S_{11}$ is not exactly zero, as \cref{fig:N8-H1-S11} shows. Indeed, in \cref{fig:H1-fits-N8:S11} we used a common colour scheme for both reflection parameters, which hides this small response. \Cref{fig:N8-H1-S11} also shows a variation of the edge mode's frequency with $\omega_m/2\pi$, suggesting hybridisation with a magnon mode.

\begin{figure}[ht!]
    \centering
    \includegraphics[width=\columnwidth]{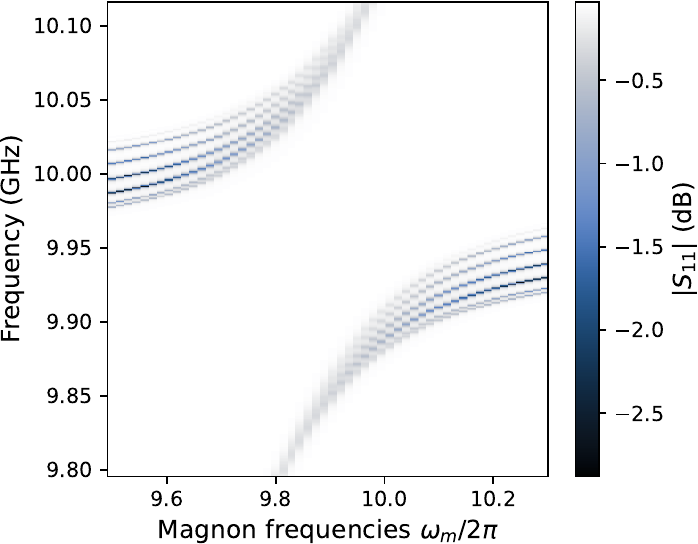}
    \caption{Simulated reflection coefficient for eight coupled cavities, each loaded with a YIG sphere, corresponding to $\mathcal{H}^{(0)}$.}
    \label{fig:N8-s-params}
\end{figure}

\begin{figure}[ht!]
    \centering
    \includegraphics[width=\columnwidth]{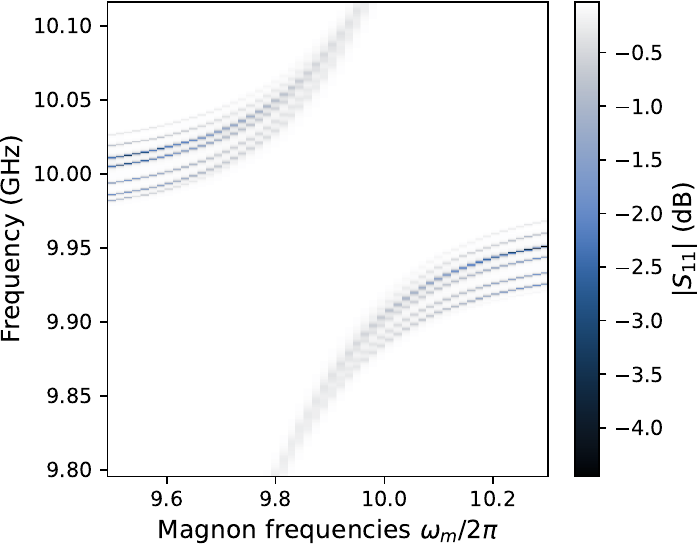}
    \caption{Simulated $S_{11}$ amplitudes for $\mathcal{H}^{(1)}$ with $N=8$ lattice sites.}
    \label{fig:N8-H1-P1}
\end{figure}

\section{\label{app:input-output}Input-output analysis}
%\subsection{Derivation}
\paragraph{Setup.}
In this appendix, we detail the input-output formalism used in the main text. Following ref \cite{1985Gardiner}, we couple the system Hamiltonian to an environment modelled as a ensemble of harmonic oscillators. The system Hamiltonian is taken to be that of a general quadratic bosonic Hamiltonian given by equation (47) of the main text, namely
\begin{equation}
    \mathcal{H}/\hbar = \sum_{i,j} \qty(K_{ij}a_i^\dagger a_j + \frac{1}{2}\Delta_{ij}a_i^\dagger a_j^\dagger + \frac{1}{2}\Delta_{ij}^* a_i a_j).
\end{equation}
In the presence of counter-rotating terms, it was shown \cite{2006Ciuti,2014DeLiberato} that the input-output theory must account for the fact that only excitations of positive frequency $\omega>0$ exists in the environment. Thus, contrary to the original formulation of ref \cite{1985Gardiner}, we consider the Hamiltonian 
\begin{equation}
\mathcal{H}_{e} = \sum_{j=1}^N \int_{\mathbb{R}^{+}} \dd[]{\omega}  \hbar\omega b_{j,\omega}^{\dagger}b_{j,\omega}
\end{equation}
to model the environment. For simplicity, we will also consider that the coupling to the environment is frequency independent, and thus we adopt
\begin{equation}
\mathcal{H}_{I,e} = \sum_{j} ih \int_{\mathbb{R}^{+}} \dd[]{\omega} \qty(\sqrt{\frac{\kappa_j}{2\pi}}b_{j,\omega} a_{j}^{\dagger} - \hc)
\end{equation}
where $\kappa_{j}$ is taken to be complex-valued and $\sqrt{\kappa_j} \equiv \kappa_j^{1/2}$. The total Hamiltonian is thus $\mathcal{H}+\mathcal{H}_e+\mathcal{H}_{I,e}$.

We will note $\widetilde{a}$ the Fourier transform of an annihilation operator $a$, defined as
\begin{equation}
    \widetilde{a}(\omega) = \int_\mathbb{R} \dd{t} e^{i\omega t} a(t).
\end{equation}
Note that for creation operator $a^\dagger$, the Fourier transform can be written $\widetilde{a^\dagger}(\omega) = \qty({\widetilde{a}}(-\omega))^\dagger$ which we will simply note $\widetilde{a}^\dagger(-\omega)$.

\paragraph{Bath degrees of freedom and input-output relations.}
The Heisenberg equation of motion for the bath operators reads
\begin{equation}
\dot{b}_{j,\omega} = -i\omega b_{j,\omega}- \sqrt{\frac{\kappa_j^*}{2\pi}} a_{j}
\end{equation}
with the formal solution
\begin{equation}
b_{j,\omega}\qty(t) = e^{-i\omega\qty(t-t_{0})}b_{j,\omega}\qty(t_{0})
-\sqrt{\frac{\kappa_j^*}{2\pi}}\int_{t_{0}}^{t} \dd[]{t'} e^{-i\omega\qty(t-t')}a_{j}\qty(t')
\end{equation}
for $t>t_{0}$. We define the input and output noise operators as 
\begin{gather}
\label{eq:iot-in-field}
b_{j}^{\text{in}}\qty(t) = \frac{1}{\sqrt{ 2\pi } }\int_{\mathbb{R}^+} \dd[]{\omega} e^{-i\omega\qty(t-t_{0})}b_{j,\omega}\qty(t_{0}), \\
b_{j}^{\text{out}}\qty(t) = \frac{1}{\sqrt{ 2\pi } }\int_{\mathbb{R}^+} \dd[]{\omega} e^{-i\omega\qty(t-t_{1})}b_{j,\omega}\qty(t_{1})
\end{gather}
where $t_{0}<t<t_{1}$, $t_{0}\to -\infty$ and $t_{1} \to +\infty$. Note that only positive-valued frequencies are considered in these definitions \cite{2006Ciuti}.

On integrating the bath operators for $t_{0}<t$ with $t_{0} \to - \infty$, we find
\begin{widetext}
\begin{align}
\int_{\mathbb{R}^+} \dd[]{\omega} \sqrt{\frac{\kappa_j}{2\pi}} b_{j,\omega}\qty(t)
&= \sqrt{ 2\pi } \sqrt{\frac{\kappa_j}{2\pi}}b_{j}^{\text{in}}\qty(t) - \frac{\abs{\kappa_j}}{2\pi} \int_{\mathbb{R}^{+}} \dd[]{\omega} \int_{-\infty}^{t} \dd[]{t'} a_{j}\qty(t')e^{-i\omega\qty(t-t')}\\
&= \sqrt{\kappa_j}b_{j}^{\text{in}}\qty(t) - \frac{\abs{\kappa_j}}{2\pi} \int_{-\infty}^\infty \dd[]{t'}  a_{j}\qty(t')  \int_{\mathbb{R}^{+}} \dd[]{\omega}  \Theta\qty(t-t') e^{-i\omega\qty(t-t')}\\
\label{eq:int-bath-op}
&= \sqrt{\kappa_j} b_{j}^{\text{in}}\qty(t) - \frac{\abs{\kappa_j}}{2\pi}\qty[a_{j} * \Gamma_{j}]\qty(t)
\end{align}
\end{widetext}
where $*$ denotes a convolution, $\Theta$ is the Heaviside function, and $\Gamma_{j}\qty(t)\equiv \int_{\mathbb{R}^{+}} \dd[]{\omega}  \Theta\qty(t) e^{-i\omega t}$. For $t<t_{1}$, we similarly find
\begin{align}
\label{eq:int-bath-op-tmp}
\int_{} \dd[]{\omega} \sqrt{\frac{\kappa_j}{2\pi}} b_{j,\omega}\qty(t)
&= \sqrt{\kappa_j} b_{j}^{\text{out}}\qty(t) + \frac{\abs{\kappa_j}}{2\pi} \qty[a_{j} * \Gamma_{j}]\qty(t)
\end{align}
and equating \cref{eq:int-bath-op,eq:int-bath-op-tmp}, we find the input-output relations $b_{j}^{\text{out}}\qty(t)=b_{j}^{\text{in}}\qty(t) - \frac{\sqrt{\kappa_j^*}}{\pi} \qty[a_{j}*\Gamma_{j}]\qty(t)$. 
Moving to frequency space, the second term in the input-output relations diverges due to the frequency-independent couplings $\gamma_{j}$. Indeed, Fourier-transforming $\Gamma$ gives
\begin{align}
\widetilde{\Gamma}_{j}\qty(\omega) &=
\int_{\mathbb{R}} \dd[]{t} \Theta\qty(t)\int_{\mathbb{R}^{+}} \dd[]{\omega'} e^{i\qty(\omega-\omega')t} \\
&= \int_{\mathbb{R}^{+}} \dd[]{\omega'} \widetilde{\Theta}\qty(\omega-\omega')\\
&= \int_{\mathbb{R}^{+}} \dd[]{\omega'} \qty[\pi\delta\qty(\omega-\omega')+i \mathrm{PV}\qty{\frac{1}{\omega-\omega'}}]\\
&= \pi\Theta\qty(\omega) + i \mathrm{PV}\qty{\int_{\mathbb{R}^{+}} \frac{\dd[]{\omega'}}{\omega-\omega'} }
\end{align}
where $\widetilde{\Theta}$ is the Fourier transform of $\Theta$ and $\mathrm{PV}$ is the Cauchy principal value. The imaginary part, responsible for the Lamb shift, diverges due to ignoring the frequency dependence of $\kappa_{j}$. In practice, $\kappa_{j}\qty(\omega)$ is peaked around some frequency $\omega_{0}$ such that 
\begin{equation}
\mathrm{PV}\qty{\int_{\mathbb{R}^{+}} \frac{\sqrt{\kappa_j(\omega)}\dd[]{\omega'}}{\omega-\omega'} }
\end{equation}
is finite. Following ref \cite{2014DeLiberato}, we will ignore the Lamb shift and simply consider the real part $\widetilde{\Gamma}_{j}\qty(\omega) =\pi\Theta\qty(\omega)$ which as expected vanishes for negative frequencies. Thus, in frequency space, the input-output relations simply read $\widetilde b_{j}^{\text{out}}\qty(\omega)=\widetilde b_{j}^{\text{in}}\qty(\omega) - \Theta\qty(\omega) \sqrt{\kappa_j^*} \widetilde a_{j}\qty(\omega)$.

\paragraph{QLEs for the system operators.}
Returning to the system operators, the Heisenberg equation of motion gives
\begin{equation}
\label{eq:eom-ak-time}
\begin{aligned}
\dot{a}_{k} &= -iK_{kk}a_{k}-i \frac{\Delta_{kk}}{2}2a_k^\dagger\\
&\quad -i\sum_{i \neq j}\qty(\delta_{ik} K_{ij}a_{j}+\frac{1}{2} \delta_{ik}\Delta_{ij}a_{j}^{\dagger} + \frac{1}{2}\delta_{jk} \Delta_{ij}a_{i}^{\dagger})\\
&\quad + \sqrt{\frac{\kappa_k}{2\pi}} \int_{\mathbb{R}^{+}} \dd[]{\omega}b_{k,\omega}.
\end{aligned}
\end{equation}
Using \cref{eq:int-bath-op} for the last term, we obtain
\begin{equation}
\label{eq:eom-ak-time-2}
\begin{aligned}
\dot{a}_{k} &= -iK_{kk}a_{k}-i \Delta_{kk} a_k^\dagger\\
&\quad -i\sum_{j \neq k}\qty(K_{kj}a_{j}+ \frac{\Delta_{kj}+\Delta_{jk}}{2} a_{j}^{\dagger})\\
&\quad - \frac{\abs{\kappa_k}}{2\pi} \qty[a_{k} * \Gamma_{k}] + \sqrt{\kappa_k} b_{k}^{\text{in}} .
\end{aligned}
\end{equation}
Remembering that $\Delta_{ij}=\Delta_{ji}$, and switching to frequency-space, 
\begin{equation}
\begin{aligned}
-i\omega \widetilde{a}_{k}(\omega)
&= -i \sum_{j} \qty(K_{kj}\widetilde a_{j}(\omega)+ \Delta_{kj} \widetilde a_{j}^{\dagger}(-\omega))\\
&\quad - \frac{\abs{\kappa_{k}}}{2}  \Theta\qty(\omega)\widetilde a_{k}(\omega) +\sqrt{ \kappa_{k} }\widetilde b_{k}^{\text{in}}(\omega)
\end{aligned}
\end{equation}
and after re-arranging
\begin{widetext}
\begin{equation}
    \label{eq:iot-mel-ak}
    \sum_{j} \qty(\qty[\delta_{jk}\qty(\omega+ i\frac{\abs{\kappa_{k}} }{2}\Theta\qty(\omega))-K_{kj}] \widetilde{a}_{j}(\omega) - \Delta_{kj} \widetilde{a}_{j}^{\dagger}(-\omega)) =i \sqrt{ \kappa_{k} }\widetilde{b}_{k}^{\text{in}}\qty(\omega).
\end{equation}
\end{widetext}
Similarly, we can find the equation of motion for $a_k^\dagger$ by taking the hermitian conjugate of \cref{eq:eom-ak-time-2}, and we find
\begin{equation}
\label{eq:eom-ak-dag-time}
\begin{aligned}
\dot{a}_{k}^\dagger &= 
i\sum_{j}\qty(K_{kj}^* a_{j}^\dagger + \Delta_{kj}^* a_{j})
- \frac{\abs{\kappa_k}}{2\pi} \qty[a_{k}^\dagger * \Gamma_{k}^*] +\sqrt{\kappa_k^*} {b_{k}^{\text{in}}}^\dagger.
\end{aligned}
\end{equation}
where the hermitian conjugate of \cref{eq:int-bath-op} has been used. Moving to frequency space, and recalling that $\widetilde{a^\dagger}(\omega) = \widetilde{a}^\dagger(-\omega)$,
\begin{widetext}
    \begin{equation}
    \label{eq:iot-mel-ak-dag}
        \sum_{j}\qty(\Delta_{kj}^{*}\widetilde{a}_{j}(\omega) + \qty[\delta_{jk}\qty(\omega+i \frac{\abs{\kappa_k}}{2} \Theta\qty(-\omega))+K_{kj}^*] a_{j}^\dagger\qty(-\omega))
        =
        i\sqrt{ \kappa_{k}^{*} } \widetilde{b}_{k}^{\text{in}\dagger}\qty(-\omega)
    \end{equation}  
\end{widetext}

\paragraph{Open-system dynamics.}
As previously mentioned, the environment only contains positive frequencies \cite{2006Ciuti}. Therefore, the damping term proportional to $\abs{\kappa_j}$ in \cref{eq:iot-mel-ak,eq:iot-mel-ak-dag} vanishes for negative frequencies, thanks to the Heaviside function. Similarly, the input noise operators vanish for negative frequencies. Indeed, taking the Fourier transform of $b_j^\text{in}$ defined by \cref{eq:iot-in-field}, we find
\begin{align}
b_{j}^{\text{in}}\qty(\omega) &= \int_{\mathbb{R}} \dd[]{t} \frac{e^{i\omega t}}{\sqrt{ 2\pi } } \int_{\mathbb{R}^{+}} \dd[]{\omega'} e^{-i\omega'\qty(t-t_{0})}b_{j,\omega'}\qty(t_{0})\\
&= \frac{1}{\sqrt{ 2\pi } } \int_{\mathbb{R}^{+}} \dd[]{\omega'} b_{j,\omega'}\qty(t_{0})e^{i\omega't_{0}}\int_{\mathbb{R}} \dd[]{\omega} e^{i\qty(\omega-\omega')t}\\
&= \sqrt{ 2\pi }  \int_{\mathbb{R}^{+}} \dd[]{\omega'} b_{j,\omega'}\qty(t_{0})e^{i\omega't_{0}} \delta\qty(\omega-\omega')
\end{align}
and hence $b_{j}^{\text{in}}\qty(\omega<0)=0$ because the integral is only on positive $\omega'>0$.

\Cref{eq:iot-mel-ak,eq:iot-mel-ak-dag} define the matrix equation 
\begin{equation}
    \label{eq:iot-matrix-eq}
    \mathcal{G} \vb{v} = i \mqty[\Gamma &0\\ 0&  \Gamma^\dagger] \vb{v}^{\text{in}}
\end{equation}
where $\vb{v}(\omega)=[\widetilde{a}_1(\omega), \widetilde{a}_2(\omega), \dots, \widetilde{a}_N^\dagger(-\omega)]^t$ is the vector of system operators, $\vb{v}^\text{in}(\omega)=\qty[b_{1}^{\text{in}}(\omega), \dots, \qty(b_{k}^{\text{in}})^\dagger(-\omega)]^t$ is the vector of input noise operators, $\Gamma = \mathrm{diag}\qty{\sqrt{\kappa_i}}$ and
\begin{align}
    \mathcal{G}(\omega) 
    &= \omega I_{2N} + \mqty[-K & -\Delta\\ \Delta^* & K^*] + \mqty[1 & 0\\ 0 & 0] \otimes \frac{i\Gamma \Gamma^\dagger}{2}\\
    &= \omega I_{2N} -\tau_3 H + \frac{i}{2}\mqty[\Gamma \Gamma^\dagger & 0\\ 0 & 0] .
\end{align}
Note that the second term of $\mathcal{G}$ is simply $-\tau_3 H$ with $H=G\tau_3$ the Hamiltonian matrix and $G$ the dynamical matrix of equation (54) of the main text. Therefore, $\mathcal{G}$ is a frequency-space ``open system'' analog of $G$. Indeed, the main difference is the presence of dissipation rates $\Gamma$, since the sign differences are simply due to the frequency space the factorisation of some imaginary numbers. Inverting \cref{eq:iot-matrix-eq} gives the vector 
\begin{equation}
    \label{eq:iot-matrix-eq:inverted}
    \vb{v} = i \mathcal{G}^{-1} \mqty[\Gamma &0\\ 0&  \Gamma^\dagger] \vb{v}^{\text{in}}
\end{equation}
which describes the response of the system operators as a function of the bath operators $\vb{v}^{\text{in}}$.

\paragraph{Reflection and transmission coefficients.} 
We now make a mean-field approximation and replace each mode with its expectation value, writing $a_i = \expval{a_i}$. Then, the reflection at mode $i$ is defined as
\begin{equation}
    S_{ii}(\omega) = \left.\frac{\widetilde{b}_i^\text{out}(\omega)}{\widetilde{b}_{i}^\text{in}(\omega)} \right|_{\widetilde{b}_{k\neq i}^\text{in}(\omega)=0}, 
\end{equation}
while the transmission from $a_j$ to $a_i$ reads
\begin{equation}
    S_{ij}(\omega) = \left.\frac{\widetilde{b}_i^\text{out}(\omega)}{\widetilde{b}_{j}^\text{in}(\omega)} \right|_{\widetilde{b}_{k\neq j}^\text{in}(\omega)=0}.
\end{equation}
By evaluating \cref{eq:iot-matrix-eq:inverted} with the appropriate boundary conditions for $\vb{v}^{\text{in}}$, and using the input-output relations $\widetilde b_{i}^{\text{out}}\qty(\omega)=\widetilde b_{i}^{\text{in}}\qty(\omega) - \Theta\qty(\omega) \sqrt{\kappa_i^*} \widetilde a_{i}\qty(\omega)$, $S_{ii}(\omega)$ and $S_{ij}(\omega)$ can be calculated, for positive frequencies $\omega>0$, as
\begin{equation}
    S_{ij}(\omega) = \delta_{ij} - \sqrt{\kappa_j^*} \left. \frac{\vb{v}_{i}\qty(\omega)}{\widetilde{b}_{j}^\text{in}(\omega)}\right|_{\widetilde{b}_{k\neq j}^\text{in}(\omega)=0}.
\end{equation}
This expression was used to compute the S-parameters in the main text.

%\subsection{Additional figures}

\begin{figure}[!ht]
    \centering
    \includegraphics[width=\columnwidth]{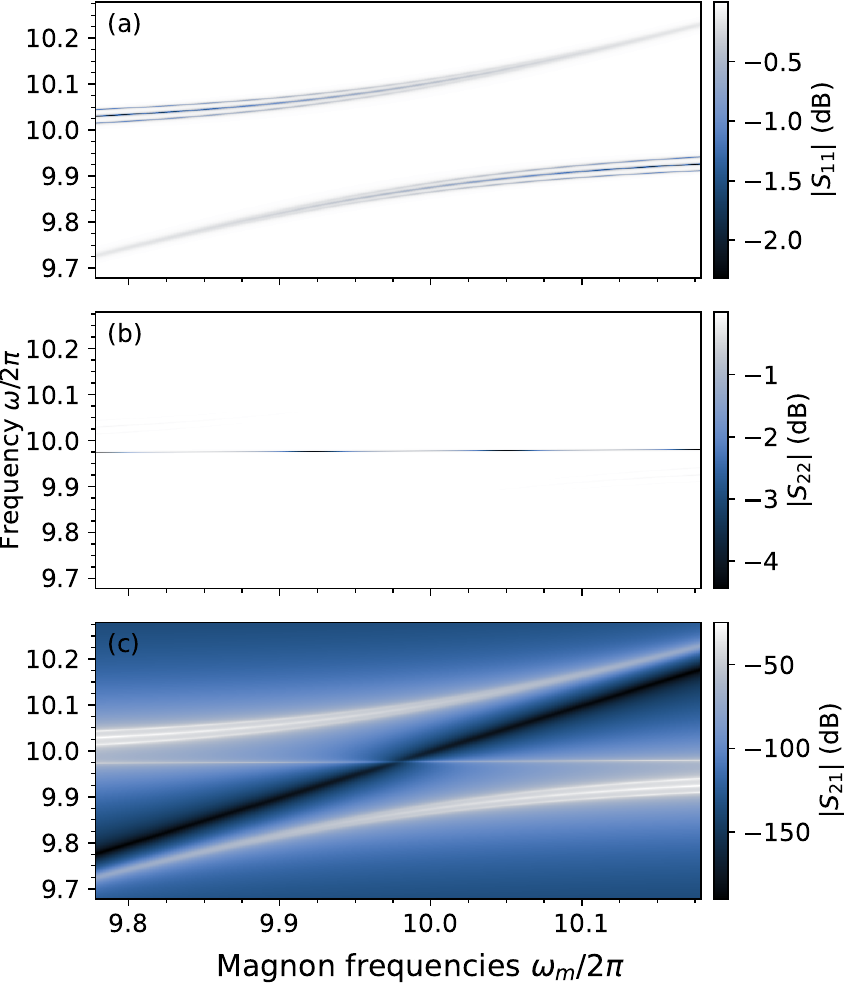}
    \phantomsubfloat{\label{fig:N4-H1:iot:S11}}
    \phantomsubfloat{\label{fig:N4-H1:iot:S22}}
    \phantomsubfloat{\label{fig:N4-H1:iot:S21}}
    \vspace{-3em}
    \caption{Reproduction of figure 8 of the main text, corresponding to a crystal of $N=4$ lattice sites with a YIG sphere removed from the last lattice site. The parameters used are identical to those used for the fit of figure 6 of the main text but employing the rotating wave approximation.}
    \label{fig:N4-H1:iot:Sparams}
\end{figure}

\section{\label{app:qbh-examples}Quadratic bosonic Hamiltonian formalism: further details}
In this appendix, we provide examples of using the formalism introduced in section V.A.1 of the main text to find the normal modes of quadratic bosonic Hamiltonians (QBHs). 
We recall that the objective is to rewrite a QBH $\mathcal H$ in a block-matrix form by identifying the dynamical matrix
\begin{equation}
    \label{eq:G:def-reminder}
    G=\begin{bmatrix} 
    K & -\Delta\\ 
    \Delta^* & -K^*
    \end{bmatrix}.
\end{equation}
in terms of the matrices $K$ and $\Delta$, see equations (53) and (54) of the main text. 
A simple two-mode example has already been discussed in section II.A of the main text. 
Here, we will focus on lattice generalisation of this example. In this appendix, we note $I_N$ the identity matrix of size $N \times N$, and $T_N$ the $N \times N$ matrix with ones on the first lower diagonal. The matrix $T_N$ will be useful for describing the hoppings between lattice sites $j$ and $j-1$ of the form $a_j^\dagger a_{j-1}$ (shifting to the left). Conversely, $T_N^\dagger$ will correspond to hopping terms of the form $a_j^\dagger a_{j+1}$ (shifting to the right).

\subsection{Photo-magnonic crystal}
The first example we consider is the photo-magnonic-crystal introduced in section III of the main text. While in section III of the main text the rotating wave approximation has been used to drop the counter-rotating terms, here we will keep them for illustrative purposes. The QBH of equation (31) of the main text reads 
\begin{equation}
\begin{aligned}
    \frac{{\cal H}^{(n)}}{\hbar}
    &=  \sum_{j=1}^{N} \qty(\omega_a a_j^\dagger a_j + \omega_m {m}_j^\dagger {m}_j)
    -\sum_{j=1}^{N}(t a_{j} a_{j+1}^\dagger +\hc)\\
    &\quad+ \sum_{j=1}^{N} (g a_{j}{m}_{j+n}^\dagger+ g a_j {m}_{j+n} + \hc).
\end{aligned}
\end{equation}
Following the operator ordering of equation (49) of the main text, we define 
\begin{equation}
\label{eq:app:def-row-vector}
\Phi^\dagger=\mqty[a_1^\dagger & a_2^\dagger & \dots & m_1^\dagger & m_2^\dagger & \dots & \dots & m_N],
\end{equation}
i.e. we stack all the photon creation operators first, then all the magnon creation operators, and so on for the annihilation operators.

We now detail a systematic method to find the hopping and pairing matrices. The idea is to rewrite the operators in the Hamiltonian in a manner allowing to directly read off the matrix elements. In particular, we will order pair of operators such that the leftmost correspond to a row vector, and the rightmost to a column vector. For instance, for the diagonal terms, we first write
\begin{align}
    \sum_{j} \omega_a a_j^\dagger a_j  
    &= \frac{1}{2}\sum_{j} \omega_a\qty( a_j^\dagger a_j + a_j a_j^\dagger) 
\end{align}
where the commutation relations $[a_j,a_{j}^\dagger=1$ has been used, and a constant energy offset has been neglected. The ordering of operators above suggests defining a row vector $\mqty[a_j^\dagger & a_j]$, and we write
\begin{align}
    \sum_{j} \omega_a a_j^\dagger a_j  
    %&= \frac{1}{2}\sum_{j} \omega_a\qty( a_j^\dagger a_j + a_j a_j^\dagger) \\
    &= \frac{1}{2} \sum_{j} \mqty[a_j^\dagger\\a_j]^t \mqty[\omega_a&\\&\omega_a] \mqty[a_j\\a_j^\dagger]%\\
    %&= \Phi^\dagger \mqty[\omega_a I_N &\\ & 0_N\\& &\omega_a I_N\\ &&&0_N] \Phi
\end{align}
This expression can be rewritten as a matrix-block multiplication involving the entire row vector $\Phi$ of operators as
\begin{align}
    \sum_{j} \omega_a a_j^\dagger a_j  
    %&= \frac{1}{2}\sum_{j} \omega_a\qty( a_j^\dagger a_j + a_j a_j^\dagger) \\
    %&= \frac{1}{2} \sum_{j} \mqty[a_j^\dagger\\a_j]^t \mqty[\omega_a&\\&\omega_a] \mqty[a_j\\a_j^\dagger]\\
    &= \Phi^\dagger \mqty[\omega_a I_N &\\ & 0_N\\& &\omega_a I_N\\ &&&0_N] \Phi
\end{align}
where $0_N$ designates the $N \times N$ whose elements are all zero. A similar calculation with the magnon terms gives 
\begin{align}
    \sum_{j}\omega_m m_j^\dagger m_j
    &= \Phi^\dagger \mqty[0_N\\ &\omega_m I_N \\ & & 0_N\\& & &\omega_m I_N] \Phi
\end{align}
Thus, we deduce that the free part of the Hamiltonian can be written
\begin{equation}
    \label{eq:app:Hn-free}
    \sum_{j} \qty(\omega_a a_j^\dagger a_j + \omega_m {m}_j^\dagger {m}_j) 
    = \frac{1}{2}\Phi^\dagger
    I_2 \otimes \mqty[\omega_a I_N\\&\omega_m  I_N]
    \Phi
\end{equation}

For the interaction terms, first note that
\begin{align}
    \mqty[a_j^\dagger & a_{j+1}^\dagger] T_2 &= \mqty[a_j^\dagger & a_{j+1}^\dagger] \mqty[0&0\\1&0] = \mqty[a_{j+1}^\dagger & 0]\\
    T_2^\dagger \mqty[a_j \\ a_{j+1}] &= \mqty[0&1\\0&0]\mqty[a_j \\ a_{j+1}] = \mqty[a_{j+1}\\ 0]
\end{align}
where we recall that $T_N$ is the $N\times N$ matrix with ones one the first lower diagonal.
In particular, we have
\begin{align}
    \mqty[a_j^\dagger & a_{j+1}^\dagger] T_2 \mqty[a_j \\ a_{j+1}] =  \mqty[a_{j+1}^\dagger & 0] \mqty[a_j \\ a_{j+1}] &= a_{j+1}^\dagger a_j,\\
    \mqty[a_j^\dagger & a_{j+1}^\dagger] T_2^\dagger \mqty[a_j \\ a_{j+1}] = \mqty[a_j^\dagger & a_{j+1}^\dagger] \mqty[a_{j+1}\\ 0] &= a_j^\dagger a_{j+1}
\end{align}
and similarly
\begin{align}
    \mqty[a_j & a_{j+1}] T_2 \mqty[a_j^\dagger \\ a_{j+1}^\dagger] =  \mqty[a_{j+1} & 0] \mqty[a_j^\dagger \\ a_{j+1}^\dagger] &= a_{j+1} a_j^\dagger,\\
    \mqty[a_j & a_{j+1}] T_2^\dagger \mqty[a_j^\dagger \\ a_{j+1}^\dagger] = \mqty[a_j & a_{j+1}] \mqty[a_{j+1}^\dagger\\ 0] &= a_j a_{j+1}^\dagger.
\end{align}
Let us see how these relations can be used for the photon-photon hopping term. As before, we will rewrite the operators in a symmetric manner, and order them by row vectors first, and then column vector. We find
\begin{widetext}
    \begin{align}
    -\sum_{j}(t a_{j} a_{j+1}^\dagger +\hc)
    &= - \frac{1}{2}\sum_j(ta_{j+1}^\dagger a_j + t^* a_j^\dagger a_{j+1})
    - \frac{1}{2}\sum_j(ta_j a_{j+1}^\dagger + t^* a_{j+1} a_j^\dagger )\\
    &=
    - \frac{1}{2}\sum_j 
    \mqty[a_j^\dagger & a_{j+1}^\dagger]
    (tT_2 +t^* T_2^\dagger)
    \mqty[a_j \\ a_{j+1}]
    - \frac{1}{2}\sum_j 
    \mqty[a_j & a_{j+1}] 
    (t T_2^\dagger + t^* T_2) 
    \mqty[a_j^\dagger \\ a_{j+1}^\dagger]\\
    &=
    - \frac{1}{2}\sum_j 
    \mqty[a_j^\dagger & a_{j+1}^\dagger & a_j & a_{j+1}]
    \mqty[tT_2 +t^* T_2^\dagger & 0_N\\ 0_N & t T_2^\dagger +t^* T_2]
    \mqty[a_j \\ a_{j+1}\\ a_j^\dagger \\ a_{j+1}^\dagger]\\
    \label{eq:app:Hn-t}
    &= -\Phi^\dagger
    \mqty[
    tT_N+t^*T_N^\dagger \\
    & 0_N\\
    &&tT_N^\dagger + t^*T_N\\
    &&&0_N
    ]
    \Phi
\end{align}
\end{widetext}
%We can clearly see that the terms $a_{j+1}^\dagger a_j$ and $a_j a_{j+1}^\dagger$ transferring an excitation to the right side of the chain are associated with the operators $T_N$. Similarly, the propagation 

Finally, note that $\Phi^\dagger T_N^n$ shifts the components of the row vector $\Phi^\dagger$ $n$ times (and similarly for ${T_N^\dagger}^n \Phi$). We deduce that interaction term is
\begin{widetext}
\begin{align}
    \label{eq:app:Hn-g}
    \sum_{j} (g a_{j}{m}_{j+n}^\dagger+ g a_j {m}_{j+n} + \hc)
    = \frac{1}{2} \Phi^\dagger
    \mqty[
    0_N & g^* {T_N^\dagger}^n & 0_N & g^* {T_N^\dagger}^n\\
    gT_N^n & 0_N & g^*T_N^n & 0_N\\
    0_N & g {T_N^\dagger}^n & 0_N & g{T_N^\dagger}^n\\
    g{T_N^\dagger}^n & 0_N & g^* {T_N^\dagger}^n & 0_N
    ]
    \Phi.
\end{align}
\end{widetext}
Combining \cref{eq:app:Hn-free,eq:app:Hn-t,eq:app:Hn-g}, we deduce 
\begin{equation}
    \label{eq:app:Hn:K}
    K = \mqty[
    \omega_a I_N -tT_N+-t^*T_N^\dagger & g^* {T_N^\dagger}^n\\
    gT_N^n & \omega_m I_N
    ]
\end{equation}
and \begin{equation}
    \label{eq:app:Hn:Delta}
    \Delta = \mqty[
    0_N & g^* {T_N^\dagger}^n\\
    g^*T_N^n & 0_N
    ].
\end{equation}

\paragraph{Examples.} Let us conclude with two examples for $N=2$ lattice sites. We first consider the case $n=0$, and up to an energy offset we have
\begin{widetext}
\begin{equation}
\label{eq:H0N2}
    \frac{{\cal H}^{(n=0)}}{\hbar}= \frac{1}{2}
    \mqty[a_{1}^{\dagger} \\ a_{2}^{\dagger} \\ m_{1}^{\dagger} \\ m_{2}^{\dagger} \\ a_{1} \\ a_{2} \\ m_{1} \\ m_{2}]^{t} 
\mqty[
{\mqty(\omega_{a}&t^{*}\\t&\omega_{a})} & {\mqty(g^{*}& \\ & g^{*})} & & {\mqty(g^{*}\\& g^{*})}\\
{\mqty(g& \\ & g)} &{\mqty(\omega_{m}\\&\omega_{m})} & {\mqty(g^{*}\\& g^{*})}\\
&{\mqty(g\\& g)} &{\mqty(\omega_{a}&t\\t^{*}&\omega_{a})} & {\mqty(g&\\ & g)}\\
{\mqty(g\\& g)} & & {\mqty(g^{*}&\\ & g^{*})} &{\mqty(\omega_{m}\\&\omega_{m})}
]
\mqty[a_{1}\\ a_{2}\\ m_{1}\\ m_{2}\\ a_{1}^{\dagger}\\ a_{2}^{\dagger}\\ m_{1}^{\dagger}\\ m_{2}^{\dagger}]
= \frac{1}{2} \Phi^\dagger G \tau_3 \Phi
\end{equation}
\end{widetext}
We recall that for a $2N \times 2N$ matrix $M$, \(M\tau_3\) has the effect of switching the sign of the $N \times N$ blocks in the second column of $M$. Thus, comparing \cref{eq:G:def-reminder,eq:H0N2}, we read off
\begin{equation}
    K^{(0)} = \mqty[
{\mqty(\omega_{a}&t^{*}\\t&\omega_{a})} & {\mqty(g^{*}& \\ & g^{*})} \\
{\mqty(g& \\ & g)} &{\mqty(\omega_{m}\\&\omega_{m})} \\
] %= I_2 \otimes \qty(\mqty(\omega_a&\\&\omega_m) \otimes I_2)
\end{equation}
and \begin{equation}
    \Delta^{(0)} = \mqty[
 & {\mqty(g^{*}\\& g^{*})}\\
{\mqty(g^{*}\\& g^{*})} &
]
\end{equation}

Let us consider the case $n=1$ next. The difference with $n=0$ is that the terms in $g$ are now shifted to the right. We have 
\begin{equation}
    K^{(1)} = \mqty[
{\mqty(\omega_{a}&t^{*}\\t&\omega_{a})} & {\mqty(&g^{*} \\ & )} \\
{\mqty(&\\ g &)} &{\mqty(\omega_{m}\\&\omega_{m})} \\
] %= I_2 \otimes \qty(\mqty(\omega_a&\\&\omega_m) \otimes I_2)
\end{equation}
and \begin{equation}
    \Delta^{(1)} = \mqty[
 & {\mqty( & g^{*}\\ & )}\\
{\mqty( & \\ g^{*} &)} &
].
\end{equation}

\subsection{Photo-magnonic crystal in momentum space}
We now describe a ``shortcut" to the momentum space formulation of the hopping and pairing matrices. The trick is to notice that
\begin{align}
\sum_j a_j^\dagger a_{j+1} &=
    \frac{1}{\sqrt{N}}
    \frac{1}{\sqrt{N}} \sum_j \sum_{k,k'} e^{ijk} e^{-i(j+1)k'} a_{k}^\dagger a_{k'}\\
    &=
    \sum_{k,k'} a_{k}^\dagger a_{k'} e^{-ik'}
    \frac{1}{N} \sum_j e^{ij(k-k')}\\
    \label{eq:app:tmp12323}
    &=
    \sum_k a_k^\dagger a_{k}e^{-ik}
\end{align}
where we used the definition of equation (40) of the main text for the momentum space operators.
By analogy with \cref{eq:app:def-row-vector}, we define the momentum space row vector
\begin{equation}
\Phi^\dagger_k=\mqty[a_k^\dagger & m_k^\dagger & a_{-k} & m_{-k}].
\end{equation}
We can then rewrite \cref{eq:app:tmp12323} as
\begin{align}
    \sum_j a_j^\dagger a_{j+1} &= \frac{1}{2}\sum_k \qty(a_k^\dagger a_{k}e^{-ik} + a_{-k}^\dagger a_{-k}e^{ik})\\
    &= \sum_k \mqty[a_k^\dagger \\ m_k^\dagger \\ a_{-k} \\ m_{-k}]^t
    \mqty[e^{-ik}\\&0\\&&e^{+ik}\\&&&0]
    \mqty[a_k \\ m_k \\ a_{-k}^\dagger \\ m_{-k}^\dagger]
\end{align}
This is to be compared with the position-space formulation,
\begin{widetext}
    \begin{align}
    \sum_j a_j^\dagger a_{j+1} &= 
    \sum_j \mqty[a_j^\dagger & a_{j+1}^\dagger]T_2^\dagger\mqty[a_j \\ a_{j+1}]
    =
    \frac{1}{2}\sum_j 
    \mqty[a_j^\dagger & a_{j+1}^\dagger & a_j & a_{j+1}]
    \mqty[T_2^\dagger & 0_N\\ 0_N & T_2]
    \mqty[a_j \\ a_{j+1}\\ a_j^\dagger \\ a_{j+1}^\dagger]\\
    &= -\Phi^\dagger
    \mqty[
    T_N^\dagger \\
    & 0_N\\
    &&T_N\\
    &&&0_N
    ]
    \Phi
\end{align}
\end{widetext}
Hence, we deduce that we have the mapping $T_N \mapsto e^{ik}$ in momentum space. Similarly, $T_N^n \mapsto e^{ikn}$.

In particular, applying this mapping to \cref{eq:app:Hn:K}, we obtain
\begin{align}
    \label{eq:app:Hn:K:momentum}
    K &= \mqty[
    \omega_a -te^{ik}+-t^*e^{-ik} & g^* e^{-ikn}\\
    ge^{-ikn} & \omega_m
    ]\\
    &= \mqty[
    \omega_a -2\abs{t} \cos(k+\phi) & g^* e^{-ikn}\\
    ge^{ikn} & \omega_m
    ]
\end{align}
with $\phi = \arg \; t$. This is the result claimed in equation (44) of the main text, after setting $\omega_a=\omega_m=0$.

\subsection{Bosonic Kitaev chain}
The Hamiltonian of the bosonic Kitaev chain (BKC) reads
\begin{align}
    {\cal H}_{BKC} = \frac{1}{2}\sum_j \qty(ita_j a_{j+1}^\dagger+ i\delta a_j^\dagger a_{j+1}^\dagger + \hc)
\end{align}
with $t,\delta$ real parameters. Since there is only one degree of freedom per lattice site, we define \begin{equation}
\Phi^\dagger=\mqty[a_1^\dagger & a_2^\dagger & \dots & a_1 & \dots & & a_N].
\end{equation}
As we have done before, we will rewrite the interaction terms in an order highlighting the matrix elements. We have
\begin{widetext}
    \begin{align}
    \sum_j(ita_j a_{j+1}^\dagger - it a_j^\dagger a_{j+1})
    &=
    \frac{1}{2} \sum_j(- it a_j^\dagger a_{j+1} + it a_{j+1}^\dagger a_j + ita_j a_{j+1}^\dagger - it a_{j+1} a_j^\dagger )\\
    &= \sum_j \mqty[a_j^\dagger & a_{j+1}^\dagger & a_j & a_{j+1}]
    \mqty[-itT_2^\dagger +it T_2 & 0_N\\ 0_N & itT_2^\dagger -it T_2]
    \mqty[a_j \\ a_{j+1}\\ a_j^\dagger \\ a_{j+1}^\dagger]\\
    \sum_j(i\delta a_j^\dagger a_{j+1}^\dagger - i\delta a_j a_{j+1})
    &=
    \frac{1}{2} \sum_j(i\delta a_j^\dagger a_{j+1}^\dagger + i\delta a_{j+1}^\dagger a_j^\dagger - i\delta a_j a_{j+1} - i\delta a_{j+1} a_j)\\
    &= \sum_j \mqty[a_j^\dagger & a_{j+1}^\dagger & a_j & a_{j+1}]
    \mqty[0_N & i\delta T_2^\dagger +i\delta T_2 \\ -i\delta T_2^\dagger -i\delta T_2&0_N]
    \mqty[a_j \\ a_{j+1}\\ a_j^\dagger \\ a_{j+1}^\dagger]
\end{align}
\end{widetext}
We can now read off the hopping and pairing matrices as
\begin{align}
    K &=\frac{it}{2}(T_N - T_N^\dagger)\\
    \Delta &= \frac{i\delta}{2}(T_N + T_N^\dagger)
\end{align}
as claimed in the main text.

\bibliography{references}